# The existence of species rests on a metastable equilibrium between inbreeding and outbreeding.

*An essay on the close relationship between speciation, inbreeding and recessive mutations.*

## Etienne Joly

Toulouse, December 2011 (V5)

Address:
1) CNRS; IPBS (Institut de Pharmacologie et de Biologie Structurale); 205 route de Narbonne, F-31077 Toulouse, France
2) Université de Toulouse; UPS; IPBS; F-31077 Toulouse, France

Tel: 33-561 17 58 70

Email: atnjoly@mac.com,

**Abstract:**

**Background**: Speciation corresponds to the progressive establishment of reproductive barriers between groups of individuals derived from an ancestral stock. Since Darwin did not believe that reproductive barriers could be selected for, he proposed that most events of speciation would occur through a process of separation and divergence, and this point of view is still shared by most evolutionary biologists today.

**Results**: I do, however, contend that, if so much speciation occurs, the most likely explanation is that there must be conditions where reproductive barriers can be directly selected for. In other words, situations where it is advantageous for individuals to reproduce preferentially within a small group and reduce their breeding with the rest of the ancestral population. This leads me to propose a model whereby new species arise not by populations splitting into separate branches, but by small inbreeding groups "budding" from an ancestral stock. This would be driven by several advantages of inbreeding, and mainly by advantageous recessive phenotypes, which could only be retained in the context of inbreeding. Reproductive barriers would thus not arise as secondary consequences of divergent evolution in populations isolated from one another, but under the direct selective pressure of ancestral stocks. Many documented cases of speciation in natural populations appear to fit the model proposed, with more speciation occurring in populations with high inbreeding coefficients, and many recessive characters identified as central to the phenomenon of speciation, with these recessive mutations expected to be surrounded by patterns of limited genomic diversity.

**Conclusions**: Whilst adaptive evolution would correspond to gains of function that would, most of the time, be dominant, this type of speciation by budding would thus be driven by mutations resulting in the advantageous loss of certain functions since recessive mutations very often correspond to the inactivation of a gene. A very important further advantage of inbreeding is that it reduces the accumulation of recessive mutations in genomes. A consequence of the model proposed is that the existence of species would correspond to a metastable equilibrium between inbreeding and outbreeding, with excessive inbreeding promoting speciation, and excessive outbreeding resulting in irreversible accumulation of recessive mutations that could ultimately only lead to extinction.

**Foreword:**

2009 was the Darwin year, celebrating the 200th anniversary of Charles Darwin's birth, and 150 years since the publication of his fabulous milestone book, 'The Origin of Species' (to which I will subsequently refer to as 'The Origin'). For a few years, I have been inhabited by a nagging ethical concern : how would humans deal with a situation where a group of individuals found themselves fertile among one another, but with limited fertility with the rest of the human race ? In other words, could speciation occur within the human race ? This concern sprouted from the idea that chromosomal rearrangements seemed to me like a very probable initial step of a speciation process, since systematic surveys of the human populations have actually shown that such rearrangements are relatively frequent ( frequency of the order of 1/1000, [1] ). Furthermore, given the success of the human race, having resulted in the huge numbers of human beings currently living on our planet, and given the amazing propensity of nature to generate new species, I felt that the chances must be quite high that speciation could occur within the human population. Most scientists concerned with evolution and speciation would probably not share those concerns because the commonly held view is that speciation is most often allopatric, i.e. it occurs when populations of individuals evolve separately from one another for a sufficiently long time that they would no longer breed efficiently with one another when they are reunited. The mobility of modern humans would thus preclude this type of phenomenon.

The year 2009 has seen the publication of a plethora of review articles on the subject of evolution and speciation, which have allowed me to start catching up on these vast subjects, and to mature my reflections on the mechanisms involved in speciation. The reading of these reviews has also allowed me to confirm that the ideas I have developed are in disagreement with the generally held views, i.e. that allopatric speciation is the most common and probable route for the appearance of new species. All the ideas developed in this essay are, however, relatively simple, and most of them are related to previously published works. But so much work has already been published on evolution and speciation that an autodidactic newcomer such as myself could not hope to read, let alone understand and remember all the primary papers published previously on evolution and speciation.

Because, as a rule, I have adopted the principle of never citing a paper that I have not read, numerous times during the writing of this essay, I have found myself unable to decide what specific paper to cite as the appropriate original source of a particular concept or observation. Although I have tried to read as many primary papers as I could rather than reviews, I found that I simply could not read everything. In addition many papers were not available to me in our institute's library or freely online (As another rule, I refuse to pay for online access, because I firmly believe that all primary research papers should be freely available to all), and this problem was even more acute for books. In such situations when I had not managed to read the primary texts (for whatever reason), I have very often chosen to cite the very comprehensive et quite recent reference book "Speciation" by Coyne and Orr (2004), and to refer to it as 'C&O', with the indication of the appropriate chapter or page number.

Probably because inbreeding does not have very good press, including among evolutionary biologists, despite reading extensively about speciation and evolution, it is only very recently, more than a year after completing the initial version of this assay, that I have finally come across certain papers which are related to populations structures and/or to the benefits of inbreeding, and were thus highly relevant to the ideas developed in this manuscript (for example, the works of W. Shields [2], S. Wright [3, 4] or H. Carson [5], which are now duly cited and discussed in the current version). If I have failed to acknowledge other previous works developing ideas related to those put forward here, the reader can be assured that this was not done maliciously but simply as a result of my relative naivety on the subject. I do, however, hold the firm conviction that, if some of the ideas developed in this essay prove to be correct and relatively novel, it was only rendered possible because of this naivety.

**Introduction:**

Among the myriad of reviews and articles that have been written about "The Origin of Species" by Charles Darwin, a very large proportion underlines the fact that, despite the title of his book, what Darwin established 150 years ago was the mechanism of adaptive evolution by the process of natural selection, but that he failed to provide answers to the many questions that surround the origin of species.

One of the important reasons for this failure was related to an issue to which he alluded to repeatedly in his book, which is that species are basically impossible to define. The main problem, which he acknowledged himself, and stays whole today, lies in the fuzzy limit between species and varieties: "*From these remarks it will be seen that I look at the term species, as one arbitrarily given for the sake of convenience to a set of individuals closely resembling each other, and that it does not essentially differ from the term variety, which is given to less distinct and more fluctuating forms. The term variety, again, in comparison with mere individual differences, is also applied arbitrarily, and for mere convenience sake.*" (The Origin, p. 52 mid Ch II).

One of the most important concepts that derives from the work of Darwin is that the process of life is one of constant evolution, which explains why so few of the life forms that occupied the earth 20 millions ago are still around today. The somewhat uncomfortable but inescapable conclusion from this is that the existence of every single one of the millions of species that surround us, including ours, must also be transitory, and this probably contributes to the difficulty that many humans have in accepting the theory of evolution, in addition to the fact that it also brings serious questions as to the existence of an almighty God. The processes of evolution and speciation are, however, very slow ones, and the 5000 years of human history (which is usually defined as starting with the invention of writing, i.e. since humans first started scribbling cuneiform signs in Mesopotamia, or hieroglyphs in Egypt) do not amount to even a tick on the clock of evolutionary times, and to our human eyes, the stability of the world thus appears as if it should stay the same for ever, and so with the species that occupy it. The fact that species are not stable entities, but in constant evolution is another factor that adds to the difficulty of defining them.

Initially, species were recognised and defined by naturalists and palaeontologists mostly in relation to their anatomical features, and it is on the basis of these features that Linnaeus opened the way to taxonomic classifications in the middle of the 18[th] century. Regarding taxonomic definitions of species, dogs are a particularly telling example of the fact that, when considering species based on morphological traits, certain organisms can differ greatly in their anatomy and still belong to the very same species.

It is some hundred years after Linnaeus, and well after Darwin and Wallace had laid down the principles of natural selection, that the biological species concept emerged, which introduced the notion of the central importance of fertility, and of the capacity to hybridize, in the definition of species. Today, the most popular definition of biological species is that proposed by Ernst Mayr in 1942, as "*groups of actually or potentially interbreeding natural populations, which are reproductively isolated from other such groups*".

The first thing to underline in this definition is that species are not defined as standalone entities, but always in relation to other species (which provides some rationale, albeit retroactive, to the fact that the singular of species is species and not specie, which refers to coined money). The second important point about the definition of biological species is about the difficulty of implementing it. Indeed, many closely related species still show some degree of fertility with one another. For example, many species which do not detectably hybridize in the wild can produce perfectly fit and fertile offspring under experimental conditions. Furthermore, even if one was to set a threshold value for the degree of hybridisation between two separate populations to consider them as separate species, the degree of mixing of populations can vary greatly depending on circumstances such as population densities, or environmental fluctuations such as clarity of waters for certain fish that use visual clues to recognise their own kin.

More recently, the amazingly fast progress in molecular biology has allowed geneticists to follow and quantify the occurrence of gene flow between divergent populations, and this is often taken into consideration when discussing whether two populations represent "good species" or not. On the subject of gene flow, one can, however, take the slightly provocative stance that gene flow can never reach the absolute zero, which is related to the fact that all organisms are based on the same genetic code. Indeed, there is more and more evidence accumulating about the prominence of horizontal gene transfer between all sorts of organisms, mediated by varied mechanisms that can involve viruses, and particularly retroviruses, or possibly by incorporation of whole organisms or just DNA. And transgenesis is another recent progress of technology which reinforces the notion that "zero gene flow" is only a theoretical limit towards which speciation can tend.

Considering the various difficulties one encounters in trying to define species, I will not engage in the somewhat sterile debate (excuse the bad pun ) of what constitutes 'good species', or rather of when two groups of animals can be considered as separate species. And even less in the consideration of whether asexual organisms can be grouped into species. Rather, I will only engage in a reflection within the 'biological species concept', as initially defined by Ernst Mayr. Furthermore, in considering only groups of organisms that reproduce sexually, I will focus on the phenomenon of speciation. Indeed, although species are well nigh impossible to define, one cannot dispute that speciation occurs, i.e. the fact that, starting from an ancestral population, some groups of animals will start breeding more among one another than with the rest of the population, and will progressively acquire a range of characters that sets them apart from the original group. This, in fact, happens everywhere and all the time around us, in wild and domestic species and is the reason for the appearance of particular characters, or traits, that lead to the definition of subtypes, morphotypes, races, varieties, subspecies, species ....

Although the possibility that speciation can occur without complete separation of two populations seems to be gaining more and more proponents [6-8], the most prevalent view about speciation today remains that geographical separation is the most likely mechanism for the origin of species: independent adaptation to different environments will push the evolution of the two populations sufficiently apart that their offspring would be unfit because outbreeding between the two populations will result in the disruption of co-adapted gene complexes. The term used to describe this type of speciation is allopatric, as opposed to sympatry, where ancestral and descendant species coexist in the same environment (or parapatry if they exist side by side, with a hybridisation zone in between). If two populations having evolved separately come back in contact later on, the intermediate phenotype of their offspring could make them unfit for either environment, and this would then provide the selective pressure for the selection of additional reproductive barriers, in a process called reinforcement, and often referred to as 'the Wallace effect'. Indeed, the earliest promoter of the view that reinforcement could occur under the pressure of natural selection was undoubtedly Alfred Wallace, who disagreed with Darwin's views that reproductive isolation could not possibly result from natural selection: "*The sterility of first crosses and of their hybrid progeny has not been acquired through natural selection*" (The Origin, Summary of Hybridism chapter). This point was a subject of written exchanges and arguments in private correspondence between the two around 1858, 10 years after their joint communication to the Linnean Society in July 1858, but Wallace formally published his views only in 1889, some twenty year later, in chapter VII of his book called Darwinism.

On the subject of allopatry versus sympatry, I do take a very divergent view to that adopted by a majority of evolutionary biologists to this day. Rather, I choose to follow Wallace's path against Darwin's in thinking that natural selection plays a direct role in promoting the reproductive isolation that defines species, and I shall actually venture some steps further than Wallace, and will advocate in the following pages that natural selection can act on the very first stages of reproductive isolation, and not just on reinforcement after divergence has taken place. Such views were also, but temporarily, those of Theodozius Dobzhansky early in his career [9], when he stated that " ...*Occurence of hybridisation between races and species constitutes a challenge to which they may respond by <u>developing</u> or strengthening isolating mechanisms that would make hybridisation difficult or impossible*". Worthy of note, Darwin must also have had a similar initial intuitions, as can be inferred from the following statement: "*At one time it appeared to me probable, as it has to others, that the sterility of first crosses and of hybrids might have been slowly acquired through the natural selection of slightly lessened degrees of fertility*" found in chapter IX of the editions of The Origin after 1866.

As for myself, I contend that, if there is so much speciation, i.e. mechanisms, be they genetic or not, causing reproductive isolation evolving everywhere, all the time, it must be because there can be basic, fundamental selective advantages for subgroups of individuals to breed preferentially among one another, and reduce their capacity to hybridize with the rest of the population. As will become clearer later on, I adopt the point of view that, if species arise as a result of direct selective pressures, then most events of speciation, even in their earliest steps, must take place as a result of the pressure of natural selection, and must therefore occur in settings of sympatry, or at least parapatry rather than allopatry since, under allopatric conditions, there can be no selective pressure to reduce breeding with individuals that are seldom encountered.

In this regard, one remarkable observation is that, inasmuch as legions of well documented examples exist where divergent types of varieties have been generated under domestication, very few, if any, examples exist where truly significant reproductive isolation has been witnessed. Thomas Huxley, one of the earliest and most dedicated advocates of Darwin's theory, actually referred to the fact that domestic varieties did not undergo speciation as 'Darwin's weak point'. But this can find an explanation within the frame of the model proposed here, since domestic varieties evolve in the absence of pressure from the ancestral stock, under what is effectively equivalent to allopatric conditions. This point of view is supported by the set of data collated by Rice and Hostert [10] from a large number of studies aimed at studying the evolution of reproductive isolation under experimental conditions. The conclusion reached by these authors is that it is neither allopatry or bottlenecks that promote reproductive isolation, but rather the occurrence of multifarious divergent selection, in conjunction, or followed by, reinforcement, as demonstrated by experiments where hybrids are experimentally eliminated.

Advocating that it can be advantageous for a handful of individuals to breed preferentially among one another rather than with the rest of the population is, however, very counter-intuitive because it is basically equivalent to advocating that inbreeding can bring on a selective advantage. And it is common knowledge to almost everyone that inbreeding can be disastrously disadvantageous, whereas hybrid vigour almost always brings your direct descendants a selective advantage.

I will, however, endeavour to demonstrate that inbreeding can have numerous advantages, particularly in the long run, and that the selective advantages brought about by inbreeding are the main driving force behind the phenomenon of speciation, whilst short term advantages of panmixia will come at a cost of accumulation of recessive mutations that will eventually represent a threat for the survival of species.

### I ) Potential advantages of inbreeding
We will hence start our reflection by asking ourselves what the advantages of inbreeding could be. If one carries out a simple literature search for the single keyword "inbreeding" on a server such as Google scholar, one can rapidly identify tens of thousands of citations. Upon rapid examination, it is actually striking to find that, in over 90% of those, the word inbreeding is systematically associated with either *depression*, *cost* or *avoidance*, compared to only a handful of papers where the potential benefits of inbreeding have actually been objectively considered. One important point to make here is that inbreeding is different from incest. Incest is the mating of extremely closely related individuals, usually sharing half of their genome (such as parent-child or brother-sister), or at least a quarter (such as grand-parent with grand-child). On the other hand, inbreeding results from the pairing of individuals that are more closely related than if they were picked at random from the surrounding population. What many studies have labelled as 'inbreeding avoidance' actually corresponded to 'incest avoidance', and we will see that, in many natural populations, although there are numerous examples of mechanisms to prevent selfing or incest, multiple strategies also exist that promote some degree of inbreeding.

I have actually identified so few papers that have constructively considered the positive aspects of inbreeding that it is possible to summarise them in just a few sentences. The notion that "selfing" is potentially advantageous can be traced back to R. Fisher in 1941 [11]. Around the same time, the works of S. Wright underlined that natural populations are seldom panmictic, but usually structured in partially subdivided, and more inbred demes. These divisions not only help to maintain more allelic and phenotypic diversity, but can also favour evolution and promote speciation [3, 4, 12]. In 1959, H. Carson put forward a model whereby speciation is promoted in small (marginal) inbred populations, whilst large, more outbred populations, will senesce, i.e. increasingly rely on heterosis, and progressively lose their capacity to evolve and to give rise to new 'young' species [5]. Many of the ideas developed in that article are very closely related to the ones I am presenting here. Because he adopted the view that speciation most often occurred through allopatry, later works by Carson focused on founder events, for which he is nowadays better known and this particular paper actually received surprisingly little attention from people trying to establish models of speciation (for example, it is not even cited in the book Speciation by C&O). Some twenty years later, based on the observation that quails mated preferentially with their cousins, P. Bateson produced the concept of optimal outbreeding [13-15], supported the following year by the work of Price and Waser on a wildflower [16]. Very soon afterwards, W. Shields put forward the theory that philopatry, i.e. the tendency of individuals of many species to breed near their birthplace, was related to the advantages conveyed by inbreeding, and in particular the capacity of inbreeding to maintain successful gene combinations [2]. Outside of the concept of crisis inbreeding developed by C. Grobbelaar in 1989 [17], and more recent works on the somewhat unexpected long term reproductive success of consanguineous marriages [18-20], I have so far failed to identify other works exploring the benefits of inbreeding that would contribute significantly to the ideas developed here (More recent but less directly relevant papers on the subject of inbreeding can be found in the 1993 book of collected works entitled 'Natural History of Inbreeding' [[21]] or in a 2006 paper by Kokko and Ots [[22]]). In the following pages, I will thus try to present and summarize the various advantages which can be found to inbreeding.

1) Inbreeding is necessary for the expression of advantageous recessive phenotypes.

This undisputable advantage of inbreeding is the one which is most central to the model presented. In the first place, I thus felt that it was important to clearly define what is meant by dominant, recessive, co-recessive and co-dominant phenotypes. The laws of genetics initially discovered by Gregor Mendl at the end of the 19th century concerned the transmission of characters in diploid organisms. Starting from homogenous stocks of peas, what he established was that all F1 had homogenous phenotypes (first law), but that those segregated in F2 generations, according to the well known ¼ - ¾ ratios for recessive versus dominant phenotypes (second law). A further observation was that different characters segregated independently from one another (third law). The considerations of linkage between genes and of genetic distance would be discovered by others, at the beginning of the 20th century, after the 're-discovery' of Mendel's results.

Conversely to Mendelian genetics, which concern genes that remain identical through successive generations, the process of evolution involves mutations, which correspond to changes occurring in the DNA. Thus, starting from an ancestral genome, a new mutation will occur one day in the cellular lineage comprising the germline of an individual, and will only affect one strand of DNA. That mutation will thus be transmitted to some of the offspring (half at the most) in which it will be heterozygous. If the new mutation leads to a new phenotype, this new trait will only surface in the first generation of offspring having inherited the mutated DNA if it corresponds to a dominant character. If it is a recessive character, some degree of inbreeding between the descendants of that individual will be necessary for it to come to light.

Evolutionary "progress" is often perceived as the acquisition of new functions, resulting from mutations driving the appearance of new genes, or at least new functions in existing genes. Since the process of evolution is blind, however, new mutations will, much more often have a detrimental effect, if only because it is much easier to brake something that works than to create a new function from scratch. As early as 1930, Fisher had indeed realised that most new mutations are detrimental [23]. But this was even before the structure of DNA was known, and we now know that this is not quite true: most mutations actually occur in the silent DNA that surrounds genes, and thus have no detectable phenotype. Today, it is commonly acknowledged that, in humans, something of the order of 100 mutations take place every generation. Of those, the vast majority will occur in silent DNA, but somewhere between 3% and 1‰, i.e. between 3 and 0.1 per individual per generation will result in a detrimental phenotype, most of them through the inactivation of genes [24]. From work on laboratory strains of knock-out animals such as drosophila or transgenic mice, at least a third, and possibly as much as 50% of the mutations that result in the invalidation of genes, such as those interrupting an open reading frame, would actually be expected to be directly lethal in homozygotes, or to have such serious consequences that the homozygous bearer of such mutations would probably not go on to breed under natural conditions of selection.

Regarding mutations that actually result in new function, the proportion of those is difficult to evaluate precisely, but textbooks classically tell us that somewhere between one for every $10^4$ and $10^5$ new mutations will lead to new or different functions, i.e. one in every one hundred to one thousand individuals.

There is, however, a very important difference between mutations that inactivate genes, and those that result in new or different functions: in diploid individuals, having just one functional copy of a gene is very often sufficient, and most mutations that inactivate genes will thus be recessive, and thus have no detectable phenotype in heterozygous individuals. In a similar proportion of cases, however, there will be an effect of gene dosage, whereby individuals having lost one copy of the gene harbour an intermediate phenotype, and those mutations are then called co-recessive. Conversely, mutations that result in a gain of function will usually be dominant. The term co-dominant does not, however apply to mutations resulting in a gain of function with an effect of gene dosage (those are still co-recessive), but to mutations resulting in a change of function of a gene, where heterozygotes will thus express both functions, but homozygotes can only express one or the other. To clarify things, I have summarized those considerations in the table below.

**Table 1: New mutations from the perspective of Mendelian genetics**

| Type of mutation | Usual phenotype | Estimated Frequency |
|---|---|---|
| Silent | None | 97 - 99.9% |
| Gene inactivation | Recessive | $3 - 0.1$ % |
| New gene function | Dominant | $10^{-4} - 10^{-5}$ |
| Gene dosage effect | Co-recessive | $3 - 0.1$ % |
| Change of function | Co-dominant | $10^{-4} - 10^{-5}$ |

Even if most recessive mutations correspond to alterations in the DNA that will result in the loss of a function, there are many cases, however, where losing a function can be advantageous for individuals. For example, losing certain patterns of colours can bring definite advantages to escape predators, such as the stripes of the African ancestor of zebras and horses. Those stripes were presumably very advantageous for remaining inconspicuous to predators in the savannah, but probably had the reverse effect for the early equidae that colonised more northern and greener latitudes and would later evolve into horses. As could already be suspected from the observations reported by Darwin in the 'Analogous Variations' section of The Origin, and later elegantly recounted by Stephen Jay Gould [25], crosses between various species of equidae, and more specifically between zebras and horses, reveal that the stripy phenotype is the dominant one. For the ancestors of horses to loose their stripes, significant inbreeding must therefore have occurred to express that recessive stripe-less phenotype, and similar reasoning could be applied for the loss of any dominant character that may have been selected for in ancestors, but was no longer beneficial, for whatever reason (climate modification, colonisation, evasion of an extinct predator or pathogen, sexual character that is no longer attractive …).

Outside of the visible external phenotypes such as those considered in the previous paragraph, the capacity to resist infections by pathogens is another type of recessive trait which I perceive as particularly likely to play a major role in the selection of relatively inbred sub-populations. Most pathogens, and in particular viruses, do show high degrees of specificity for their hosts. This is due to the fact that pathogens use particular receptors to penetrate the body and/or the cells of their hosts. Infections by harmful pathogens will therefore eliminate individuals expressing that receptor, and select for organisms able to resist invasion because they carry mutated receptors to which that pathogen can no longer bind. Such characters of natural resistance are, however, usually recessive because heterozygous individuals will still carry one gene for a functional receptor, which will suffice to render those individuals susceptible to invasion by that pathogen. One particularly relevant example of this is the case of humans carrying the CCR5-Δ32 mutation, which, when homozygous, provides complete resistance to HIV infection, and an increased survival of a couple of years when heterozygous [26]. This delayed sickness would, incidentally, favour the spreading of HIV rather than be beneficial to the population, and thus bring a further advantage to the homozygotes for the CCR5-Δ32 allele. The geographic distribution of the mutant CCR5-Δ32 allele does suggest that this mutation arose several hundred years ago in northern Europe, and it is hypothesized that it was probably selected for because it provided resistance to a pathogen different from HIV, because the HIV epidemic only arose much later, in Africa [26].

Although the pressure of a particular pathogen can provide a very definite advantage to those individuals that can resist infection by that pathogen, the fact that this resistance will only be found in homozygotes would be a major hindrance for the spreading of that resistant allelic form to a whole population (something often referred to as Haldane's sieve), but would hugely favour particular subgroups where that allele would be homozygous, which could only occur through inbreeding. In addition to the fact that natural populations tend to be fragmented [4, 12], increased inbreeding will also result from increased selective pressures such as abrupt environmental changes or epidemics caused by very virulent pathogens, via a reduction in the effective size of populations. Under such conditions of increased strain, the individuals issued from groups harbouring advantageous recessive mutations will be endowed with a massive selective advantage. But the recessive nature of the characters that would be selected for under those conditions would provide the grounds for reinforcing breeding within the group rather than with members of the ancestral stock. Pushing this concept even further, Chris Grobbelaar actually proposed, over twenty years ago, the interesting idea that a mechanism of crisis inbreeding would be advantageous, whereby situations of stress would result in a shift from sexual preferences towards inbreeding [17].

2) Reducing the recombination load:

One important concept in evolutionary genetics is that the fitness of individuals is not the result of a simple sum of functions harboured by each one of their genes, or loci, but that complex relationships exist between these different loci. For example, many phenotypes are epistatic: they result from particular associations of alleles carried by different genes. One of the major advantages of sex is that it will favour the shuffling of alleles between individuals, and thus promote the formation of such functional allelic combinations. It is commonly accepted that, if such associations of alleles from different genes are particularly advantageous, this can lead to the selection of co-adapted genomes. But, as outlined by S. Wright , "*in a panmictic population, combinations are formed in one generation only to be broken up in the next*" [3]. This dissociation of functional gene combinations is what is called the recombination load. And inbreeding is the only strategy that will reduce it, by allowing the maintenance of particular allelic combinations, albeit in only a portion of the offspring.

These aspects have been extensively developed and thoroughly documented by W. Shields in his book on the relationship between philopatry and inbreeding [2] : "*One potential advantage of inbreeding, then, is that its genomic consequence of maintaining interlocus allele associations may permit more faithful transmission of coadapted genomes than would be possible with wider outbreeding*".

From the point of view of the ideas developed here, advantageous allele associations are actually quite similar to recessive phenotypes, even if they are based on the association of dominant phenotypes. Indeed, once they have become fixed in a population, their fate will be threatened by hybridisation with an outside population that would not harbour those particular alleles. The threat would be less direct because, contrarily to recessive phenotypes, the advantageous association of two dominant alleles would still be present in all F1 individuals, but it would only be maintained in 9/16 of an F2 offspring, and in just 25% if the F1 matted with an individual from the outside population. On the other hand, the advantageous epistatic combination will be maintained in all future generations if the hybrid offspring backcrossed with the isolated population. If 'invaders' were rare, this would represent a very effective way for the introgression of genetic diversity into the isolated group, but under a more sustained presence of outsiders, we can see how the recombination load could promote the selection of reproductive barriers.

Alterations in the chromosomal structure also contribute very significantly to the recombination load (for example the case of a reciprocal translocation which will be depicted later (see Fig. 2 and text relating to it). For such translocations, the general rule is basically the same as for epistatic combinations, with healthy F1 offspring. And the reduced fertility of those F1 effectively correspond to an extremely reduced fitness of those F2 individuals that do not inherit the right genetic combination. And similarly to advantageous gene combinations, once a particular chromosomal rearrangement has become fixed in a population, usually through inbreeding, the most effective way for the descendants of hybrid offspring to recover complete fertility will be by backcrossing with the isolated group. In cases where populations differ by several chromosomal rearrangements, however, hybridisation would become a real threat because the fertility of hybrid would be dramatically affected.

3) Fighting Muller's ratchet: A third advantage of inbreeding is that, for diploid organisms, it is the only effective way to fight off the accumulation of recessive deleterious mutations in their genomes. The notion that mutations accumulate inexorably in genomes over the course of generations is commonly referred to as Muller's ratchet [27]. Muller advocated that a major reason for the prominence of sexual reproduction among all animal species was due to the need to eliminate these mutations through genomic recombination. Following the views initially expressed by Fisher [23], Muller, in his early work on Drosophila, had documented himself that most new mutations tended to have recessive phenotypes. When it came to persistence of those in the genome over generations, however, he considered that all mutations were partially dominant (i.e. co-recessive, see table 1), and that even the most recessive deleterious mutations must have some slight effect ( 2 to 5 % ) on reproductive fitness [28]. Those weakly deleterious mutations would therefore be eliminated progressively over successive generations. Muller, however, carried out all of his work before the discovery of the structure of DNA and of how genes worked. Although his arguments were clearly valid for weakly deleterious co-recessive mutations, we now know that a very large proportion of deleterious mutations will be perfectly recessive and that mut/WT heterozygotes will show very little, if any, reduction in fitness compared to WT homozygotes [29, 30]. At any rate, even if inactivation of a fair portion of genes leads to co-recessive phenotypes through an effect of gene dosage, the frequency of deleterious mutations giving rise to completely recessive phenotypes will still be much higher than those leading to dominant, or co-dominant traits. Inbreeding, by promoting the conditions whereby recessive mutations can find themselves in a homozygous state, will hence allow the expression of those deleterious effects resulting from recessive mutations.

The adjective "inbred" has clear derogatory connotations when referring to human beings and the commonly held perception about inbreeding is that it promotes degeneracy of the genome. Somewhat ironically, inbreeding actually results in "improving" the genome, and the fact that inbreeding results in elimination of recessive deleterious mutations from the population is actually well known, at least by animal or plant breeders and scientists : the extent of inbreeding depression decreases over successive generations of inbreeding [1][31]. Via this type of phenomenon, the consequence of inbreeding will be that the allelic frequency of recessive mutations will be lower in the offspring than in their parents. For each mutation, the efficiency of the process is, however, remarkably low.

---

[1] I owe the notion that inbreeding is bad for your offspring, but good for their genomes, and hence for future generations, to a conversation I had several years ago with my former colleague Geoff Butcher regarding the criticisable habit of certain scientists of using outbred rodents for their experiments on the grounds that those are usually healthier and fitter than inbred ones. This practice indeed introduces genetic variability in the experimental samples, which can lead to results that are either too variable to be significant, or even sometimes plain artefactual. On this subject, Geoff Butcher expressed the extremely wise point of view that, if a scientist wants to work with very healthy rodents, he/she should be using F1 animals obtained from crossing two separate inbred strains. Those types of animals all have strictly the same genetic background, and are indeed extremely healthy because they benefit from remarkable luxuriance, in other words hybrid vigour that is seen in individuals that carry almost no recessive or partially recessive deleterious mutations.

Indeed, in the case of a heterozygous breeding pair, the allelic frequency for the mutated copy of the gene would only pass from 0.5 in the parents (each heterozygote for the deleterious allele), to 0.33 in the offspring (see box 1). But inbreeding is the only practical way for the members of a species with an obligatory diploid genome to cleanse their genomes off the recessive mutations that will otherwise inexorably accumulate over successive generations until they reach an equilibrium, when the average number of recessive mutations in the genomes of individuals is sufficiently high that the rate at which they accumulate in the genome is balanced by a rate of elimination by random chance rather than by consanguineous descent (see box 1). The reason why I have used the word "practical" in the previous sentence is because of the bdelloid rotifers, the one undisputed example of asexual diploid organisms, that seem to have adopted an alternative strategy to sex to cleanse their diploid genomes from recessive mutations, but as discussed in addendum 1, it calls upon such extremes that it would be impractical for most other organisms. Haploid organisms such as prokaryotes do not have this problem of keeping their genome from accumulating deleterious mutations, because in haploids, all mutations are dominant, and deleterious ones will hence be eliminated very rapidly. Multiple cases exist in nature of the use of a haploid state by otherwise diploid eukaryotes, and in addendum 2, I have developed three such examples that I find particularly eloquent i.e. the cases of organisms that go through haploid stages, of the sexual chromosomes and of the endosymbiotic organelles.

********************************************************************************************

**Box 1** : **Comparing the effects of accumulation of recessive deleterious mutations in populations undergoing various degrees of inbreeding, and with a theoretical completely outbred population.**

Fig 1A: Mendelian laws predict that when a crossing occurs between two individuals heterozygous for a recessive deleterious mutation, allelic frequency for that mutation drops from 0.5 in the parents to 0.33 in the offspring.

| | **WT** | **Recessive deleterious mutation** |
|---|---|---|
| **WT** | WT/WT | WT/<u>mut</u> |
| **Recessive deleterious mutation** | WT/<u>mut</u> | 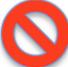 <u>mut/mut</u> 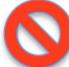 |

When breeding takes place between two individuals each carrying one copy of a defective essential gene, one quarter of their offspring will be either non viable, or very unfit because they will be homozygous for the deleterious mutation. If the mutation is truly recessive, the other three quarters will be perfectly viable, and two out of three among that viable offspring will be heterozygous for the mutation. The allelic frequency of that deleterious allele will hence pass from 0.5 in the parents to 0.33 in the offspring, and the mutation load from 1 to 0.66 mutations per individual.

As a rough estimate based on the simplistic case of a single gene, one could therefore say that a rate of spontaneous mutation of 0.17 per generation (0.5 – 0.33) will be compensated by a reduction of 0.25 in fertility. This value of 0.17 is rather compatible with the various estimates of the rate of spontaneous mutations, which are, for humans, between 0.1 and 3 new deleterious mutation per genome per generation [24]. Although I realise that those figures are probably inaccurate for the additive effect of multiple genes, it was beyond my limited mathematical capacities to perform more precise calculations. I am confident, however, that others will later find such calculations rather straightforward, and it will then be particularly interesting to evaluate what types of equilibriums are reached for various mutations loads, various rates of mutations, and various effective sizes of population (i.e. various degrees of inbreeding).

Fib 1B: Evaluation of the fertility as a function of mutation loads and inbreeding coefficients.

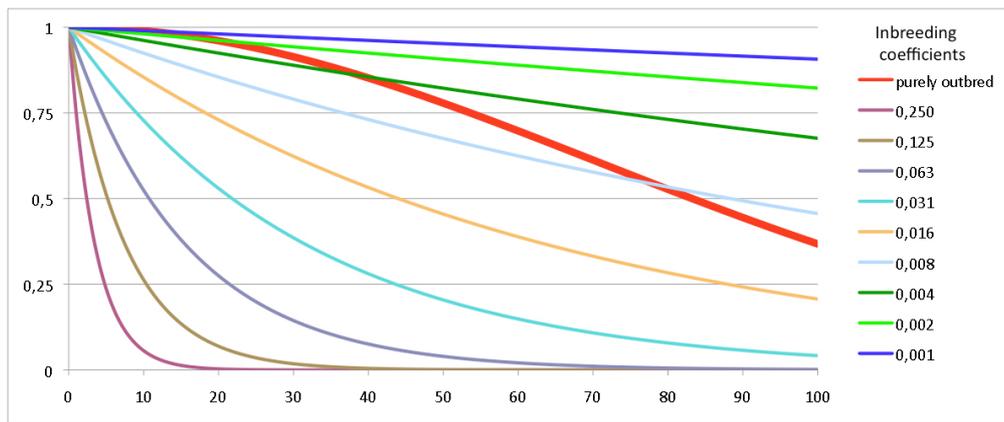

The theoretical fertility of breeding pairs in a population can be calculated as a function of M, the average number of recessive mutations per individual ( i.e. the mutation load) and of I, the overall average inbreeding coefficient in that population ( i.e. the probability that a locus taken at random in the genome will be homozygous by descent, corresponding to half the average degree of consanguinity of parents). The average fertility will then be $(1- I)^M$. The different coloured curves were calculated for the indicated inbreeding coefficients, and we can see that fertilities only start to be significantly affected for populations with inbreeding coefficients > 0.01, corresponding to parents with degrees of consanguinity of 0.02, i.e. roughly that of third or fourth cousins.

In parallel, one can also evaluate the fertility of breeding pairs in an panmictic infinitely large population, where there is effectively no inbreeding (in real populations, the average fertility would actually be a factor of the two degrees of fertility). For mammals, if one estimates that roughly one third of genes are essential, this would amount to a total of approximately $10^4$ essential genes. If the mutation load in the population is M, the probability of any locus being mutated will be $M/10^4$, and the probability of carrying two mutated alleles of any given gene will be $(M/10^4)^2 = M^2.10^{-8}$ and hence the effect on fertility would be $(1- M^2.10^{-8})^{10.000}$ overall since the threat applies for every single one of the 10.000 essential genes. This is represented as the thick red curve on panel B. We can see that, whilst the chance of carrying two inactivated copies of the same gene remains extremely low for mutation loads below 20, it starts becoming quite significant for mutation loads over 30, and fertility will drop below 75% when genomes have accumulated, on average, over 50 recessive mutations. For populations harbouring levels of consanguinity superior to 0.02, the reduction in fertility is, as could be expected, much more sensitive to mutation load, and for a population with an inbreeding coefficient of 0.06, a drop of fertility to 75% will occur with a mutation load between 5 and 6, but this figure climbs to nearly 30 mutations for an inbreeding coefficient of 0.01.

We have seen in the previous paragraph that, based on calculations for a single gene, a drop of 0.25 in fertility would keep up with the rate of 0.17 new deleterious recessive mutation, i.e. one per genome every six generations. The figures would possibly be slightly different if one considered the additive effect of multiples recessive deleterious mutations affecting different genes, each with lower allelic frequencies, and clearly different with different mutation rates. The mutation rate of 0.17 per generation cannot, however, be very far from reality since the most extreme estimates go from 0.1 to 3, and a decrease in fertility of 0.25 does not seem a completely unrealistic figure to keep up with new recessive mutations occurring once every six generations. For humans, this does not, however, mean that one in four newborn babies would come to the world with mental retardation or grievous physical defects. Indeed, most recessive mutations that touch essential genes would be expected to cause spontaneous premature abortions at very early stages of pregnancy, and many even before they would be recognised as miscarriages. From this point of view, it is actually rather striking to note that, in modern humans, miscarriages occur at a rate of somewhere between 10 and 40 %. Whilst the occurrence of these miscarriages is clearly also related to other factors such as the age and the health of the mother, these figures suggest that it is not unreasonable to envisage that the price to pay to fight Muller's ratchet is that a fair proportion of the zygotes (say 20 to 30 %) will have to be lost to compensate for the occurrence of one new recessive mutation every six generations. And these figures also seem compatible with what one sees in mice. Indeed, although mice can have as many as 10 to 12 pups in a litter, inbred strains are much less prolific, with litters often limited to 4 to 6 pups. When I have had to sacrifice pregnant female mice for experiments on embryonic tissues, I have often been struck by the proportion of aborted foetuses one can find in the uterus of a gestating female mouse, which is often near 50%. Thus, even in inbred mice in which the inherited mutation load must be close to zero, the rate of abortions suggests that de novo recessive mutations occur at a rate that is probably superior to one in six zygotes, or one in six generations.

Diploid genomes must have contributed greatly to the adaptive 'explosion' which took place among eukaryotes 1,5 billion years ago. The most important factor for this must have been the robustness of organisms, i.e. their newfound tolerance to new mutations that would have been instantaneously deleterious in haploid organisms. Conceivably, this may even have allowed the diploid organisms to "lower their guard", i.e. to reduce the fidelity of the replication of their DNA, and favour mechanisms of recombination [32], thereby favouring the appearance of novel adaptive mutations, helping them in particular to combat pathogens more efficiently, or to adapt to new environments. This view is supported by the fact that the vast majority of metazoans of today are obligatory diploids. The drawback of relying only on diploid genomes is that this also gives rise to the insidious type of Muller's ratchet I have just discussed, whereby recessive deleterious mutations can start accumulating silently in the genome of outbred individuals. Without sex, the benefits of a diploid genome would, thus, be very short lived, especially on the evolutionary time scale, and genomes would ultimately reach a mutational meltdown [33]. But sex without inbreeding is fraught with even more insidious, and thus far greater dangers that, as we will see, can ultimately lead to species extinction.

DNA replication is far from being a perfectly faithful process, and the rate of appearance of mutations in the genomes of vertebrates is commonly recognised to be of the order of $2.10^{-8}$ per nucleotide for every generation, although the complete sequencing of the whole genomes of a family of four suggests it may be half as high [34]. For mammals, since their haploid genomes comprises roughly $3.10^9$ base pairs, each diploid newborn will thus carry, on average, around 100 nucleotides that will differ from those it should have inherited from its parents if DNA replication was perfectly faithful, and if DNA was perfectly stable and completely resistant to damages by radiation and chemicals. Among those mutations, the vast majority will be silent, but, as summarized in table 1, some will modify or inactivate gene functions, and most of those will be deleterious, but recessive.

In the long run, the phenomenon of evolution will be based mostly on the acquisition of new characters, corresponding to dominant mutations. But this can very easily be obscured by the much higher prevalence of recessive mutations. This can be ascertained by the repeated observations that the particular characters selected for in domestic species prove almost systematically to be recessive against the phenotype of the wild stock.[2] Even if DNA replication could be selected to become completely faithful, this would not be a solution, because, as famously underlined by Leigh van Valen [35], organisms have no choice but to evolve continuously in the face of natural selection, just like Lewis Carol's Red Queen, who needs to keep running just to stay in the same place.

But because evolution is blind, and occurs only by random mutations, in order to have a chance to see adaptive mutations arise, be they new functions or the advantageous loss of existing ones, there will be no avoiding the hundred fold excess of deleterious mutations, which will need to be eliminated by natural selection. As alluded to earlier, most of those deleterious mutations will, however, be perfectly recessive, i.e. they will have no phenotype in heterozygotes. Hence, within a large out-breeding population, the chance that one individual will carry two copies of an inactivated gene will be very low. But those will consequently be transmitted to half of the offspring, and over successive generations, since such mutations will keep accumulating, the mutation load will inexorably increase. Even at the lowest rate of the range envisaged above, i.e. one additional recessive mutation every ten generations, the mutation load will thus still increase rather rapidly until, as proposed by Muller [28], it reaches an equilibrium where as many mutations are eliminated at every generation than arise due to new spontaneous mutations. This process of elimination, which correlates directly with infertility, will, obviously, be greatly dependant on the inbreeding coefficient, i.e. on the effective size of the population. In box 1, I have tried to evaluate how the accumulation of recessive mutations in a population can affect the fertility of individuals as a function of the inbreeding coefficient in that population. From rather simplistic calculations, I conclude that, if the rate of accumulation of recessive mutations is of the order of one every six generations, this will be compensated by a drop in fertility of the order of 0.25. These figures, although rather speculative, seem to be compatible with the rates of spontaneous abortions one sees in human and mice, of which a fair proportion (I would guess between one and two thirds) are probably due to genetic causes. As already underlined by Muller 60 years ago [28], the proportion of miscarriages due to genetic defects necessary to keep the mutation load in a steady state will be principally dependent on the rate with which new mutations appear in the genome at every generation. The process of outbreeding will indeed reduce the initial frequency at which recessive mutations are found on both copies of a gene, but this advantage will only last for a while, until the mutation load has increased to levels where the decrease in fertility due to mutations once again compensates for the rate at which they appear. The advantage of outbreeding is thus very short lived on the evolutionary time scale. And, as mentioned earlier, I contend that it opens the door to a much greater threat. Indeed, if a large population undergoes extensive outbreeding for hundreds of generations, the equilibrium will only be reached when each individual carries, on average, several dozens of recessive mutations in its genome. If that population undergoes a sudden increase in selective pressures, for example because of a novel pathogen, of competition with another species, of a recrudescence in predators or of abrupt changes in the natural environment, the effective size of that population will shrink, and the inbreeding coefficient among the survivors will consequently become very significant[3]. If we imagine that the mutation load in such a large population had reached 40, and that the reduced numbers of individuals causes the inbreeding coefficient to rise to 0.03 in the remaining population, this will result in only 30 % of viable zygotes. If we consider that this would happen under conditions where natural selection would be particularly harsh, the delayed cost of having avoided inbreeding for the short term benefits provided by outbreeding may well, in the long run, play a major role in the rapid extinction of that species, as well as reducing their capacity to colonise new environments (in the section 'convergence of character' of The Origin, Darwin himself remarked that '*When any species becomes very rare, close interbreeding will help to exterminate it*'). In the face of Muller's ratchet, as Muller himself very rightly stated 60 years ago, "*We cannot eat our cake today and have it tomorrow*" [28]p150.

---

[2] Since they did not know about Mendel's laws, the capacity of certain mutations in both pigeons and dogs to complement one another to restore a wild type phenotype after many generations of 'true' breeding did contribute greatly to confuse both Darwin and Wallace about the durability of acquired recessive traits.

In cases where there is a relatively sudden shift in the pressures of natural selection, such as those caused by natural catastrophes (volcano, meteorites …), or by a global change in the earth's temperature, the resulting shrinkage in effective populations sizes would thus be expected to be less well tolerated by the more prominent populations, i.e. probably those having taken full advantage of extensive outbreeding. Incidentally, such a mechanism would provide an explanation for the phenomenon of punctuated equilibrium proposed by Gould and Eldredge [36, 37]. Indeed, over periods of stability, the individuals of the most successful species will proliferate and colonise ever increasing territories. They will thus be the ones most likely to be found in the fossil record. But with this increase in effective sizes of populations will come the insidious consequence of increased mutation loads, and consequently the least chances to survive when unrest arises, causing dramatic reduction in the sizes of the populations. From this point of view, it is thus not surprising that, during periods when the natural scene changes, it should be the most numerous species, those found in the fossil record, that would struggle the most in the face of imposed inbreeding caused by population shrinkage, and become extinct with an apparent simultaneity.

4) <u>Reducing the cost of sex</u>: Another advantage of inbreeding is that it reduces the cost of sex. Indeed, in sexual reproduction, each parent passes only half of its genome to each of its offspring, which is directly related to the consideration that the cost of sex is two-fold [38], as compared to asexual reproduction, where each offspring inherits all of the parent's genome. But this factor of two is not quite a completely accurate measurement, if only because for most metazoans, sexual reproduction is obligatory and not an option. Furthermore, if we consider a hypothetical species with the most outbred population possible, each individual of that species will still be more genetically closely related to all the other individuals of the same species than to any other individual of a closely related species. In other words, all individuals of a given species share more common ancestors than they do with those of a closely related species. Hence when they breed within their own species, individuals do share some significant level of relatedness with their sexual partner compared with that of an individual of another species. So, even in a completely outbred population, because individuals of the same species will necessarily share some common ancestors, the cost of sex is never quite as high as two. And the more closely related an individual is to it's partner, the less that cost will be, for both of them.[4] Consequently, any evolutionary step that will favour inbreeding rather than producing offspring with more distantly related individuals, even of the same species, will thus reduce the cost of sex.

5) <u>Inbreeding promotes population fragmentation, which can, in turn, promote collaborative or altruistic behaviour</u>:

From the point of view of the 'selfish gene' hypothesis [39], individuals should always favour their own interests, or at least those of closely related individuals [40, 41]. On the other hand, mathematical modelling has led certain population biologists to conclude that group level selection cannot work, and that for any behavioural trait to be selected, that trait must have a direct selective advantage for the individual. Such views are, however, much less prominent today, and anyone who is not convinced that group-level selection can play a major role in evolution should read the excellent recent review by Wilson and Wilson [42].

---

[3] There is a rather counterintuitive potential further advantage for inbreeding in times of harshness since increased inbreeding coefficient usually causes individuals to become smaller. Indeed, smaller individuals require less nutrients for their survival, and size is also well known to be inversely proportional to population density. Hence, a rather intriguing possibility lies with the idea that, under conditions of increased natural selection, small sizes caused by inbreeding depression may actually bring on a selective advantage in the struggle for survival.

[4] Rather than relying on coefficients of consanguinity, I perceive that a much simpler and accurate way of calculating the cost of sex is by simply counting the sheer number of nucleotide differences between parents and offspring. With this type of approach, one can easily see that mating with a member of the same race or variety will be less costly than with a more remotely related individual. This also provides the simple means to incorporate the accumulation of neo mutations over successive generations in the calculations, or to compare parthenogenesis with self-fertilisation.

The type of reasoning which led to the rejection of group-selection was always based on the assumption that populations consist of large numbers of individuals breeding freely with the rest of the population. But, as underlined by Wright himself [12], natural populations are not like that. If we only look at the human population, although all individuals can theoretically breed with all those of the opposite sex with apparently equivalent efficiencies, we can see that the total human population is structured in ethnic groups, races, types, families … and that certain characters are more prominent in certain groups of individuals than in the rest of the population. In addition to the well recognised and very significant advantage of slowing down the spread of pathogens, and of favouring the maintenance of genetic diversity [4], population fragmentation has the other, much less direct and less obvious benefit of favouring the evolution of altruistic behaviours, by making group-level selection possible [42]. On the subject of group selection, I choose to adopt the view that, in fragmented populations, each group effectively becomes equivalent to a multi-cellular organism (see [43] for recent views on organismality). In metazoans, the fact that all the cells share the very same genetic makeup makes it possible for the vast majority of cells to sacrifice themselves either directly by apoptosis, or by differentiating into somatic cells that have absolutely no hope of generating offspring, for the benefit of the very few that will be destined to the germ line. Similarly, if a population is comprised of many small groups of individuals that are more closely related to one another than to the rest of the population, I firmly believe that it then becomes possible for natural selection to favour the evolution of collaborative or altruistic behaviours, because, in the end, even if those behaviours do not directly benefit the individuals that undertake those altruistic behaviours, the members of that group, and hence, on average, all the genes of the gene pool of that group, will fare better than those of the "group next door" that may have stuck with strictly selfish behaviours. On this subject, in 1871, Darwin himself made the following statement in his book "The Descent of Man":

*It must not be forgotten that although a high standard of morality gives but a slight or no advantage to each individual man and his children over the other men of the same tribe . . . an increase in the number of well-endowed men and an advancement in the standard of morality will certainly give an immense advantage to one tribe over another.*

Although, when one looks at natural populations, scores of examples can be found in all the kingdoms of life where altruistic, or at least collaborative behaviours have apparently been selected for, the questions linked to group level selection remain very contentious issues today. I know of no better example of cooperative altruistic behaviour than that of the lowly slime mould, Dictyostelium discoideum, and I contend that it is promoted by the ability of single cells to colonise new niches, resulting in fragmented populations. One of the reasons for which I find the example of Dictyostelium particularly telling is that it is not complicated by the intervention of sexual reproduction ( see addendum 3 for more details).

In some cases, speciation could conceptually correspond to the need for populations having developed cooperative/altruistic strategies to fend off more selfish invaders. The issue of altruism is, however, really a side issue to the main focus of this essay. All I wish to say here is that, from an admittedly ultra-Darwinian point of view, the only realistic way to explain the evolution of cooperativity and altruism in natural populations is via group level selection, and this selection can only occur in populations that are fragmented into small groups of genetically inter-related individuals, or in other words, by natural selection acting on groups undergoing more inbreeding than if the population was considered as a whole. The fact that inbreeding can have the additional characteristic of providing a selective advantage at the levels of populations simply reinforces the view that inbreeding can and will occur and will not always be avoided. This will result in structured populations, which will, in turn contribute to the phenomenon of speciation.

6) Disadvantages of inbreeding: For the sake of fairness of argument, it seems necessary to counterbalance our arguments here, and underline that inbreeding also has several very significant disadvantages. Indeed, when starting from an outbred population, inbreeding depression will result in a high proportion of completely unfit offspring, and in most of the offspring being less fit than those from outbred breeding pairs. Another consequence of excessive inbreeding is that, by reducing the gene pool available for generating varied combinations of genotypes, it will result in less diversity, and thus in a more limited adaptability of the populations. Hence populations that undergo excessive inbreeding will be less likely to develop new functions than large populations undergoing outbreeding, where new functions bringing selective advantages can rapidly spread to the whole population, and can further combine with other advantageous functions that will have arisen independently in other individuals. Inbreeding may thus result in a slower rate of evolution.

This last argument does, however, need to be balanced by several counter-arguments. First, as we have seen previously, advantageous traits are not necessarily dominant, and those that are recessive can only come to light under some level of inbreeding. Thus, although inbreeding will reduce the probability of dominant traits spreading to whole populations, it will increase the frequency at which recessive traits appear, and since the mutations causing such traits are much more frequent than those causing novel functions this may balance the effect of inbreeding on slowing evolution. Second, when it comes to epistatic phenotypes resulting from advantageous gene combinations, we have seen that inbreeding is, once again, the only way to maintain them. Finally, as has been recognised for a long time, the rate at which characters can become fixed in populations is inversely correlated to the size of those populations [44]. By reducing the effective size of populations, the slower rate of evolution caused by inbreeding may thus also be compensated. As we will see later on, I actually contend that excessive inbreeding, leading to excessive speciation, will consequently result in the shorter lifespan of individual species, and thus in an accelerated rate of the species' turnover, which is not the same thing as the rate of evolution, although the two are too often considered equivalent.

Another potential disadvantage of excessive inbreeding is that it could result in reductions in the levels of polymorphism in a population, by provoking what would effectively amount to repetitive bottlenecks. For jawed vertebrates, which rely on polymorphism at the level of the MHC (Major Histocompatibility Complex) for fighting and eliminating infectious pathogens, this would be expected to have particularly nefarious consequences. As we will see later, however, comparing MHC polymorphism between related species reveals that inbreeding, and speciation, can apparently take place without losing healthy levels of polymorphism over the MHC region [45, 46], and presumably over most of the genome.

**II) Focusing our reflections on what the ORIGIN of species could be**. Or how can it sometimes be beneficial for a few individuals to breed preferentially among themselves rather than with the rest of the population, in others words with the ancestral stock?

We have thus underlined how inbreeding can have numerous advantages, and how systematic outbreeding is actually a strategy which has mostly short-term advantages, but that can lead to great drawbacks in the long run. I now propose to follow the path laid out by Darwin in the title of his book, and to focus on the very **origin** of species, i.e. to try to imagine what initial genetic event could eventually lead to the separation of a subgroup of individuals that will breed preferentially with one another rather than with the rest of the population.

Outside of the rather anecdotic cases of one step speciation via polyploidy (see C&O, p321), for the vast majority of metazoans, successive steps of progressive separation appear as more likely scenarios to reach speciation. But even if it does not result in instant speciation, an initial mutation must occur at some stage which will eventually result in promoting the interbreeding between individuals carrying that mutation rather than with the rest of the population. I have chosen to call such a process '**saeptation**', from the latin word saeptum : barrier, envelope. In other words, saeptation will be the consequence of a mutation that will promote increased inbreeding within a group inheriting that mutation, and thus in a reduction of the gene flow between this new group and its immediate ancestral stock.

Lets us now envisage what type of mutation could eventually lead to saeptation. As seen earlier, this new mutation will, one day, occur on one strand of DNA of one cell belonging to the germline, and hence be present in up to half of its gametes, and go on to be present on one chromosome of all the cells of some of its offspring.

1) Saeptation scenarios caused by a recessive mutation: As alluded to repeatedly in the previous paragraphs, I think the most likely scenario involves a recessive mutation as the very first step, i.e. the initial saeptation, which will end up promoting partial reproductive isolation of its bearers. The first reason for this is that, as outlined in table 1, outside of silent mutations, new mutations will most frequently lead to loss of functions, and will usually be recessive. But, as we have seen in the previous section, a loss of function does not necessarily mean a selective disadvantage.

Let us go back to the example of the horse precursors, and how they could have lost the stripes carried by their zebra-like ancestors. In the first place, to reveal the non-striped recessive phenotype, some significant inbreeding must have taken place. That inbreeding could actually have been promoted by the very fact that the group for which the stripe-less phenotype was advantageous was in the process of colonising more northern latitudes. Colonising populations, having smaller effective sizes, have consequently higher inbreeding coefficients [47], and we will see later that this is particularly relevant for the situations of island colonisation. Another consequence of the small size of such a group is that it will greatly facilitate the fixation of an advantageous recessive phenotype [44]. This isolation of a small relatively inbred group would hence result in reduction of the gene flow with the ancestral group because the adapted group would occupy a different territory. This would not, however, really represent a step of biological speciation, i.e. bona fide reproductive isolation, because if one individual of that adapted group ended up among individuals of the ancestral stock, it would probably breed with them very happily and efficiently, and the defining stripe-less phenotype would be diluted and only surface on very rare occasions[5]. But it would lay the grounds for the evolution of further isolating characters because, in the context of their isolated group, it would be very disadvantageous for individuals to breed with stripy partners from the ancestral pool since all of their offspring would then end up with the dreaded stripes on their back, and thus be much more susceptible to becoming eliminated by predators.

Consequently, if an additional mutation took place in a member of that adapted group that led to more effective reproduction with kin than with individuals not carrying that second mutation, the inherent disadvantage of such a mutation due to the reduction of fertility with the rest of the adapted group would be balanced by a very significant advantage to its bearers because it would help prevent that sub-group of individuals from being re-invaded by the dominant but disadvantageous trait. This preferential mating with kin would also amount to promoting further inbreeding. This may be further facilitated by the fact that, when populations have previously gone through stages of significant inbreeding, the cost of inbreeding depression is very much reduced because most recessive deleterious mutations will have already been cleansed from the genome. Hence, from the above reasoning, we see that, in the context of an outbred population, a mutation that simply results in promoting inbreeding will struggle to become established because it would have many disadvantages to weigh against the advantage of reducing the cost of sex. But in the context of a group having undergone significant inbreeding, the safeguard of the mutation load against further inbreeding will have become much weaker, and under the selective pressure of the persistent threat posed by invasion by the ancestral stock, the probability of additional steps of saeptation within that group would thus be much higher.

---

[5] As we will see later on, this type of phenomenon actually happens in sticklebacks, which gain a selective advantage by losing their armour plates when they colonise freshwater environments.

2) Scenarios involving two mutations (Dobzhansky-Muller model ): To explain how mutations promoting reproductive isolation could ever appear in natural populations, Bateson (1909), Dobzhansky (1936) and Muller (1942) all came up with a similar hypothetical model, which is nowadays unjustly referred to as the Dobzhansky-Muller model (see C&O, p 269). This model calls upon the existence of two completely separated groups (allopatry), where two separate mutations take place that would each have no effect on the reproductive fitness in the group in which they arise, but that would result in incompatibility between the groups if and when those two groups are brought back in contact with one another. Such models are, however, not in line with Darwin's views that each step along the very long path of an evolutionary process must carry its own selective advantage. In the context of a group carrying a recessive advantageous mutation, however, we can see how the pressure of the outside populations, carrying dominant but disadvantageous alleles, could promote the selection of a mutation favouring reproductive isolation from the ancestral stock. At the end of the previous paragraph, I have argued that this selective pressure may be sufficient to promote further steps of saeptation, i.e. isolation from the other members within the adapted group, because the disadvantages of this mutation promoting inbreeding would be overcome by the advantage of resisting invasion by the dominant disadvantageous phenotype. And this modified tilt of the balance would be further favoured by the reduced inbreeding depression resulting from the relatively high level of inbreeding already present within that group.

Another scenario is, however, possible, which is to a certain degree related to the Dobzhansky-Muller model in that it would involve multiple steps, but those would occur in sequence, and not independently: the secondary steps of isolation would target traits specific to the saeptated population which could quite possibly be the one having driven the saeptation, but not necessarily. Indeed, during the initial phases of saeptation, inbreeding among a limited number of individuals would result in a high proportion of other genes becoming homozygous, and could thus reveal additional recessive phenotypes only rarely encountered in the ancestral population. In addition, in other genes than the one having driven the saeptation, certain alleles would have become much more frequent, either because they were genetically linked to the advantageous mutation, or simply because the smaller size of the population had favoured their drift towards fixation. For these three types of genes (additional recessive phenotype, genetically linked to the advantageous recessive mutation, gene having reached fixation by chance), the allelic frequencies would therefore be very different in the saeptated inbred population and in the ancestral one. And those would then represent as many potential targets for the selection of isolating mechanisms that would prevent the individuals of the saeptated group from mating back with the ancestral group. Technically speaking, this would, however, not represent saeptation, but reinforcement, because the mechanism of isolation would specifically target the outsiders, and not the direct ancestral stock, i.e. the isolated group. This type of scenario would thus involve two or more steps like the Dobzhansky-Muller model, but the fundamental difference with the Dobzhansky-Muller model is that selective pressures would be driving the isolation, rather than rely on chance for the separate evolution of two traits that will, at a later stage, turn out to be incompatible. One of the predictions inferred from the Dobzhansky-Muller model is that the rate of accumulation of reproductive barriers should increase with time, the so called "snowball effect" [48-50]. But this prediction does not actually allow to discriminate with the sympatric scenario described above. Indeed, if the threat of hybridisation is maintained throughout the speciation process, one would expect a similar snowball effect: once some degree of reproductive isolation has started accumulating between the two populations, resulting in reduced inclusive fitness of the hybrids further than the simple initial loss of the recessive advantageous phenotype (for various reasons including reduced fertility, intermediate maladaptive phenotypes, poor health, increased recombination load or even lethality), the cost of mating and/or breeding with the ancestral stock will have increased even more. Consequently, the pressure for selecting further mechanisms of reproductive isolation will also be increased, and one would thus expect the rate at which such traits are selected to go up, until such times when the two populations are sufficiently isolated that neither represents a significant threat for the other one.

3) Scenarios involving a dominant mutation : Lets us now consider whether a scenario can be envisaged whereby a dominant mutation would promote saeptation. The most obvious type of such a mutation would seem to be one that modifies the actual niche of the population, a phenomenon often referred to as ecological speciation. Indeed, if individuals carrying a novel mutation can start occupying new territories (geographical, seasonal, nutritional ...) they will, in this new territory, naturally find themselves in the presence of those other individuals carrying the same mutation, which will, by definition, be descended from the same ancestor, and will therefore be their close relatives (sibs or cousins). Since we are now talking about a dominant mutation, to allow the first individuals with the new mutation to find mates to reproduce, the initial separation between the adapted subgroup and the ancestral stock can, however, only be partial, and the possibility of hybridisation between the two groups must therefore be preserved. Although inbreeding among colonisers may carry an initial cost because of inbreeding depression, this could easily be offset by the advantage of the lack of competition in the new territory, and the inbreeding depression would only be transient, and recede after a few generations. Although, as we will see later, dominant mutations could play important roles in further steps of the speciation process, i.e. in reinforcement, it is thus hard to envisage how they could, on their own, promote the selection of reproductive barriers with the ancestral stock. In the case of a dominant mutation leading to the colonisation of a new niche, the increased inbreeding among the individuals carrying the mutation would, however, greatly increase the probability of revealing some additional recessive characters, of which some may turn out to be adaptive to the newly colonised environment. And those recessive mutations could, in turn, provide the grounds for a selective advantage to stop breeding with the ancestral stock.

### 4) The special cases of co-recessive characters, chromosomal translocations and reinforcement.

#### 4a) Co-recessive characters.

Within the frame of the analyses carried out in the previous paragraphs, mutations that lead to hybrids harbouring intermediate co-recessive phenotypes (see table 1) would seem particularly prone to promoting speciation. Indeed, if such a mutation brings about an adaptive phenotype, such that the partial gain or the partial loss of a function makes it possible to colonise a new niche (warmer or colder climates, higher altitude, different food, different breeding time…), the heterozygotes of the first few generations would be closely related to one another, but would be expressing intermediate phenotypes that would not separate them too much from the ancestral stock, and hence allow for the generation of multiple individuals. Crossing of those semi-adapted individuals with one another would be favoured by the fact that they would occupy that new niche. This would result in a quarter of their offspring becoming homozygous for the adaptive trait, which they would hence express more strongly, and would possibly be restricted to occupying only the newly colonised niche, with little or no possibility of contact with the ancestral one. The intermediate phenotype of the heterozygotes could thus be likened to some sort of stepping stone for the assembly of an isolated, necessarily more inbred group of individuals homozygous for the adaptive trait. Once that group has been constituted, in addition to the fact that the cost of sex would be higher with the outside group than within the group, a further advantage would be that additional adaptations to the new niche would probably be selected for quite rapidly, and the phenotype of the offspring that would result from encounters with the ancestral stock would very possibly make them unfit for either environment. This would thus provide the grounds for the Wallace effect, i.e. for the selection of further mutations reinforcing the reproductive isolation between the two populations. We can thus see how co-recessive traits could conceptually promote reproductive isolation even more rapidly than completely recessive ones.

Importantly, whether the mutation driving the saeptation is completely recessive or co-recessive could have significant consequences on the size of founder populations. Indeed, in the case of completely recessive mutations, those could stay completely silent for long periods of time within a population, and hence surface when crossings occur between individuals that are not necessarily very closely related to one another. In the case of a co-recessive mutation, however, the new intermediate character will be expressed in half the offspring of the founding individual, and the founding population will thus necessarily be comprised mostly by brother-sister matings, or close cousins at best. We will come back later to considerations regarding the size of founder populations and preservation of heterogeneity in the population.

#### 4b) Chromosomal translocations.

Chromosomes can be either circular, as in most bacteria and in endosymbiont organelles, or linear, as in all eukaryotes and a few bacteria. As far as I know, there are no known organisms with circular chromosomes that can carry out meiotic sexual reproduction, and all eukaryotes also have multiple chromosomes. Multiple linear chromosomes thus appear as a prerequisite to meiosis, with three chromosomes being the smallest number documented, in the fission yeast Schizosaccharomyces pombe (most species have several dozens, and up to several hundreds, or even over one thousand in certain ferns). One of the main reasons having driven the arrangement of the genetic information on such multiple and linear structures is almost certainly to promote one of the main purposes of sex, i.e. to achieve an efficient shuffling of the genes between individuals having evolved in parallel, via both inter- and intra-chromosomal recombination. Another commonly recognised advantage of this arrangement in metazoans is that the maintenance of telomeres provides a certain level of safeguard against the rogue selfish multiplication of cells that will lead to cancer. Outside of these two obvious advantages, I perceive that the arrangement of genomes on multiple linear chromosomes is also likely to play a central role in the phenomenon of speciation. Indeed, in line with the observation that even closely related species almost always differ in their chromosomal architecture, the role of chromosomal rearrangements in speciation has long been hypothesized (see C&O p 256-267, citing White 1978).

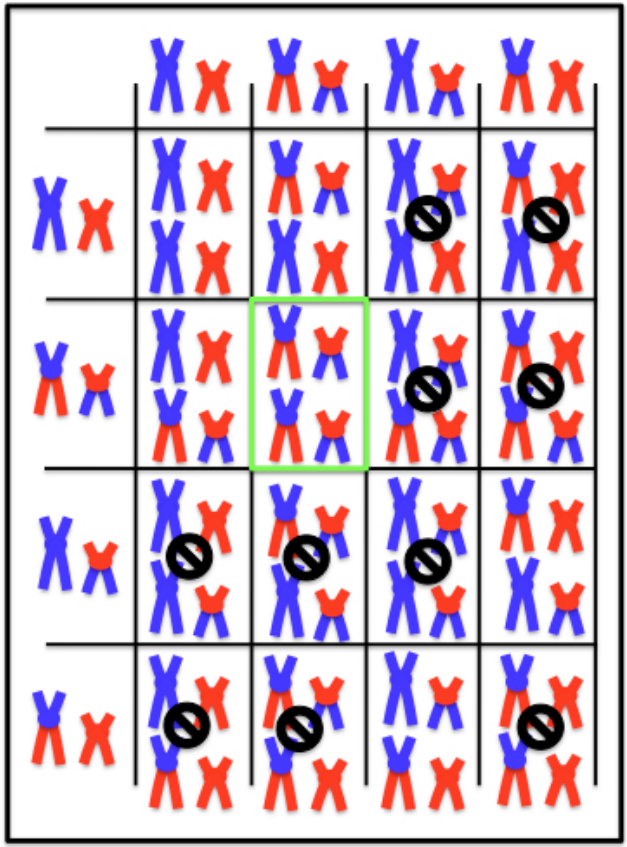

**Figure 2: Predicted chromosomal structures in zygotes issued from individuals carrying a whole arm reciprocal chromosomal translocation.** In an individual carrying a reciprocal chromosomal translocation, only 50% of the offspring is viable (first line). If the cross takes place between two heterozygotes, the proportion of viable offspring drops to 6/16 (= 3/8). Once the translocation has become fixed in a population, crosses with the ancestral stock will generate a first generation (F1 ) that will be 100% viable, but those F1 individuals will be back to the situation of reduced fertility faced by the individuals who first carried the translocation, and this will be true whether they cross to individuals from the ancestral stock, or to individuals homozygous for the translocation.

One hurdle to this hypothesis, however, is that a chromosomal rearrangement such as the textbook example of a whole arm reciprocal translocation pictured in figure 2 will result in a significant decrease of the fertility of the individuals in which this translocation occurs in the first place, with half of the zygotes predicted to be non viable when mating occurs with individuals of the rest of the population, which would not carry this translocation. Once the translocation has become fixed within a group, complete fertility will be restored to all individuals of that group. But for this to happen, heterozygous individuals carrying the same mutation will first have to mate with one another, and under such circumstances, the proportion of viable offspring is predicted to drop even a little bit more, from 1/2 to 3/8 ( Fig. 2), and this is without accounting for the inbreeding depression that would necessarily occur since those individuals would, logically, have to be closely related to one another. Furthermore, the translocation would then become homozygous in only 1/6 of their viable offspring (corresponding to 1/16 of the zygotes). Although other types of chromosomal remodelling, such as inversions or centromeric fusions, may not affect the proportion of viable offspring to the same extent as reciprocal translocations, some effect on the proportion of viable gametes would still be expected since such modifications are known to disturb the phenomenon of chromosomal pairing that takes place during meiosis [51].

Given the above considerations, it is difficult to see how chromosomal translocations could ever take hold in any population and reach fixation unless they were directly associated with a phenotype endowed with a very significant selective advantage. If that advantage corresponded to a dominant phenotype, the remodelled chromosomes could spread to the whole population. Many phenotypes associated to chromosomal remodelling would, however, be expected to get fixed via inbreeding rather than through a selective sweep. For example, a chromosomal modification could bring loci corresponding to an advantageous gene combination near to one another on the same DNA strand, and thus reduce the recombitional load. Many such genetically linked sets of genes can actually be found in the genome, for example in the MHC [52]. This genomic architecture can only have been the fruit of successive events of genomic remodelling, and the fixation of most of those must have required very significant inbreeding. Alternatively, one of the breakpoints may disrupt a gene, and this would be expected to lead to a recessive phenotype, which, once again, would only be expressed in the context of inbreeding.

In addition to the argument that even very closely related species usually do show significant differences in their chromosomal architecture, the view that chromosomal remodelling plays a significant role in speciation is also supported by the relatively high frequency at which chromosomal rearrangements do occur, and could thus conceivably be sufficiently frequent to occur even in small isolated groups undergoing saeptation. Indeed, systematic studies of human karyotypes have revealed that detectable neo-rearrangements occur at a frequency of approximately one in a thousand [53]. Whilst many of such rearrangements may result in spontaneous abortions (as many as 50 % of human reproductive failures could be due to chromosomal abnormalities), many others will be viable, as testified by the fact that as many as one in 625 phenotypically healthy human beings carries a reciprocal chromosomal translocation [54]. Because those translocations do provoke significantly reduced fertility, unless they are linked to an advantageous phenotype, they are expected to get progressively eliminated from large outbreeding populations over successive generations. But finding them at such a sizeable frequency vouches for the fact that individuals carrying chromosomal rearrangements will occur quite often in humans, and hence probably in all species.

Another possibility to consider is that chromosomal rearrangements could be selected for as secondary saeptation steps, i.e. simply because they would reduce fertility of a saeptated group when they breed with the ancestral group, even if it would initially also involve some reduced fertility with the rest of that founder group. Conceptually, this decrease in fertility may sometimes represent a sufficient advantage to be selected for its own sake, as suggested by the observation that chromosomal rearrangements are more frequent between sympatric than between allopatric species of drosophila [55]. The recessive beneficial advantage would then be one of maintaining optimised fertility, but the process would certainly be much more direct, and thus favoured if the chromosomal translocation were directly associated to a mutated gene leading to an advantageous phenotype.

4c) The Wallace effect: Secondary steps towards speciation, i.e. reinforcement .

Once a small group of individuals has 'sprouted' from the ancestral stock, if they have to keep expressing the recessive advantageous traits that drove the constitution of that group, breeding with the ancestral stock will represent a permanent threat for the welfare of their offspring, and the different sizes of the two groups will be a factor that greatly increases the weight of this threat (see [56]). If the initial mutation was directly linked to a chromosomal rearrangement, this would limit the gene flow between the two groups, but would actually further increase the threat because the hybrid offspring would be viable, but less fertile.

After an initial step of saeptation, further steps of reproductive isolation from the ancestral stock would therefore be clearly advantageous for that new, but much smaller group. Within the saeptated group, any further mutation that would increase reproductive isolation from the ancestral stock would therefore be expected to carry a very significant advantage, and could thus rapidly spread to the whole group, which the small size of the saeptated group would further favour.

We can now ask ourselves what sort of mutations and/or traits could intervene in the progressive establishment of completely separated populations, i.e. undetectable gene flow, such as what one witnesses between closely related groups recognised as separate species, although living side by side in natural environments. And I contend that, once a saeptated group has been constituted, in which individuals are all more closely related to one another than to the rest of the ancestral group, further steps of reproductive isolation will not necessarily have to rely on recessive mutations. In the previous paragraphs, I have argued that, in some circumstances, the selective pressure from the ancestral stock may be sufficient to promote further steps of saeptation within the isolated group, based on additional recessive mutations, which would be favoured by the increased inbreeding coefficient, and consequent low mutation load within that saeptated group. On the other hand, dominant traits would presumably spread to the group very rapidly, and would have the added advantage that the process would not require the elimination of the rest of the group. In the long run, as long as hybridisation with the ancestral stock remains a threat, any additional trait that significantly reduces the chance of producing offspring with members of that ancestral population could bring on a sufficient advantage to be selected for. As such, mechanisms that prevent either mating or the formation of zygotes (and hence called prezygotic isolation) such as sexual preference, occupation of niches more remote from the ancestor, gamete incompatibility or even culturally acquired traits could all contribute to protecting the newly formed group from the threat of breeding with the ancestral population. This type of reasoning, which assumes an asymmetric relationship between a newly formed group and a more numerous ancestral stock, provides an explanation for the observation first underlined by Muller in 1942 that incompatibilities between closely related species are very often asymmetric (C&O, p274).

When prezygotic isolation is not complete, and closely related species can still mate and produce zygotes, those hybrids are often found to be either non-viable, or fit, but sterile. Scenarios for the development of this type of barrier between species, which is called postzygotic isolation, are slightly more difficult to envisage because one needs to explain how, although mating has occurred and gametes used to generate zygotes, it can still be more advantageous not to produce offspring at all than to produce hybrids. For explaining this, however, I find one observation particularly useful: whilst problems of viability usually affect offspring of both sexes, problems of sterility usually follow Haldane's rule, and almost always affect only the heterogametic sex ( C&O, p311-312). We can thus consider the problems of explaining hybrid lethality and hybrid sterility as completely separate cases of postzygotic isolation.

Regarding hybrid lethality, I can see two obvious reasons whereby it would be better not to produce offspring at all than to produce hybrids. First, if there is a significant cost to one or both parents for the rearing of offspring that will ultimately be unfit, it will be advantageous to save those resources for the subsequent rearing of "purebred" offspring. And second, if the hybrid offspring occupies a niche that overlaps with that of the purebred offspring, those two types of offspring would then be competing with one another. Sometimes, a further threat for the more inbred offspring could lie with the fact that the hybrids would be particularly fierce competitors for the occupation of the niche because they would benefit from hybrid vigour, and it would thus be best not to produce that hybrid offspring at all.

Regarding the phenomenon of hybrid sterility, I can see three ways whereby it can be promoted, which are not mutually exclusive.

1) The first one lies with chromosomal rearrangements. As already mentioned in the previous pages, chromosomal rearrangements are very often associated to phenomena of speciation, and even closely related species are often found to diverge by several chromosomal structural differences.

Although hybrids carrying a single chromosomal translocation such as the one depicted on figure 2 will only see their fertility drop by 50 % when they mate with homozygous individuals of either type, this proportion will drop further for every additional chromosomal rearrangement and soon reach figures approaching zero. A factor further contributing to sterility is the observation that chromosome pairing has been found to be a necessary step for the proper completion of meiosis, at least in eutherian mammals ( C&O p 262-264, citing Searle1993 ). As we have seen in the previous pages, the fixation of such rearrangements would be most likely to occur when they are directly linked to an advantageous phenotype. The observation that there are more differences in chromosomal architecture between drosophila species living in sympatry than in allopatry [55] does, however, suggest that the reduced fertility provided by such rearrangements may sometimes represent a sufficient advantage per se.

2) The second reason lies with the haploid nature of the sex chromosomes in the heterogametic sex (see addendum 2). As already discussed earlier (section II-3 ), following a process of saeptation, the allelic frequencies of many genes in the newly formed group would be expected to be significantly different from that in the ancestral population. Similarly to what was discussed above, those genes, whether carried by autosomes or sexual chromosomes, would thus represent potential targets for the selection of new mutations carried by the sexual chromosomes: newly mutated genes would still function well with the genotypes frequently present in the isolated group, but would no longer work in combination with the genotypes prominent in the ancestral stock. This would be particularly likely for the heterogametic sex because any mutation carried by one or the other of the sex chromosomes, even those corresponding to a loss of function, would be immediately dominant, as already underlined by Muller in 1940, and formalised as the dominance theory put forward by Turelli and Orr [57]. Since sexual chromosomes are, necessarily, endowed with many genes related to sexual reproduction, a likely phenotype resulting from such a selective process would be one affecting the sexual capacities, and hence result in the sterility of the heterogametic sex. Alternatively, the genes involved in the reproductive isolation may be part of the large number of genes carried by the chromosomes which are diploid in half the individuals (X in mammals and flies or Z in certain insects, fish, reptiles and birds. For the sake of clarity and simplicity, I will use X as an example for the rest of this paragraph, but I could just as well have used Z). Lets us now envisage that a mutation takes place on a gene carried by the X chromosome, such that the gene product will still function well with the allelic form of some other gene found at high frequency in the saeptated group threatened by hybridisation, but will no longer function with the allelic form(s) found in the ancestral group. As long as the individuals of the group breed among one another, that mutation would have no detectable effect, and would thus not really have any reason to spread to the whole group.

But if hybridisation with the ancestral stock took place, because this mutation corresponds to a loss of function, it will most of the time result in a recessive phenotype, and it would thus have the typical characteristics of X-linked deficiencies, i.e. be silent in diploid female offspring, and expressed in the hemizygous males. The X chromosome carries many genes involved in vital functions, and disabling those would presumably result in lethal phenotypes. Under the threat of generating hybrid offspring with an outside group, the individuals carrying such mutations would then be endowed with a definite advantage that would explain how, although neutral within the saeptated population, such mutations could be driven to fixation in the group undergoing speciation. The above scenarios would thus explain why phenotypes of reproductive isolation are often asymmetric, why they are often stronger in situations of sympatry, and provide potential explanations for Haldane's rule, i.e. why, when inter-species crosses take place, if only one sex is affected, it is usually the heterogametic one that is either non-viable [58], which I contend could often occur by recessive mutations of vital genes on the X chromosome, or sterile, by mutations of genes involved in sexual reproduction carried either by the Y or the X chromosome.

3) The third reason for which hybrid sterility may be selected for lies with the fact that sexual reproduction is usually much more costly for females than for males, with the latter having the capacity to produce virtually unlimited numbers of offspring. In the case where a population undergoing speciation competes with the ancestral stock for the occupation of a niche, I contend that the generation of hybrids where females are fit and fertile, but males are unfit can represent an extremely advantageous strategy. These aspects will be developed further in section IV.

**III) There is probably seldom such a thing as truly allopatric speciation** :
In the previous section, we have seen how advantageous recessive traits could promote the formation of small saeptated groups within large populations, and how the need to keep expressing those recessive phenotypes could subsequently drive reinforcement, i.e. further steps of reproductive isolation, based on a whole array of different mechanisms. The recurring theme of the reasoning developed in the previous pages is that reproductive isolation would not arise as a bystander effect of divergent evolution, but would be directly selected for under the pressure of an outside group, most frequently the immediate ancestral population. Even if today, the majority of evolutionary scientists believe that most events of speciation must have occurred in allopatry, I do actually believe that if truly allopatric speciation ever happens, i.e. for whole populations to drift apart sufficiently to become infertile with one another, it must be an extremely slow process, and consequently a very rare occurrence. Indeed, if populations of individuals are completely separated, there will be no selective pressure for evolving features that will further reduce gene flow between the two groups, because the gene flow will already be non extant. If the geographical barrier is later lifted, the features of the individuals in each group will almost certainly be quite different because they will have adapted to their respective environment. Some mechanisms of preference between similar phenotypes may favour reproduction among the individuals having co-evolved, but since there will have been no selective pressure, I contend that there would be no reason why the individuals from either group should have become infertile with those of the other group. This is in fact in complete agreement with what has been very recently described for Caribbean Anoles lizards. Those have evolved independently for millions of years on separate islands that only joined relatively recently to form the large island of Martinique, and more reproductive barriers appear to have been selected for between populations that have evolved side by side to adapt to coastal or mountainous conditions than between those that have evolved on separate islands [59].

This is also exactly what happens with domesticated species. Under conditions of domestication, species can diverge to become very noticeably different, and reproduce for scores of generations under very divergent conditions of selection, yet they do not become infertile with one another. In this regard, I find the example of dog breeds particularly telling. Upon comparing the skeletons of a great Dane and of a Chihuahua, or of a Dachshund and a Saint-Bernard, no taxonomist in their right mind would ever place them as belonging to the same species. Yet, when my steps take me to public parks or other places where people go to let their four legged friends relieve their natural needs, I am often struck (and amused) to see how dogs of very different sizes and appearances can still recognise one another as potential sexual partners. And we do know that they do indeed belong to the same species. They all share exactly the same chromosomal architecture as wild wolves. In fact, if all these dogs of different sizes were placed in a giant enclosure and fed regularly, some sexual preferences between certain types may surface (see long citation of Wallace's book in section V), pregnancies between small females and large males may turn out to be fatal for the mothers, and the smaller males would probably not fare too well in fights with larger ones, but in the end, all those dogs would produce extremely fit offspring that would certainly be much more homogenous than the starting population, and would almost certainly contain genes inherited both from the Chihuahuas and the great Danes. I contend that, if domesticated species do not undergo speciation, it is because the process of selection is carried out by the breeders, and not by natural selection. Under natural conditions, individuals, and groups of individuals, compete directly with one another for the production of offspring and the occupation of a niche, and loosing this competition means dying with no offspring.

In settings of domestication, even if most characters that are selected by the breeders are recessive, and could even sometimes be associated to chromosomal rearrangements, there is never any direct pressure for individuals to stop breeding with the ancestral stock, and there can thus be no selection for either saeptation, or reinforcement. The fact that different domestic breeds, including dogs and pigeons, have now been maintained in effective allopatry, i.e. in complete separation from one another for hundreds of generations without any discernible sign of speciation ever being witnessed is, in my eyes, one of the stronger arguments against the possibility that allopatric speciation, resulting from divergent selection and/or genetic drift, could play a significant role in the phenomena of speciation that are clearly taking place continuously in the natural world.

Another argument against the role of intrinsic genetic incompatibility resulting from a random process in the evolution of reproductive isolation can be found in comparing the estimations of lifetime of species, and of the time it takes for such incompatibilities to develop. Indeed, for both mammals and birds, the fossil record tells us that the average time of existence of a species is around one million years [35], whereas the time it takes for the genomes of mammals to diverge sufficiently to become genetically incompatible is estimated to be around 2-4 million years [60], and well over 10 million years for birds [61]. Given those numbers, one can note that there is a flagrant inconsistency between the biological data and the fossil record since one would have to envisage that most taxonomic species would become extinct before they would have a chance of evolving into genetically incompatible species. I perceive this as a strong argument against the idea that allopatric (and hence passive) genetic divergence could be the main factor responsible for speciation.

Detractors of the views expressed in this essay would not fail to point out that there are many documented examples of allopatric speciation, i.e. where groups of individuals that were geographically separated have become "good species", i.e. completely infertile with one another. But to counter this argument, we only need to think back to the ancestral species, the one which is presumed to have occupied the ancestral territory, and colonised the new one (or, as proposed by Darwin, become split in two by a rising mountain range). If the two modern species cannot breed with one another, then we can safely assume that at least one of the two would also have been infertile with the ancestral species. But, by definition, individuals of that ancestral species were initially present on the two territories, and that species cannot have disappeared before the appearance of a subgroup of individuals that were less fertile with the ancestral individuals, and would eventually lead to the modern species. The logical consequence of this point of view is that, when allopatric speciation appears to have occurred, it actually probably corresponds to several successive steps of 'sympatric' saeptation, with the new, better adapted group replacing the ancestral intermediate.

The most striking examples of speciation often occur on islands, and when Charles Darwin visited the Galapagos in the course of his voyage on The Beagle, the observation of all the very unusual specimen found on those remote islands would later on help him greatly to formulate his theory of evolution, as well as to consider the idea that geographic isolation could contribute to speciation because of the independent evolution of populations that would progressively become infertile with one another.

Let us now consider the phenomenon of island speciation from the point of view developed in the previous paragraphs, i.e. that speciation occurs mostly as a consequence of natural selection, in other words in a context where it is advantageous for subgroups of individuals to stop breeding with the ancestral stock. Colonisation of islands are, inherently, very rare events, and even more so for an obligatory sexual species because this implies that at least two individuals from opposite sexes find themselves on the same island at the same time, which could, quite often, be brothers and sisters descended from a single pregnant female. The initial population will, consequently, go through a very tight bottleneck, with extreme degrees of inbreeding. The resulting reduced fitness of the individuals may, however, be well tolerated because, in the newly colonised territories, those few individuals will have no competition from kin, and presumably very few predators and pathogens adapted to them. Because of this initial episode of inbreeding, however, the cost of subsequent inbreeding will be expected to become much reduced after just a few generations, and this population of colonisers would then presumably multiply quite rapidly to occupy its newfound niche. But the characters of the ancestral stock would probably not be best adapted to their new environment, and conditions would thus seem very favourable for the selection of new characters allowing them to adapt. As we have seen before, mutations leading to recessive characters are much more frequent than dominant ones. And these would be even more likely to come to light in the envisaged conditions, where inbreeding would be favoured both by the small size of the population, and by the fact that inbreeding depression would be minimal. Hence, if a recessive mutation occurred that brought on an adaptive advantage to the new environment of the colonised island, there would be a very significant advantage for the individuals carrying the adapted, recessive, phenotype, to reduce their breeding with the rest of the colonising group. Any mutation coming to reinforce that saeptation would thus be advantageous, and would not necessarily have to be recessive itself. Hence, mechanisms reinforcing the isolation of the adapted group from the rest of the population, such as traits of genetic or post-natally inherited sexual preference, gametic incompatibility, genomic incompatibility or chromosomal rearrangements could evolve within that group, whereas the initial selection of such traits is normally not favoured in larger, more outbred populations, where inbreeding depression is high.

The picture we get from the above scenario is one where, when a secluded niche, such as an island, is initially invaded by very few individuals, successive steps of saeptation and/or reinforcement among a few adapted individuals will be greatly favoured by the initial inbreeding episode. And at every step, the better-adapted descendants of that group would most probably wipe out the less-well adapted stock of their immediate ancestors.

For every one of these steps, the reduction of gene flow with the immediate ancestors would not necessarily be very high but, although that ancestral stock would have long been eliminated from the island, each one of those steps would reduce the fertility between the population of adapted individuals and their immediate ancestors, and consequently would be expected to have a cumulative effect on the fertility between the adapted population and the ancestral stock. Hence, if the population of individuals that have adapted to the island through successive steps of saeptation and/or reinforcement was ever brought back in contact with the more numerous, outbreeding population which stayed on the continent, individuals from those two groups would very probably be completely infertile with one another, even if the latter one had not evolved away much from the ancestral stock. The speciation process so witnessed would, however, not really have occurred in allopatry, but as a succession of sympatric steps which can only occur under the selective pressure of the immediate ancestral stock. An argument that supports the validity of this type of reasoning is the recurrent observation that events of speciation seem especially prone to occur in the context of small populations, such as those promoted by small islands. The size of the niche itself (for example a small island, or a small lake) could indeed be the main factor contributing to the maintenance of a relatively high degree of inbreeding, and hence to the reduced level of inbreeding depression that can promote speciation. Thus, even in the context of islands that are not completely isolated from the regular invasion by individuals from the mainland (such as the Baleares, the Caribbean or the Canaries), or from other nearby islands (such as the Galapagos), small islands have been found to be particularly propitious to speciation in all sorts of genera (birds, lizards, mammals, insects…).

To conclude this section, I would say that, for most cases considered as undisputable examples of allopatric speciation, the times of separation are often much longer than the expected lifetime of the species considered. Also, since in most cases ancestor and speciating groups probably co-exist for much less time than the lifetime of species, it is not surprising that so few cases of speciation appear sympatric. But it is not because we do not see it happen that sympatric speciation does not happen. Thus, contrarily to the stance proposed by Coyne and Orr, I contend that allopatric speciation should not be considered as the default mode (C&O, p84). Rather, to prove that truly allopatric speciation has ever taken place, I advocate that one would have to demonstrate that no step of saeptation has taken place during the evolutionary process, whereby one sub-population would have become reproductively isolated from its immediate sympatric ancestor, and subsequently eliminated it.

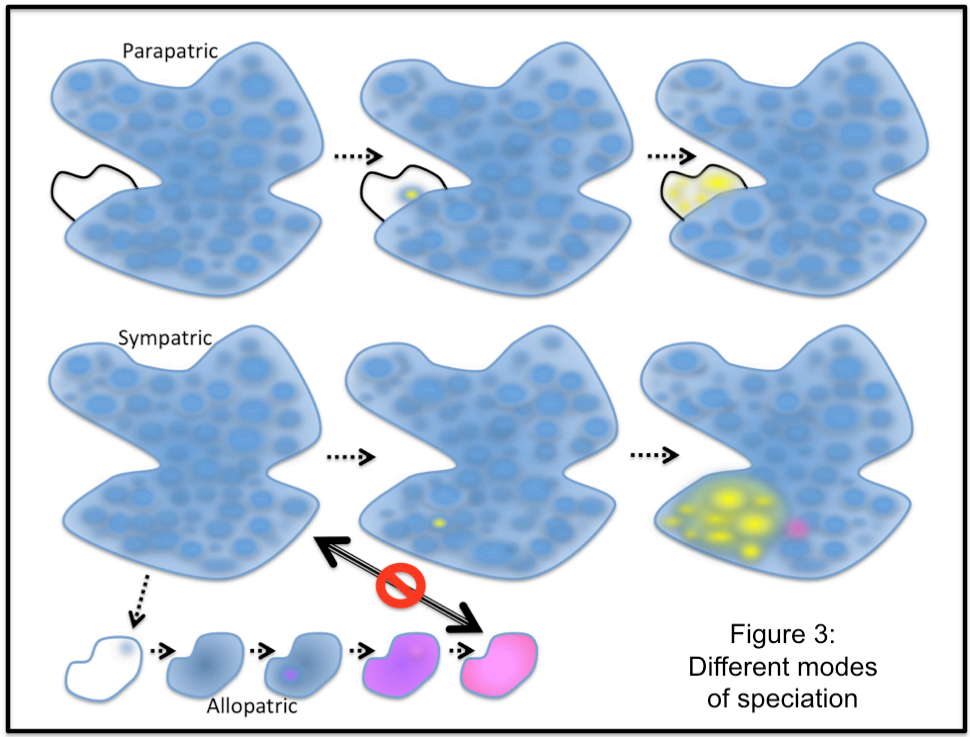

Figure 3:
Different modes
of speciation

## IV) What relationship can be expected between the different modes of speciation, the mechanisms of reproductive isolation that are being selected for, and the diversity of the newly separated population?[6]

Despite the arguments presented in the previous section, there is no denying that the conditions under which speciation occurs (sympatry, parapatry, allopatry) would be likely to play important roles on both what types of reproductive isolation mechanisms are being selected for, and on the size and diversity of the founding population that will ultimately result from the speciation process. In figure 3, I have drawn simplistic sketches that would correspond to scenarios of speciation occurring in those three conditions. In this drawing, the shapes represent the niche occupied by a population. I feel that an important point to underline regarding the nature of niches is that they are not solely linked to geographical constraints, but to many other factors such as the nature of the nutrients, the timing of the life cycle, the identity of other partner species such as pollinators for plants, or hosts for parasites, etc… All in all, I perceive that the defining point between parapatry and sympatry is whether the niches of two populations undergoing speciation are sufficiently non overlapping that neither could ever wipe out the other one. On the other hand, even if two groups have such different life styles or life cycles that they seldom breed with one another, but still compete for the very same food, or for the same territory, one could fully expect that one of the two protagonists will, sooner or later, inherit a new character allowing it to eliminate the other one completely. In short, when occupation of the niche equates to competition for survival, I will call this sympatry; if the two populations can exist side by side without one ever being wiped out by the other one, I will call this parapatry; and when the two populations have so few interactions that neither is a threat for the other one, I will call this allopatry.

In figure 3, within the niches, I have not represented populations as uniform entities, but as fragmented in subpopulations, where the less intense areas correspond to reduced densities of population, and hence higher degrees of inbreeding. Under conditions of parapatric speciation, the group undergoing speciation will colonise a different, adjacent niche (new territory, different nutrients, different breeding period …). For the reasons exposed in section II, the process of speciation will be much more likely to be triggered if the character that allows this colonisation is recessive, and hybridisation between the two groups would thus represent a much bigger threat for the members of the newly formed and less numerous group than for the ancestral stock. Under such conditions, one would thus expect reinforcement, or further saeptation, to be selected for essentially in the younger group.

Particularly interesting examples of parapatric speciation are those provided by ring species, whereby new species arise in successive steps around a circumventable geographic barrier such as a mountain (Greenish Warbler around the Himalaya), an ocean (Herring Gulls around the Atlantic Ocean) or a valley (Ensatina Salamanders around the central valley in California) [63]. In the end, although some gene flow persists between direct neighbours, i.e. between ancestral stock and new populations having colonised a new parapatric niche, the species that end up meeting at the opposite end of the ring are completely infertile with one another. The simplest explanation for this type of phenomenon seems to be that the additive effect of incomplete reproductive barriers will finally result in truly isolated species. With regard to the ideas proposed here, it will be particularly interesting to see if characters can be identified that have contributed to the progressive adaptation of the species along the barriers, and when those are due to recessive characters, whether this is accompanied by more significant reproductive isolation from the ancestral stock.

---

[6] The views developed in this section are somewhat related to the considerations on founder effects developed as models of 'genetic revolution' by Mayr (1954) (see C&O p 387-393), 'founder-flush theory' by Carson (1975) and 'genetic transilience' by Templeton [62] , but contrarily to those, I do not believe that drift under conditions of true allopatry would suffice to promote the fixation of characters of reproductive isolation other than on extremely rare occasions.

In a context of sympatric speciation, the younger group having undergone saeptation will have to compete directly with the individuals of the ancestral stock for the occupation of the niche. Whilst the speciating group would have the advantage of the newly acquired, but recessive, advantageous trait such as the resistance to a pathogen, the ancestral group would have the important advantage of a much more numerous starting population, presumably harbouring more diversity. The counterbalance of this would be, however, that this larger and older group would probably also carry a heavier mutation load than the speciating group. In the context of a competitive struggle between the two groups, population densities would presumably thin out for both groups, leading to increased inbreeding. Whilst this would not be a problem for the younger group, it would most probably result in a very significant drop in fertility for the older and more numerous ancestral stock because it would carry a heavier mutation load. This view is supported by a recent report showing that the fitness of an invasive species of ladybirds is actually increased by bottlenecks having resulted in a decrease of their mutation load [64]. In such circumstances, because of both the newly acquired selective advantage having driven the saeptation, and its lighter mutation load, the odds would thus seem very likely to tilt towards the younger population most of the times.

The lower part of figure 3 sketches the scenario of island colonisation developed in the previous section, whereby a handful of founding individuals give rise to a completely isolated population, and the high inbreeding conditions, resulting in low mutation load, favour successive steps of sympatric saeptation that will ultimately result in complete infertility between the population occupying the island and the ancestral stock.

The common point between the last two scenarios of speciation is that the newly formed groups have to compete with their direct ancestors for the occupation of the niche. In line with Darwin's views, the stakes in this struggle are 'the survival of the fittest', which implies the ultimate elimination of the other kind. Hence, for a population undergoing sympatric speciation, to paraphrase General Philip Sheridan, "the only good ancestor is a dead ancestor". For achieving this, I perceive that post-zygotic mechanisms, which will often affect only the heterogametic sex, are particularly effective strategies which can have, as we will see, multiple types of advantages. Indeed, in the context of a newly formed group, even if the members of the group somewhat benefit from the advantageous recessive character they express, they may also be affected by more inbreeding depression, and the much smaller effectives of the newly founded population could easily be overwhelmed by the sheer number of the competitors. Lets us now envisage the consequences of generating hybrid offspring where one sex is fertile and the other one is either sterile or dead. For the next generation, this will result in a deficit in potential partners of the heterogametic sex, and that situation will have several consequences: i) it will free up some space in the niche that the purebred members of the saeptating group can then move into without competition. In a further elaboration, one could even envisage that there could be an advantage to the sterile hybrids being very fit because of hybrid vigour. They would thus occupy a large portion of the niche, but would eventually die with no offspring, and leave all that space vacant for the offspring of their fertile neighbours. ii) In mammals and flies, where the males are heterogametic, a further advantage would be conferred by the fact that the males can produce offspring with numerous partners at very little cost. In conditions where hybrid females remain fertile and hybrid males are sterile, the males from the saeptated group would thus find themselves with more potential partners. Subsequently, the offspring resulting from mating with those hybrid females would generate more fertile females, and, if the sterility was due to only one locus, presumably only 50 % of fertile males. Although this type of reasoning could also apply to species where the females are heterogametic (certain insects, fish, reptiles and birds), this effect of the process would be somehow restricted by the fact that females are, by nature, restricted in the number of eggs, and hence offspring that they can generate. This could, however, be compensated for by monogamous behaviours, because a sterile hybrid female would effectively neuter the sexual activity of her fertile male partner. In this regard, it is quite remarkable to note that, whilst 90% of bird species are monogamous, only an estimated 3% of mammals are[7]. iii) An important consequence of the process of 'sleeping with the enemy' will be that, among the offspring resulting from crosses between the purebred stock and the hybrid homogametic offspring, 50 % will become homozygous for the advantageous recessive trait, and could thus formally join the saeptated group. Through this type of process, the saeptated group, which may initially have been endowed with rather limited genetic diversity, may thus progressively incorporate a significant portion of the diversity present in the ancestral stock.

This last point brings us to consider the question of the evolution of genetic diversity through the process of speciation. In this regard, great insights can be gathered from comparing the diversity of the major histocompatibilty complex (MHC) between closely related species. The MHC, which is found in all jawed vertebrates, is the most polymorphic region of their genomes. The reason for this is that it is involved in many aspects of immunity, and thus under very strong selection, with the diversity of MHC molecules being used to fight off the amazing capacity of pathogens to adapt to their host[8]. Comparisons of allelic diversity between closely related species such as human and chimpanzee [45], or mouse and rat [46], have revealed that certain polymorphisms of MHC molecules have survived all the successive steps of speciation that have separated each species from their common ancestor. Such observations thus strongly suggest that speciation, even if it involves inbreeding, does not necessarily have to occur via very tight bottlenecks, and thus tend to support the validity of the types of scenarios proposed at the end of the last paragraph.

---

[7] On the subject of bird monogamy, in The Origin, Darwin himself underlines several times the fact that it has been possible to derive and keep so many different breeds of pigeons because those can be paired for life, and then kept in the same aviary. His report of the common observation of sudden reversion of certain phenotypes towards wild type phenotypes does, however, vouch for the fact that even among birds, some adultery still occurs regularly.

In the case of human and chimps, the presumed last common ancestor is called Nakalipithecus, who lived some 10 million years ago. Since then, although the precise details of our ancestry are stilled hotly debated, it is clear that our family tree must have counted at least half a dozen successive species, first belonging to the gender Australopithecus ( anamensis, afarensis, africanus …), and then to the gender Homo ( habilis, erectus …). Over that time, 30 million sequence differences have accumulated between the human and chimp genomes, corresponding to 1% divergence, as well a 10 chromosomal modifications (9 inversions and 1 centromeric fusion ), of which one can reasonably expect that about half must have taken place in the branch leading to humans, and the other half in that leading to chimps. Incidentally, although it is interesting to note that the number of chromosomal rearrangements is roughly of the same order as the number of speciation steps on the presumed path between Nakalipithecus and the two modern species, it does not prove in any way that each of those rearrangements was necessarily correlated to a phenomenon of speciation. Indeed, some of those chromosomal rearrangements could have become fixed in the population because they were directly linked to dominant beneficial characters, and would thus have undergone selective sweeps.

Another intriguing recent observation is that the evolution of humans has involved the loss of more than 500 stretches of DNA which are otherwise found in chimps and in many other mammal species [66]. Since most of these DNA sequences are located in non-coding regulatory regions, such alterations would be more likely to result in intermediate  phenotypes in hybrids than in purely recessive traits. Following the reasoning developed in the previous pages, most of these mutations may thus have spread to the whole populations, but the fixation of some may have involved and/or contributed to the isolation of relatively small groups of individuals from their direct ancestors.

When two separate human genomic sequences are compared, one allegedly finds around 0.2% divergence, which would amount to 6 million mutations per haploid genome. Our species has only been around for 250.000 years, and thus approximately 10.000 generations. As we have seen previously, new mutations accumulate at the rate of approximately 60 per haploid genome per generation. One would thus expect only 600.000 new mutations to have accumulated in each genome since the appearance of Homo Sapiens. The level of divergence seen between human genomic sequences thus provides additional support for the fact that events of speciation, even if they implicate a process of inbreeding, do allow for the conservation of high levels of genomic diversity[9].

Genomic divergence between populations tends to be highly variable across the genome, and divergent selection has been proposed as the main reason for this observation [67]. This unevenness of genomic diversity would, however, also occur with the various scenarios envisaged in the previous pages: the genomic regions surrounding the loci having contributed to driving reproductive isolation would be expected to have reached fixation very rapidly, and hence to show very limited diversity. Furthermore, the rate of fixation would be very different if they corresponded to recessive or to dominant characters. Indeed, if a recessive character leading to saeptation is being selected for, it will necessarily be fixed very rapidly in the saeptated population, and one would thus expect a few centimorgans of the genomic region surrounding the recessive allele to become fixed with it, and hence to harbour very limited diversity, and this would be even more true for co-recessive traits. Conversely, whilst the allelic frequency of an advantageous dominant character will rapidly increase to 70 or 80 % in a population, it will take a very long time to reach complete fixation, i.e. to eliminate all the non-advantageous recessive alleles. Somewhat ironically, it is actually inbreeding that would allow the elimination of the last ancestral, recessive and less advantageous alleles, via a mechanism equivalent to the one described in section I-3. Consequently, during all that time before complete fixation of the dominant allele, there will be many chances for crossing-overs to occur around the gene coding for the advantageous dominant trait, and the size of the region of reduced diversity should therefore be much more limited than in the case of the selection for a recessive trait.

---

[8] Regarding the relationship between speciation and immune responses, an extremely recent paper (published during the refereeing process of this manuscript) suggests that the loss of certain antigens expressed in either sperm or placenta may be contributing to the establishment of reproductive barriers because females lacking that particular antigen could then develop an immune response against it, The antigen studied in that report is the Neu5GC glycan, which is present in primates and not in humans, due to the loss of the CMAH enzyme which occurred in early hominins [65].

---

[9] The model of Transilience developed by Templeton [62] addresses similar issues from the point of view of population genetics.

The prediction that follows this reasoning is that this may actually provide the means to identify the regions carrying the genes involved in events of speciation, and conceivably even the very genes having driven the speciation[10], as well as a reasonably accurate estimation of the dates at which it happened. Indeed, as is already well under way for humans with the 1000 genome project [68], if one documented the levels of diversity of silent intergenic DNA over the whole genome for a good number of unrelated individuals belonging to the same species, this would not only provide the means to really evaluate the degree of inbreeding within a population, as well as the inbreeding coefficient for each individual, but one would also expect to be able to rapidly identify regions of limited diversity. Although the occurrence of chromosomal rearrangements may confuse the interpretation [51, 69], the gene responsible for driving the fixation would be expected to be at the centre of such regions, and the level of divergence of intergenic sequences within those regions would provide a relatively precise estimate of the time of fixation.

Finally, the slope with which the level of diversity decreases with genetic distance from the centre would provide an indication of whether the character that drove the fixation was recessive, and was hence probably involved in a phenomenon of saeptation, co-recessive, or dominant, and hence corresponded to adaptive evolution (including mechanisms of reinforcement). If such an exercise was carried out for tens of thousands of markers distributed over the whole genome in hundreds of unrelated individuals belonging to the same species, this could, I predict, provide a very informative picture of the successive steps of speciation in the evolutionary history of that species[11].

**V) The existence of species can only be transitory because it corresponds to a metastable equilibrium**.
The field of Taxonomy was initiated by the Swedish zoologist, Carolus Linnaeus, who, in his book *Systema Naturae* (first edition published in 1735, tenth and last in 1758), recorded some 9000 species of plants and animals. Today, this number has reached several millions, and it is estimated that around ten millions species of plants and animals of more than one millimetre inhabit our planet

[70], and this number probably corresponds to less than 1% of the species that have existed since metazoan life started on earth 1.5 billion years ago, with an estimated average lifetime of a species around 4 million years, based, obviously, on morphological data from the fossil record rather than on biological ones [71, 72][12].

In this regard, the estimated number of 5000 extant mammal species represents only a tiny portion, and mammal species are particularly short lived, with an estimated average lifetime of just one million years, whilst reptiles, and species of higher plants and trees can last over 20 million years. All in all, it is pretty clear that very few of the species that we can find on earth today were there 20 million year ago. As already underlined in the introduction, the somewhat uncomfortable, but inescapable conclusion from this observation is that all the species that surround us, including our own, are bound for extinction.

The theory developed in the previous pages can actually lead us to suggest an explanation for this observed inherent tendency of species to disappear over time. Indeed, we have seen that the mutation load in a population is inversely related to the degree of inbreeding in this population, and the existence of species thus appears to rely on a fragile, metastable equilibrium, which I find very appropriate to represent in the context of the Yin Yang symbol to evoke the balance between degrees of inbreeding and outbreeding (Fig. 4).

On the one hand, increased inbreeding will initially be costly, but once the safeguard of a sizeable mutation load has been lifted by a few rounds of inbreeding within a small group of individuals, further inbreeding is likely to have more advantages than disadvantages: favouring the expression of adaptive recessive phenotypes, keeping the mutation and the recombination loads down, reducing the cost of sex, and promoting collaborative behaviours by population fragmentation. But, as we have seen, this increased inbreeding will at the same time favour the appearance of saeptated groups, for whom the way of existence will equate to the elimination of the ancestral stock, and hence the disappearance of the original species.

---

[10] If the selective force driving the selection was a particularly nefarious pathogen, however, it may well be that it would have disappeared with it's host, and all that would be left would be an allelic form of a gene that was once used as a receptor for a now long vanished pathogen.

[11] In this regard, I would not be surprised if a locus having driven saeptation in the ancestors of the laboratory rat, Rattus Norvegicus, was one day found to lie near the MHC because the rat MHC has been found to have a much more restricted diversity of MHC haplotypes than those found in Mouse or Human. Alternatively, it may be that the ancestors of the rat population have gone through one or several tight bottlenecks, resulting in limited diversity of sequences through the whole genome.

[12] If we consider that sexually reproducing eukaryotes have existed for 1500 million years, and if the average lifetime of a species has been 4 million years since then, this amounts to an average number of approximately 400 steps of speciation separating the species of today from the first metazoan ancestors. If, along the way, every species had speciated into two descendants every 4 million years, this would give a number of species equal to $2^{400}$ which is so big that my desktop calculator refuses to calculate it, but which I make out to be something near $10^{120}$, which is a number vastly superior to the number of atoms on earth (ca. $10^{50}$). From this type of calculation, we can see that the struggle for existence highlighted by Darwin and Wallace for individuals must also apply to species, and that the destiny of most species is either to disappear, or sometimes to yield one, and seldom more descendants.

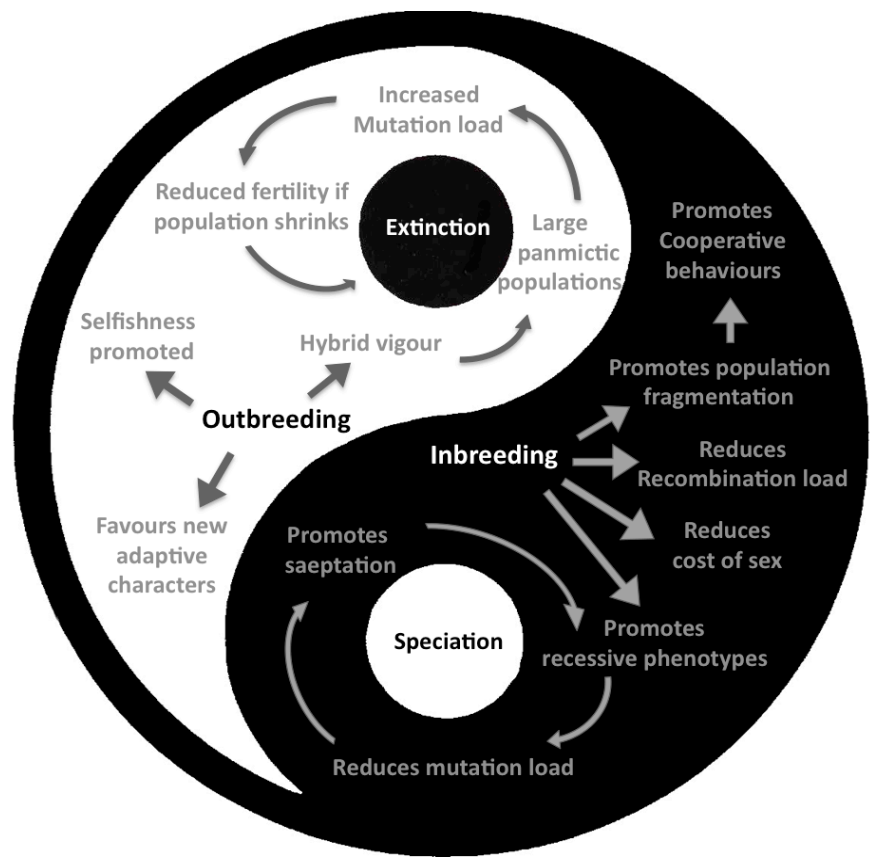

**Figure 4: The existence of species rests on a metastable equilibrium between inbreeding and outbreeding.**

On the other hand, extensive outbreeding will bring hybrid vigour, and delay the appearance of reduced fertility due to the accumulation of recessive deleterious mutations. This type of phenomenon may be particularly prominent for very successful species that end up effectively panmictic rather than being fractionated into smaller subpopulations. The evolution of individuals within such population would then favour the strongest, longest lived, largest individuals. In this regard, van Valen underlined that, for mammals, *"Occasionally, a small mammal becomes a large one, but a large mammal never becomes a small one"* [73] [13]. Regarding the fossil record on which van Valen based most of his work, it may in fact be worth to consider the possibility that it's composition may be biased towards species that, having adopted a panmictic strategy, would see the size of the populations swell very rapidly to very large numbers, but would also, as proposed by Carson [5], be going down an evolutionary dead end. Indeed, after only of few dozens of generations, the accumulation of recessive mutations would subsequently prevent any chance of any significant degree of inbreeding, and hence any possibility of a fresh start via an event of speciation. Because of their large populations, and their persistence over relatively long periods because of their decreased capacity for evolution, such species would thus have a high chance of "making it" into the fossil record. But, at some stage down the road, such populations would inescapably fall victims of their own success because they would have a very poor capacity to respond to crises triggered by increased levels of selective pressures by outside factors such as pathogens, predators, competitors, natural catastrophes or shifts of the climatic conditions. For example, this type of situation may well have applied to the Multituberculates, which were very common mammals during the paleocene, but underwent complete extinction during the Eocene [74], probably because of the competition of the newly arisen rodent competitors. If there are 10 million species on our planet, and the average lifetime of a species is 4 million years, then the turnover rate should be under three species per year. This may appear as a clear underestimate, especially in our modern era, which has been dubbed the anthropocene, since ecological changes due to human activities provokes the disappearance of thousands of species every year. We should, however, bear in mind that the extinction rates measured by paleontologists are those which correspond to the disappearance of organisms based on the anatomical features detectable in fossils, and that species differentiated by colours, timing of life cycle or breeding habits would not be registered. Similarly, events of speciation corresponding to the loss of one or a few recessive traits would almost certainly not de detected by the fossil record. Based on the arguments raised above, I perceive that most events of extinction identified by paleontologists probably correspond to those of panmictic species having succumbed to increased selective pressures which initiated a process of irreversible decimation because of high inbreeding depression resulting from important mutation loads.

---

[13] It should be noted, however, that these rules do not seem to apply to island mammals that are larger than rabbit size, which tend to become smaller there. Leigh van Valen called this the island rule.

Hence, one major difference between the outbreeding and inbreeding strategies is that the former leads to a very high probability of ultimate extinction, whilst the latter would lead to an increased probability of formation of saeptating group(s) within the population, that will ultimately cause the elimination of the ancestral group by one or more descendant new species. The outbreeding strategy, however, is probably the one that takes place most frequently in natural populations because it brings on much more immediate advantages. Darwin and Wallace's theory of evolution, which is concerned with the acquisition of new adaptive traits, is in fact based on considering this type of strategy. And it is indeed by relying on the flexibility and variability of the genome taking place in parallel in the numerous individuals of a large population that one can hope to see surface the very rare events that will correspond to new adaptive functions. Since such new traits will, most of the time, be expressed in a dominant fashion, they will thus rapidly spread to the whole population.

On the other hand, there are many instances where it is advantageous to get rid of a character, and the susceptibility to pathogens seems to be particularly relevant here. But, as discussed at length in the previous pages, the loss of a function usually corresponds to a recessive trait, and the expression of recessive traits necessarily calls for some degree of inbreeding. Another important consideration is that inbreeding will be necessary to maintain, and ultimately fix certain gene combinations, and this will also be true of chromosomal rearrangements. On this subject, W. Shields offered the interesting point of view that one can consider that individuals belong to separate species when the intensity of outbreeding depression is so high that no long term descendants can result from their crossing [2].

The degree of inbreeding necessary to keep mutation loads in check is probably much less than that required to promote speciation, and if we consider the very divergent outcomes of the two strategies, and the timescales involved in evolutionary processes, we can easily see why most natural populations are so seldom panmictic, as outlined by Wright over 60 years ago [4, 12]. Extensive outbreeding may indeed be endowed with short term advantages for individuals, but in the long run, there is not really a choice between the two strategies in the struggle for survival. And I thus contend that, if so many of the species that surround us are not panmictic, it is because they derive from a long line of ancestral species that have not succumbed to the short term benefits of excessive outbreeding. From the above arguments, I conclude that, even if inbreeding is not immediately advantageous, it is an absolute requirement, an unavoidable price to pay, for long term survival of the descendants. This probably provides the ultimate example of group level selection because species that fall for the short sighted advantage of extensive outbreeding will relatively rapidly have to face the cost of unmanageable mutation loads, leading to unavoidable extinction.

Many factors contribute to the fact that natural populations do not become panmictic. First, the world is so vast that most species are necessarily fragmented into myriads of small groups, with every event of colonisation providing an opportunity for episodes of increased inbreeding, resulting in a reduction of the mutation load. And there is also a natural tendency for individuals to associate with kin, as Wallace himself underlined in the following paragraph taken from his book, 'Darwinism', in Chapter VII's section entitled 'The Isolation of Varieties by Selective Association', (1889), which I do not resist the pleasure of sharing with you:

*But there is also a very powerful cause of isolation in the mental nature—the likes and dislikes—of animals; and to this is probably due the fact of the comparative rarity of hybrids in a state of nature. The differently coloured herds of cattle in the Falkland Islands, each of which keeps separate, have been already mentioned; and it may be added, that the mouse-coloured variety seem to have already developed a physiological peculiarity in breeding a month earlier than the others. Similar facts occur, however, among our domestic animals and are well known to breeders. Professor Low, one of the greatest authorities on our domesticated animals, says: "The female of the dog, when not under restraint, makes selection of her mate, the mastiff selecting the mastiff, the terrier the terrier, and so on." And again: "The Merino sheep and Heath sheep of Scotland, if two flocks are mixed together, each will breed with its own variety." Mr. Darwin has collected many facts illustrating this point. One of the chief pigeon-fanciers in England informed him that, if free to choose, each breed would prefer pairing with its own kind. Among the wild horses in Paraguay those of the same colour and size associate together; while in Circassia there are three races of horses which have received special names, and which, when living a free life, almost always refuse to mingle and cross, and will even attack one another. On one of the Faroe Islands, not more than half a mile in diameter, the half-wild native black sheep do not readily mix with imported white sheep. In the Forest of Dean, and in the New Forest, the dark and pale coloured herds of fallow deer have never been known to mingle; and even the curious Ancon sheep of quite modern origin have been observed to keep together, separating themselves from the rest of the flock when put into enclosures with other sheep. The same rule applies to birds, for Darwin was informed by the Rev. W.D. Fox that his flocks of white and Chinese geese kept distinct.*

*This constant preference of animals for their like, even in the case of slightly different varieties of the same species, is evidently a fact of great importance in considering the origin of species by natural selection, since it shows us that, so soon as a slight differentiation of form or colour has been effected, isolation will at once arise by the selective association of the animals themselves; and thus the great stumbling-block of "the swamping effects of intercrossing," which has been so prominently brought forward by many naturalists, will be completely obviated.*

Such types of preference for closely related individuals may not need to be based on purely genetic factors, but could be culturally inherited, i.e. transmitted as memes [39], as has been documented many times with the phenomenon of imprinting in birds raised in nests of different species. In addition, there is probably also simply a natural tendency of individuals with similar phenotypes to breed more willingly and effectively with one another.

The concept that social structures and altruism are more likely to arise between genetically related individuals was initially developed by Hamilton [40, 41], and this was later coined as the *green beard altruism effect* by Richard Dawkins [39], to describe a hypothetical gene that would result in both a detectible trait and in altruistic behaviour among those bearing it. The occurrence of such a gene seems, however, rather unlikely, and there are, indeed, very few reported occurrences of such possible green beard genes. Moreover, the green beard hypothesis posits that the green beard would be a dominant character, i.e. a gain or a change of function, which, as underlined repeatedly in the previous pages, is far less likely to arise through mutations than a loss of function.

More recently, however, mathematical modelling of *beard chromodynamics* yielded the conclusion that the most stable arrangement for the maintenance of altruism was for a situation where beard colours are polymorphic, and the genes for altruism only loosely coupled to those for beard colours [75]. In other words, populations are most likely to get organised into groups of individuals that behave altruistically towards one another if the polymorphism of characters in the global population allows individuals to recognise those that are most likely to be genetically related to themselves, i.e. the ones that look like them, and the social genes do not have to be the same as those used to evaluate kinship. In French, we have a proverb that says '*Ce qui se ressemble s'assemble*', and the existence of races and varieties in the natural world vouches for the spontaneous occurrence of structuration of natural populations which can only be the result of some preferential association, and reproduction, between individuals that are more closely related to one another than to the rest of the population. The recent finding that, even in the fungus Neurospora, some degree of reproductive isolation could be observed between stocks that had been grown for relatively short periods in different selective environments [76] indicates that a tendency for preferential mating with individuals bearing similar phenotypes can occur even in microscopic organisms.

For species that live exclusively on land, most niches would naturally have patchy distributions, providing an automatic enforcement of a fragmentation of populations. But for species that live in the sea, or that can take to the air such as birds or insects, there will be no enforced limitation to taking advantage of the short term benefits of extensive outbreeding. In this regard, it is actually remarkable to note that many species of birds, fish, or marine mammals not only show strong preference for kin characters, but, as outlined by Shields [2], also show strong philopatry. On this subject, I am in complete agreement with his views that the tendency of these animals to come back to breed to the very same place where they were born was most probably selected for because it promotes a significant degree of inbreeding. This would once more be the result of group selection, with the groups or species adopting more outbreeding strategies succumbing rapidly to unmanageable mutation loads. Individual examples of various mechanisms promoting inbreeding will be developed in the next section.

The picture we come to at the end of this section is indeed one of a Yin Yang equilibrium between outbreeding and inbreeding, in line with the notion of optimal outbreeding proposed by Bateson [13], whereby outbreeding is necessary for the acquisition of new characters favoured by the parallel evolution of the many individuals in whole populations, and inbreeding is necessary to eliminate not only the deleterious recessive mutations, but also to maintain certain favourable gene combinations, and lose certain functions that have become undesirable, and in particular the susceptibility to pathogens. Like many things in biology, including life itself [77], the existence of species has all the characteristics of a metastable equilibrium because departing from it will promote further distancing from the equilibrium, either of outbreeding, destined for extinction, or inbreeding, which will favour speciation. The observation of the constant rates of extinction within genera reported by van Valen [35] does find an explanation in this model because the occurrence of events of destabilisation of the equilibrium would correspond to presumably rare stochastic events, as was concluded by a recent study [78].

**V) Many classical examples of speciation appear to fit the model proposed.**

For a scientist, one of the main problems in trying to understand the phenomenon of speciation is that it is basically impossible to perform experiments that will lead to bona fide speciation, i.e. complete reproductive isolation between two groups of individuals. The first reason has been dubbed a 'methodological contradiction' by Lewontin in 1974. Indeed, studying the genetics of speciation involves experiments that cannot be done, i.e. cross species that are, by definition infertile with one another.

The second reason is one of the time scale, and/or of the size of the samples required. Indeed, because the mutations that lead to evolution and/or speciation occur purely by chance, they will only ever occur very rarely. In the previous sections of this essay, I have argued that speciation is promoted by inbreeding, i.e. by the small size of breeding groups. But the probability of a new mutation occurring in the very few individuals of that breeding group is consequently infinitely small, and if one started from just one such breeding group, one would probably have to wait for thousands of generations for saeptation to occur. This is probably similar to what happens on isolated islands, where the time scales estimated to reach speciation are of the order of tens of thousands of years (and no one would ever get funding for an experiment on this time scale ;-). Of note, in the Park Grass Experiment performed in East Anglia, adjacent plots of meadow have been continuously subjected to different fertilizer treatments since 1858, and some signs of reproductive isolation, with reinforcement via different flowering times, have been identified between populations derived from one type of grass [79]. I could, however, find no reported data regarding the reciprocity of this reproductive isolation, or about the character which may have been selected for (and hence even less about its dominant or recessive nature).

Another approach to test the validity of a model of speciation is to make predictions regarding the type of results that could be expected from the model proposed, and then to check whether those predictions hold up when the genetic source of reproductive isolation is dissected between closely related species. Many such predictions can thus be made from the model(s) developed here, with the main ones being that i) When speciation occurs, i.e. when a new group gets reproductively isolated from an ancestral population, this should very often be due to one or more recessive advantageous mutation(s) occurring in the new group. ii) Genera that undergo a lot of speciation should be those that carry the lowest mutation loads, correlating to lifestyles favouring inbreeding such as frequent self fertilisation, or very fragmented populations.

As we will see in the following pages (summarized in table 2), those predictions match the situations of many (and possibly most) of the most extensively documented examples of speciation in wild populations, and this holds true for all branches of plants and animals, including our own [14].

### Fish
The first ancestral vertebrate was a fish that lived in the ocean over 500 million years ago, and today more than 30,000 species of fish occupy our planet's waters, having adapted to the many diverse habitats found in oceans, seas, rivers, lakes… Since all these different species of fish occupy niches of very different sizes and architectures, one would not expect the same inbreeding/outbreeding strategies to have been selected in all fish, for example those breeding in the open ocean and to those living in lakes or in small streams.

Among fishes, the most often cited and discussed case of speciation is that of the haplochromine cichlids found in the African lakes, which are hosts to hundreds of closely related species. Cihlids actually represent a very large family of over 3000 species which are widely distributed over the lakes and rivers of Africa and South America (some of their best know representatives are the Tilapia, which are used extensively for aquaculture, and the Angelfish, of Amazonian origin, which is commonly found in domestic tropical aquaria). For many years, speciation of cichlids has been the subject of various heated controversies. The first one over whether the speciation seen in African lakes really occurred in sympatry or allopatry, and a very strong argument for the fact that sympatric speciation was possible for cichlids recently came from work on cichlids found in a very small crater lake of central America [80].

Another contentious issue has been whether some of the sympatric populations found in the African lakes represent varieties or 'good species'. This latter question is particular relevant for the diverse groups of Lake Victoria which were initially behaving as completely isolated groups, because of assortative mating mostly based on colour patterns. In recent years, as a result of human activity, the waters of the lake have become progressively more and more turbid, and the various vibrant colours which characterised the males of the various types of cichlids one hundred years ago disappeared progressively because the members of the various 'species' hybridised with one another, to result in a much more homogenous and duller population [81]. The first point to make from this observation is that, once again, it underlines the difficulty of defining species, since it shows that effectively good species can later on become varieties when their living conditions change, even in the wild. From the point of view of the ideas developed in this essay, the observation that hybridisation resulted in the disappearance of the vibrant colours is very reminiscent of the colours of Darwin's pigeons, and does thus strongly suggest that the expression of many of these vibrant colours corresponds to recessive mutations, and that the expression of these must therefore require inbreeding. Conversely, another bright phenotype know as orange blotch is a dominant trait expressed only in females, to whom it brings a selective advantage [82]. This phenotype is encountered in certain individuals of several species of the lakes Malawi and Victoria, but has not led to speciation. Thus, as predicted by the model, bright colours associated to recessive mutations lead to reproductively isolated groups, whilst those associated to dominant mutations do not.

---

[14] Regarding the predictive value of the proposed model, my relative naivety on the subject of speciation, to which I have already alluded to in the foreword of this essay, has proven to be a great advantage. Indeed, it not only contributed to my capacity to have ideas that seem to diverge quite significantly from the currently accepted dogmas, but once I had formally developed these ideas, it allowed me to gather data which already existed in the literature, but of which I was not aware, to test the validity of the model. In this respect, I think it is worth underlining that my writing of all the previous sections preceded the writing of this last section, which coincided with my acquisition of the knowledge about the precise details corresponding to the various models of speciation. I consider this to be very significant because the model therefore has the added strength of having proven to be predictive rather than being adapted to explain the evidence.

We will now turn to the reason why cichlids may have such a tendency to undergo explosive speciation, and I contend that an explanation can be found in another recent study, based on *Pelvicachromis taeniatus*, a river-inhabiting African cichlid. In this report, Thünken and collaborators look at the mating preferences of this cichlid fish, and find a strong preference for kin over non-kin, and, importantly, no sign of inbreeding depression in the offspring resulting from sib mating [83]. These observations strongly suggest that, in cichlids, there can be an inherent tendency to prefer mating with very closely related individuals. In rivers, where the context is presumably easily disruptive for population structures, this preference for kin may be important for the maintenance of low mutation loads, and for the expression of recessive phenotypes. When cichlids with extensive inbreeding habits colonise lakes, however, the disrupting of groups of closely related animals would no longer occur, and this could promote frequent saeptation and result in the explosive speciation that is witnessed in African lakes. The existence of those species will, however, be particularly fragile, because of the segmentation of the whole population into a myriad of very small groups, that will be reproductively isolated from one another, and will thus not benefit from sharing their gene pool with a very large number of individuals. And the high probability of saeptation of some subgroup will, in addition, represent another permanent threat for the occupation of the niche. The situation of cichlids thus represents an inherently unstable situation, with an excessive tendency for speciation. Ultimately, a more stable equilibrium would presumably be reached if some sub-population lost its inherent preference for kin, and started adopting a more outbreeding strategy.

Although visual clues clearly play a pivotal role in the preferences of cichlids for kin, another important component could possibly involve the MHC since, in other fish genera, MHC discrimination has been shown to be involved in inbreeding avoidance, or more correctly, incest avoidance. In this regard, the case of salmon provides a particularly interesting example of equilibrium whereby the MHC is used both by the mating adults for promoting MHC diversity (but not for inbreeding avoidance) [84], but conversely, fertilisation is apparently favoured when eggs and sperm share MHC similarity [85]. Since, under normal conditions, salmon copulation is usually polyandrous, and thus provides the grounds for sperm competition and/or sperm selection by the ovum, the choice of mates would seem to play a relatively minor role, especially since, as alluded to earlier, the tendency of salmon to return to the very same place where they were born must further enhance their tendency for inbreeding. In the study by Landry et al., however, a relatively small number of males (41) and females (35) were placed in the arm a river that is not usually occupied by salmon because of impassable waterfalls. Under such conditions of unfamiliar grounds and low populations numbers, the results obtained by typing offspring for the MHC class II β chain suggested that matings had occurred to favour offspring that was heterozygous. In arctic charr, which is closely related to salmon, Olsen et al. found that ancestry, whereby sibs were preferred to non-sibs, had a more important influence than MHC preference [86]. The fact that these two latter studies looked at the MHC class II locus whilst the study of Yeates et al. looked at MHC class I may be of some importance because, in teleost fish, the regions for class I and class II are not linked [87], and cues conveyed by one class of MHC molecules could have different consequences to those of the other class, which may have contributed to the apparent discrepancy of some of those results.

Sticklebacks are another type of fish which are found both in the sea and in fresh waters and have been very useful for the study of speciation. Sticklebacks, which are also known as tiddlers in English, are very efficient colonisers, and are widespread in the northern hemisphere, over much of Europe, Asia, and America. Regarding the role of the MHC in mating preference, an intriguing observation is that sticklebacks appear to favour mating with individuals of intermediate MHC divergence, to yield offspring with an optimal number of 5 to 6 MHC class II β alleles [88], which they apparently achieve by smelling the peptides that can bind to MHC molecules [89]. In another study, however, Frommen and Bakker found some signs of inbreeding avoidance in groups of fish raised separately, but with no data on the correlation to MHC similarity [90]. From the point of view of the ideas developed in this essay, the data from salmon and sticklebacks contribute to the drawing of a picture whereby clues from the MHC allow fish to mate preferentially with individuals with which they share some genetic similarity, but not too much, which is entirely compatible with the idea of a balance between inbreeding and outbreeding.

Another fascinating observation made on sticklebacks is the phenomenon of parallel speciation. Indeed, when populations of sticklebacks colonise freshwater environments, and lakes in particular, they have a strong tendency to evolve into adapted forms that lack certain features that characterise the anadromous (sea dwelling) sticklebacks:

- In shallow lakes, one finds mostly a form called benthic, which is larger, with smaller eyes, and feeds mostly on invertebrates found on the lake's bed. The benthic form has a great reduction in the number of armour plates and of pelvic structures.
- In deep lakes with steep sides, the favoured adaptation tends to be a form called limnetic, which corresponds to a smaller fish which feeds mostly on plankton at the lake's surface, and is, overall, less different to the ancestral marine fish.

In at least 6 lakes of British Columbia, Canada, evolution has repeatedly driven the apparition of both benthic and limnetic forms from the same ancestral stocks. Those benthic and limnetic forms, although capable of producing perfectly fertile offspring when given no choice of partner, cohabit in those lakes in apparent complete reproductive isolation from one another. For those various sympatric pairs, size and colour were shown to be the two main phenotypes that contribute most to the reproductive isolation [91]. A truly remarkable finding was the observation that, for all these populations, benthic individuals from one lake mated preferentially with benthic partners from other lakes than with limnetic ones from any lake and the converse was true for limnetic fish [92]. As underlined by the authors, the observation that, under similar conditions, evolution can lead to the parallel selection of similar sympatric reproductively isolated populations is a very strong argument in favour of the idea that natural selection is involved in the speciation process. This was in fact further supported by their observation that reproductive isolation between benthic and limnetic individuals seemed even more pronounced between fish from the same lakes than between fish obtained from different lakes. It thus suggested that reproductive barriers could be selectively raised against the population that represented the most direct threat, i.e. the fish of the same lake, but of the other morph.

Further work, orchestrated by the group of David Kingsley, has led to the dissection of the genetic mechanisms responsible for the loss of armour plates or of pelvic structures which are both particularly prominent in benthic sticklebacks. As predicted by the model proposed in this essay, both phenotypes were found to be due to mostly recessive mutations. The loss of armour plates was mapped to the ectodysplasin gene ( EDA) which, in mammals, is known to be involved in many ectoderm features such as teeth and hair [93]. Remarkably, the same allele of the EDA gene, which carries just four amino acid differences compared to that found in fully plated fish, was identified in all the low plated morphs obtained from Europe, and from both the American coasts. That same allele was also identified in fully plated fish caught in river estuaries, albeit at low frequencies (3.8% in California and 0.2% in British Columbia). Another allele was, however, found in Japanese stocks, which shows no changes from the wild type in the protein coding sequence, but falls in the same complementation group as the other low-plate phenotypes. These results suggest that the allele responsible for plate loss in sticklebacks has been around for several million years, and has spread widely over the northern hemisphere, probably because it is associated to a very significant advantage in freshwater populations, where it would thus get amplified, and then fed back into the marine population by episodes of hybridisation. Because it is essentially recessive, this allele can remain 'hidden' at low frequency in the marine populations. When marine stocks colonise freshwater niches, however, this must favour some degree of inbreeding, which would rapidly reveal the recessive phenotype, and the selective advantage would then rapidly increase the allelic frequency in the isolated population. In conditions where the threat of the fully-plated allele persists, this will provide the grounds for selection of reinforcement via mechanisms such as reproductive isolation, which could ultimately result in proper speciation.

The loss of pelvic structures was also very recently shown to be due to recessive mutations corresponding to deletions in the promoter regions of the Pitx1 gene [94]. Remarkably, characterisation of the promoter regions of this gene in nine different populations of benthic sticklebacks revealed that the same 488-bp segment was missing in all nine populations, but this was due to nine different events of deletion. This observation thus testifies that the advantageous phenotype of losing pelvic structures arose repeatedly and independently in all those completely separate benthic populations as a result of selective pressures, and, contrarily to the previous example, was not 'hidden' as a recessive trait in the ancestral marine population. The other point that can be made from this observation is that, under the right conditions of selection, recessive mutations due to loss of existing genetics materials are sufficiently common that they can be repeatedly obtained in completely independent populations. Another remarkable observation contained in that article is that, as predicted at the end of section IV, there is a considerable reduction in the heterogeneity of sequences focused on the region surrounding the Pitx1 gene [94]. Amazingly, this reduction only spreads over a few kilobases, which suggests that events of DNA recombination such as crossing-overs must occur very frequently over the region carrying this gene. As discussed in section IV, the tightness of the region of reduced polymorphism may actually be related to the fact that a sizeable proportion of the Pitx1 mutations have a completely recessive phenotype, which would increase the delay with which the genomic regions carrying the mutation would become fixed, and thus provide plenty of opportunities for crossing-overs to occur in the close vicinity of that region.

Altogether, the picture that shapes itself regarding speciation in sticklebacks adapted to lake environments is one where either hidden recessive phenotypes, or relatively probable inactivating mutations initially result in recessive advantageous phenotypes, promoting successive steps of saeptation from the ancestral stock of fish of anadromous origin. Subsequently, once separate groups have been formed, reinforcement based on sexual preferences will then follow, driven either by the ancestral stock or by the other morph, based on a variety of phenotypes, among which size and colour are particularly prominent.

So far, in this section, we have only considered fish species that are naturally structured and/or have been recognised as prone to undergo speciation (both factors actually going hand in hand if we accept the proposed model). In the oceans there are, however, many other types of fish populations that are extremely numerous, and hence probably much more prone to panmictic reproductive strategies. Those that spring to mind are, for example, mackerels, sardines, anchovies or cods, and for all of those, great fluctuations of effectives have been witnessed over the years, with recovery rates that often prove difficult to predict. This is particularly true for the cod populations, which are proving very slow to recover from the overfishing that has taken place over the past decades. In line with the model proposed in this essay, I contend that, for fish populations that are sufficiently numerous to adopt a panmictic strategy, the variations in numbers, and in particular their episodic slow recovery after population shrinkage, could partly be due to reduced fertility caused by high mutation loads in the context of increased inbreeding coefficients caused by population shrinkage.

To conclude about fish, the currently available data suggest that, in species that tend to have a certain structure imposed by the niche they occupy and/or their breeding habits, mechanisms exist that would ensure a balance between inbreeding and outbreeding by favouring mating between individuals of relative relatedness. When circumstances change, however, such as when cichlids or sticklebacks find themselves in the more stable and secluded environment of a lake rather than in streams or the ocean, this will tilt the balance towards inbreeding, and favour speciation.

**Birds**
For birds, the capacity to take to the air potentially opens an almost limitless capacity for dissemination. Many bird species are, however, rather sedentary, with a strong tendency for territoriality. And for those that are migratory, similarly to fish, there is strong philopatry, i.e. a very strong tendency to return to the very place of their birth when they reach sexual maturity.

Contrarily to fish, however, there is no clear sign that the MHC plays a strong role in regulating the relatedness of mating partners, probably because the sense of smell is less developed in birds than in fish. Rather, visual and auditory clues are used extensively in the establishment of the usually monogamous breeding pairs. Remarkably, rather than being innate, sexual preferences of birds are actually mostly cultural, i.e. mainly acquired via a mechanism called imprinting, which takes place during the first few weeks of life. One must, however, underline that there must also be some level of innate capacity of certain birds to recognise kin. Otherwise, how would the cuckoo ever recognise it's mate? During the imprinting period, birds learn to identify various characters such as the song of their parents, as well as the size, shape and colours of their parents' or siblings' anatomical features such as beaks or plumage. Imprinting has been demonstrated in too many bird species to cite them all here, with varying degrees of importance put on song or anatomical features depending on each species. The most picturesque and best know example is certainly that of the experiments performed with geese by Konrad Lorenz where he showed that the goslings became imprinted on him (or more precisely on his gumboots) during the first few hours after their hatching. When it comes to choosing a mate, those preferences would hence promote pairing between closely related individuals. Working with Japanese quails, Bateson actually demonstrated that cousins were the preferred partners, i.e. individuals that differed a little bit from the parental picture, but not too much [14]. Based on his observations, Bateson proposed the notion of 'optimal outbreeding' [13], which could not possibly be more in line with the ideas put forward in this essay.

One could not possibly evoke the subject of speciation in birds without mentioning the most emblematic case of Darwin's finches. I will thus briefly discuss those as a final example. Those famous finches were collected by Darwin (or more precisely shot and preserved by his servant, Syms Covington) on the Galapagos islands during the second voyage of The Beagle, and only identified later by the ornithologist John Gould as a new group of twelve separate species of finches which seemed most related to ground finches found on the south American continent. Today, Darwin's finches are classified into fourteen different species that have different distributions on the different islands of the archipelago, and for which the most telling anatomical difference lies in the size and shape of the beaks, which are variously adapted to feed on different nutriments (different size of seeds, different parts of cactuses, or various other sources such as insects or larvae). Molecular characterisation of those different species has led to the conclusion that all those species derive from a common ancestral stock which probably comprised at least 30 founders ( C&O, p 403). There is clear reproductive isolation between the various species, with imprinting documented to occur both on songs and on beak shape (incidentally, the shape of the beak has by itself a strong influence on the song). The main factors that control the shape of the beak have actually been indentified as bmp4 (depth and width ) [95] and Calmodulin (length) [96]. Both factors act independently from one another, and in a dose-dependent manner. The various beak phenotypes are thus expected to behave as co-recessive traits since hybrids would express intermediate, less suitable phenotypes.

Since some hybridisation (of the order of a few %) between certain species can still occur [97], a dogmatic evolutionist could argue that those populations thus do not represent true species. For the purpose of the ideas developed in this essay, Darwin's finches simply provide a very telling example of a population of individuals founded by a very limited effective. In the restrained context of those small islands, inbreeding coefficient were thus necessarily increased, and, given the natural propensity of birds to prefer mating with close kin, and the co-recessive nature of the traits selected, this situation has led to one of the most impressive examples of adaptive radiation documented to date.

**Mammals:**
Contrarily to fish and birds, most mammals are restricted in their dispersion (the technical term for this is limited vagility), and most populations of mammals are thus naturally fragmented, and this is particularly true for those that live in relatively small groups, such as horses or certain primates, or are active colonisers, such as murine rodents (rats and mice)[15]. When the natural tendency of a species is for a small number of individuals to find themselves repeatedly isolated into separate colonies, thus imposing high inbreeding coefficients, it is expected that the natural instincts should evolve to compensate for this, and thus favour outbreeding whenever possible rather than further inbreeding.

---

[15] The fact that it has been possible to generate consanguineous lines of rats and mice has proven extremely useful for scientific research. For other species such as rabbits, hamsters or guinea pigs, this has, however proven much more difficult. I contend that this could in part be explained by the natural tendency of muridae to colonise new environments, which must have kept their mutation

Such behaviours have indeed been documented in many mammalian species, and in particular in the house mouse, *Mus Musculus domesticus*. For many years, experimental evidence has been accumulated showing that there was indeed inbreeding, or rather incest avoidance between mice from different inbred strains, and documenting that the MHC was playing a pivotal role in this phenomenon. More recently, however, the group of Jane Hurst used wild mice rather than inbred strains to document the mating behaviours of mice, and identified that major urinary proteins ( MUPs) had a much more potent influence on kin recognition, and incest avoidance, than did the MHC [98]. The discrepancy between those results and those obtained previously by other groups finds an explanation with the fact that the process of deriving inbred mice has yielded strains with very limited inter-strain variability of the MUPs [99]. Furthermore, in an extremely recent paper, the group of Jane Hurst actually characterises Darcin, an invariant urinary protein found in the urine of male mice, which behaves as a pheromone by inducing contact-dependent imprinting of females to prefer the males harbouring the other smells found in that urine [100]. The observation that diverse MUP complexes undergo parallel evolution in different species suggests that polymorphic MUPs, as well as other polymorphic factors [101], may play an important role in regulating the mating behaviour in many species [102], which may call for revisiting some of the results obtained regarding the pivotal role of the MHC in regulating the degree of inbreeding between individuals in vertebrate species, including fish.

The precise mechanism(s) driving incest avoidance is, however, of little relevance to the ideas discussed here. Rather, we can find multiple arguments that provide strong support for the ideas proposed here in the study published by Bush et al. more than 30 years ago [103].

Firstly, they underline that the effective size of mammal populations (which is inversely correlated to the average inbreeding coefficient) appears to be inversely correlated to the rate of speciation: Whilst speciation is very rapid in horses and primates, which have very structured populations, it is much slower in marsupials and carnivores, which have much more diffusive breeding strategies, and slowest in bats and whales, probably because of their high vagility. The various altruistic behaviours frequently witnessed in certain colonies of bats is, however, often viewed as being due to high levels of relatedness between the individuals comprising those colonies. One could thus envisage that the remarkable longevity of bat species may relate to the stability of the equilibrium between outbreeding (due to their high vagility) and inbreeding (due to the structure of their colonies, promoted by the importance of cooperative behaviours for their survival [104]).

Second, the rate of speciation is also shown to be strongly correlated to the rate of chromosomal evolution, and horses, primates and rodents are indeed genera where many instances of chromosomal rearrangements have been documented between closely related species, which can sometimes produce hybrids that are either infertile (as for the equine species) or of limited fertility, as for the chromosomal species of alpine mice [105, 106].

Third, the authors also underline that the organisation of populations into clan or harems, where a single dominant male sires most of the females is another mechanism which reduces the effective size of populations, and thus increases the average inbreeding coefficient. There are a few notable exceptions where it is actually the female that gives rise to most of the offspring of a colony. One of them is the African wild dog, which lives in pack of 20 to 40 animals, and inbreeding is reduced by males and non-reproducing females emigrating away from the population. Another one is the eusocial naked mole rat, which is found in east Africa and is actually more closely related to porcupines than to rats. Those live exclusively underground, in colonies of 50-100 individuals where all the offspring descends from one single 'queen'. Although inbreeding coefficients have been found to be extremely high in those animals, this must be a consequence of their lifestyle rather than by choice since outbreeding was found to be preferred when available [107].

Inbreeding is, however, not avoided to the same degree in all mammal species, and there are also numerous examples of kin preference in mammals, which are often the result of imprinting, in other words a cultural rather than a genetic heritage. On the whole, one finds that mammals in which inbreeding avoidance is the most prominent are those for which their natural lifestyle would most often provoke the isolation of small groups. Yet, they should presumably be those carrying the smallest mutation loads. Hence, they should be the ones for which inbreeding is the less costly. For mammals as for fish and birds, the overall picture therefore seems to match a model of balance between inbreeding and outbreeding, in line with Bateson's optimal outbreeding model [13] rather than outright and systematic inbreeding avoidance. One study carried out in wild American Pikas (which are related to rabbits and hares) actually found that, much like Bateson's Japanese quails, the preferred partners were those of intermediate relatedness [108].

**Insects**

With more than one million species identified, the class of the insects is, by far, the most numerous one of the whole kingdom of eukaryotic life, and basically comprises half of the metazoan species recorded to date. Insects are thus clearly very prone to speciation. Although insect populations are often very large, they are also very frequently fragmented into very restricted and diverse niches, which often exist only transitorily, and which must thus be repeatedly colonised by a handful of individuals.

loads very low, and also shaped their genomes to cope with repeated episodes of extreme inbreeding.

To my knowledge, no behavioural inbreeding avoidance has ever been described in insects, and it is only very recently that some level of outbreeding preference has been reported, in polyandrous female field crickets, via a process of preferential sperm-storage [109], and this despite similar success of mating with sibs or non-sibs [110]. Conversely, numerous instances have been documented whereby insects show kin preference, based on a whole range of processes which include preferred mating protocols, acoustic and visuals clues and pheromone detection. Repeated episodes of colonisation, and the absence of inbreeding avoidance must contribute to keeping the mutation loads down, and thus promote the phenomenon of speciation in insects. As developed in addendum 2, the haplodiploid mode of reproduction of insects such as the hymenoptera (ants, bees, wasps … ) corresponds to a very effective way of eliminating recessive mutations, and it is quite remarkable that the hymenoptera represent more than 30% of all insect species. Another factor which may contribute significantly to the tendency of insects to undergo speciation is that the selective pressures due to predation are particularly significant for insects, and traits that can reduce detection by predators are quite often recessive.

Among all insects, Drosophila has proven a particularly useful tool for many aspects of biology, and particularly for genetics and the study of speciation. So much data has been published on speciation in Drosophila that it would be unrealistic to attempt to summarise it here (there are more than 50 sections discussing Drosophila in the book *Speciation* by Coyne and Orr (2004), many of them several pages long). I will therefore restrict myself to outlining a few points that seem to be most relevant to the ideas developed in this essay.

- Regarding genetic loads in fragmented populations, as early as 1964, Dobzhansky was underlining the observations made by several groups that "*the heaviest genetic loads are found in common and ecologically most versatile species of Drosophila, and the lightest ones in rare and specialized species and in marginal colonies of common ones*" [111].

- Reproductive isolation between different species of Drosophila relies mainly on two mechanisms: choosiness of females for the males of their own species, and hybrid sterility.

- When crossings occur between different species, mating preferences almost systematically disappear in F1 females ( C&O, chapter 6), which testifies for the recessive nature of those phenotypes. If we follow the type of reasoning developed in the previous pages, this would suggest that such characters leading to behavioural isolation must have arisen in the context of saeptation, which could have been either primary, or secondary to the constitution of two populations. The repeated observation that stronger assortative mating is found between populations of flies that are in close contact in the wild (C&O, p 357-365) brings very strong support to the idea that reproductive isolation is a phenotype that is selected for, and not just the result of divergence between populations that are not in contact with one another.

- Regarding hybrid sterility, it follows Haldane's rule since it is almost always the males that are sterile. A large body of evidence from various studies suggests that this sterility is often asymmetric (ie concerns the males obtained through only one of the two types of crossings), and results from the accumulation of multiple small effects mapping to various genes rather than to the large effect of major genes ( see C&O, p299-319). This is in complete agreement with the scenarios proposed in section III and sketched in figure 3, whereby reproductive isolation arises as a succession of small steps, most often selected for under the threat of hybridisation with the ancestral population expressing a dominant but deleterious trait.

- Multiple studies, of which many come from the group of Mohamed Noor, underline the implication of chromosomal rearrangements in the reproductive isolation seen between closely related species of drosophila. When they have been mapped, the genes for female preference and hybrid male sterility were found to be associated with chromosomal rearrangements [55, 112], and furthermore, such rearrangements are much more prominent between sympatric species than between allopatric ones ( [55] C&0, p309 ). The explanation most commonly offered for these observations, in line with the Dobzhanski-Muller model, is that the chromosomal rearrangements prevent recombinations between multiple genes having co-evolved. As proposed in section II 5c, if reproductive isolation evolves as a response to the threat of hybridisation with a neighbouring distinct population, chromosomal rearrangements could also have two additional effects contributing to the isolating phenotypes: first induce some level of infertility in hybrids, and second be endowed with an intrinsic phenotype, either by the inactivation of a gene leading to an advantageous recessive phenotype, or by modifying the genomic context of the genes surrounding the rearrangement.

To conclude with the most important, it is based on reviewing a large number of studies carried out by himself and by many other groups working on Drosophila that, as early as 1959, Carson proposed his model whereby speciation is promoted in small, more inbred populations [5].

Altogether, the masses of data accumulated with various species of drosophila seem to be in perfect agreement with the model proposed, whereby the flies' lifestyle, which involves repeated colonising of isolated habitats by a few individuals, results in very fragmented populations, with high inbreeding coefficients and thus much smaller $N_e$ than would be inferred from their large numbers [3]. Consequently, such Drosophila populations will carry low mutation loads, which must increase the probability of both the appearance of advantageous recessive phenotypes, and fixation of chromosomal rearrangements. Whilst the resulting groups would not initially be strongly infertile with the ancestral population, hybridisation will be detrimental to the fitness of the offspring, which would promote reinforcement mostly in the more threatened, less numerous newly arisen group, thus explaining the asymmetry of the isolation phenotypes often observed between drosophila populations.

Another fly species which is an old favourite for the study of speciation is the apple maggot fly, *Rhagoletis pomonella*, which one finds in North America and which adapted very rapidly from its native hawthorn host to cultivated apples after those were introduced in north America in the 1800's, and the first report of this speciation can be traced back to Walsh in 1867, less than 70 years after apples were introduced, and very soon after Darwin's publication of The Origin. Although some gene flow can still occur, there is very significant reproductive isolation between the two species, based on a combination of factors which include the fact that larvae have different timings for their emergence from their diapause (i.e. the larval life), leading to adults having reduced overlapping periods for hybridisation, and also that Rhagoletis mate on or near the fruit of their host plant. The data on Rhagoletis fits the proposed model very well: Firstly, preferential responses to specific fruit odours are recessive since they have been shown to disappear in F1 hybrids [113]. Second, multiple loci related to diapause have actually been mapped to regions of chromosomal rearrangements which have been shown to have introgressed from an isolated population of Mexican Rhagoletis. The overall picture is thus one where recessive odour-based fruit preference would drive saeptation, and chromosomal rearrangements associated to different diapause phenotypes would reinforce the isolation both by favouring intra-group synchrony, and presumably also by reducing inter-group fertility.

Chromosomal rearrangements are found in closely related species, or sub-species in many other types of insects, and the best documented example is probably in the Australian wingless grasshoppers, which were studied by White, and which led him to propose the model of stasipatric speciation [114](C&O, p16), and more recently that of chains of chromosomal changes [115], whereby sequential chromosomal rearrangements progressively reinforce the genetic isolation of a population from the ancestral one, in conjunction with other mechanisms of reinforcement such as hybrid sterility [114].

If we now turn to butterflies, we can find two examples that underline the correlation between the recessivity of phenotypes and the phenomenon of speciation. The first example is that of the peppered moth, *Biston betularia*, which was first reported by J.W. Tutt in 1896, and has since become an emblematic example of adaptive evolution. Originally, the populations of those moths were light colored (peppered), which provided very good camouflage against the barks of trees. During the industrial revolution, however, many lichen died, and the average color of tree trunks turned much darker because of soot deposits. This made the light colored peppered moths much more conspicuous for their bird predators, and led to the selection of a darker phenotype, the black-bodied moth, which initially represented less than 2 % of individuals, but raised to around 95% over the five decades between the middle and the end of the 19th century. With the color of tree trunks progressively returning to a more natural light color, the frequency of dark moths has since been decreasing slowly. The rapidity of the initial selection process is explained by the fact that the darker phenotype is due to a dominant mutation. In fitting with the model, this did not, however, lead to any detectable process of reproductive isolation.

Conversely, in another butterfly, a recent report describes that a recessive phenotype is associated with the type of mating preference expected to correspond to the early steps of speciation [116]. In western Ecuador, one finds *Heliconius cydno alithea*, which is a mimetic butterfly which follows the models of other Heliconius butterflies, *H. sapho* (white) and *H. eleuchia* (yellow). Those two latter species produce toxic chemicals that protect them against predation. Within the population of *H. cydno alithea*, depending on the region, one finds white and yellow butterflies in various proportions, which correlate with the relative abundance of the respective white and yellow models in that same region. Whilst white and yellow *H. cydno alithea* are not reproductively isolated, Chamberlain and colleagues found that the yellow males showed a marked preference for yellow females, whilst white males were indiscriminate. Crosses between yellow and white butterflies also revealed that white is the dominant phenotype. Remarkably, male preference was found to segregate with the K locus coding for wing colour, which may be explained by the fact that the same pigments dictating wing colour are also used as filtering pigments in insect eyes.

Thus, in butterflies as in cichlids, selection of advantageous recessive colour patterns can lead to some degree of reproductive isolation, whilst dominant ones do not.

**Flowering Plants**

With close to 300.000 species recorded, flowering plants compete with arthropods for the second place for the phylum with the most species [70, 117]. Among those, the rate of speciation appears to be particularly prominent in plants capable of self-fertilisation. This is in part related to the phenomenon of speciation by polyploidy, which is actually relatively rare, and occurs over just one or very few generations, and is thus not really relevant to the mechanisms we are trying to dissect in this essay ( see C&O, chapter 9). As first proposed by Baker in 1953, the higher number of species among selfing plants is often interpreted as related to their higher capacity to colonise new environments (see[118]), and this does indeed fit the para- and/or allopatric scenarios proposed in section III.

An additional factor may, however, be that the capacity to self fertilise, which is the ultimate form of inbreeding, would be very effective at reducing the mutation load, which would, in turn, favour speciation. In a more recent report Heilbuth concluded that it was not so much the capacity to self-fertilise that increased speciation, but dioecy (i.e. the complete separation of the population between males and females) that was associated with lower number of species [119], which is in complete agreement with the observation that dioecious plants only comprise 6% of all flowering plants, among which one finds Holy, Willow, Ash, Juniper and Gingko biloba (one of the longest lived species know to date, C&O, p425). To reach this conclusion, Heilbuth compared multiple plant families for species richness among three types of plants : those capable of selfing, those where selfing could be possible but is prevented via various self-incompatibility mechanisms, and dioecious plants, and found comparatively low numbers of species only in the latter. This puzzling observation can, however, find explanations in the light of the model proposed here. Indeed, the prediction from the model is that, under conditions of excessive inbreeding, populations will undergo very frequent speciation, but the durability of these species will, consequently, be much reduced because most new species will tend to eliminate their immediate ancestor. The incapacity to self fertilise may thus reduce the rate of speciation, but would increase the lifetime of the species, with a net result of equivalent numbers of species. Furthermore, the diversity of mechanisms used for self-incompatibility in various plant species suggests that those have been repeatedly and independently selected for, probably because they represented a selective advantage in populations that had an excessive tendency to undergo rampant speciation. These views are supported by very recent report in which Goldberg and collaborators documented that, in solanaceae, self-incompatibility has been maintained for over 30 million years in 40 % of species because, although self-fertilising species undergo more speciation, they also go extinct more rapidly, which the authors suggest could be due to a conjunction of their smaller effective population sizes, decreased polymorphism, narrower geographic distribution, decreased capacity to select for advantageous gene combinations and to eliminate the deleterious ones [120].

In addition to underlining the correlation between the selfing capacity of plants and their propensity to colonise remote grounds, Baker was also among the first to propose that sex could have evolved as a mean to reduce inbreeding [121]. Multiple arguments exist to suggest that dioecy is a much more efficient guard against inbreeding than mechanisms of self incompatibility such as gametophyte incompatibility ( see [119]), and switches between dieocious and selfing modes of reproduction must also be much less likely than between self-incompatible and -compatible ones, such as recently described for the annual plant Capsella [122]. Given this, it is thus not surprising that dioecious species should be guarded against inbreeding via higher mutations loads, and thus have a much lower tendency towards speciation, and thus be much less numerous than those with complete or partial hermaphrodism. One should not, however, make the mistake of equating speciation with adaptive evolution. Evolution is related to the acquisition of new characters, which is much favoured by the exchange of genetic material among numerous individuals in large and long established populations. Speciation, according to our model, is mainly due to the loss of some undesirable trait(s), which can only occur via inbreeding between a necessarily restricted number of individuals, which will almost always result in some loss of diversity, and thus reduce further adaptability.

To conclude on plants, we will turn towards a couple of examples which are common favourites of speciation specialists.

The first case is that of two closely related species of monkeyflowers, *Mimulus lewisii* and *cardinalis*. Those are found in the hills and mountains of California, with the pink M. lewisii occupying higher altitudes, and the bright red M. cardinalis occupying the valleys. Although the ranges of the two species overlap between 1500 and 2000 meters, and despite the fact that they are capable of producing viable and fertile offspring in the greenhouse, hybrids are almost never found in the wild, which is highly related to the fact that the red *M. cardinalis* is pollinated mostly by hummingbirds, whilst the paler *M. lewisii* is pollinated almost exclusively by bumble bees. Phylogenetic comparisons have established that the ancestral phenotype was the paler colour of *M. lewisii*, and the bright colour of *M. cardinalis* is actually a recessive phenotype due to a mutation in the YUP gene, which prevents carotenoid deposition in the petals [123]. By deriving near-isogenic lines of *Mimulus* for various characters, Bradshaw and Schemske managed to demonstrate that, although the two species have diverged by many other detectable traits that segregated diversely, the preference of either hummingbirds or bumblebees was primarily controlled by this single locus. Hence, we have here an example where a recessive mutation has led to what I consider a clear case of parapatric speciation by provoking a switch to a different pollinator, which presumably allowed *M. cardinalis* to colonise the lower ranges and valleys where hummingbirds are found.

The last plant example I have chosen to discuss is that of oaks (*Quercus*), for their extraordinary capacity to resist speciation, since complete reproductive isolation still has not been reached between many of the approximately 400 'species' of oaks recorded to date. Those correspond to very different types which are distributed over very spread and diverse habitats of the northern hemisphere, but many of those 'species', and particularly the group of white oaks which are most prominent in north America, can still intercross and yield perfectly fit offspring, with clear signs of hybridisation and gene flow having been documented in wild populations [124-126]. Whilst these observations have led to numerous and lengthy debates on the appropriateness of such or such definition of species, and left evolution biologists puzzled for many years, the very limited tendency of oaks to undergo speciation may also find an explanation in the model proposed here. Firstly, although oaks are monoecious, gametophytic self-incompatibilty has been found in many types of oaks. Second, oaks have a very significant capacity to spread both via pollen and via acorns that can be transported over rather short distances by animals such as squirrels, but also over much longer distances by floating down streams, or conceivably even across sea waters (Darwin spends several pages of The Origin discussing the resistance of various seeds to seawater). This capacity to diffuse must greatly favour hybridisation, and thus the spread of dominant advantageous traits. At the same time, oaks have a clear tendency to congregate in forests that are comprised mostly of oaks, and this must surely provide the grounds for a certain degree of inbreeding, which must result in the mutation load remaining reasonably low. Another factor that must favour the selection of very vigorous hybrids lies with the number of acorns that an oak produces. Indeed, during its very long lifetime, a single oak produces hundreds of thousands of acorns, which will give rise to several thousands of seedlings, and maybe a few dozen young trees, but of which only a handful (two, statistically) will go on to produce progeny themselves. This tremendous level of selective pressure probably contributes significantly to preventing the survival of those suffering from any significant degree of inbreeding depression. Altogether, I would surmise that the situation of oaks probably hovers near the equilibrium between the yin of inbreeding and the yang of outbreeding, with a mutation load sufficiently low to prevent insurmountable inbreeding depression when selective pressures rise, sufficient outcrossing to share dominant advantageous phenotypes, and at the same time maintaining a mutation load sufficiently high to prevent the degree of close inbreeding that promotes the successive steps of saeptation leading to speciation.

**White Sand lizards: an experiment in progress**:
In the first paragraphs of this section, I underlined why it was so difficult to carry out experiments related to speciation. The example I have chosen to conclude this section is one where nature may actually have provided us with such an experiment, including the indispensible internal control. In White Sands, New Mexico, USA, dunes of white gypsum formed less than 6000 years ago. In those dunes, one finds several types of lizards harbouring very light colours, which are each descended from their darker found in the nearby Chihuahuan desert. Amazingly, in three separate species, the group of Erica Rosenblum has recently mapped the cause of this albinism to different mutations of the very same gene : the melanocortin-1 receptor (which is, incidentally, also associated to red hair in human). Even more remarkably, the mutation leading to a white phenotype is dominant in two species, Sceloporus undulatus and Holbrookia Maculata, whereas it is recessive in another, Aspidoscelis inornata [127]. In line with the ideas developed here, in previous versions of this essay, I had predicted that, if there was asymmetry in the mating preferences, those should be stronger in the morphs with the recessive form (white or brown), and most prominent in Aspidoscelis inornata, for whom the white phenotype is recessive, and the threat of breeding with the more numerous ancestral stock of brown lizards thus much more significant. Although the mating preferences between white and brown lizards have not yet been documented for those species, Rosenblum and Harmon have since combined data from nuclear and mitochondrial genotyping with morphological assessments to evaluate the progress of each of these ecotypes towards speciation [128]. Their results suggest that, contrary to my above prediction, A. inornata seems to have progressed less towards speciation that the other two species. The results did, however, support the model presented here: a correlation was found between the degree of ecological speciation evaluated in these three sets of lizard populations and their population structure: The most consistent signs of speciation were found in H. maculata, which lives in small isolated groups. Conversely, very few signs of speciation were found in A. inornata, for which populations adopt a much more continuous distribution, and intermediate degrees of speciation and of population structure were found for the third one, S. undulatus. For those lizards, it thus seems that initial populations structures had much more influence than simply the dominant or recessive nature of a single trait being selected.

## VI ) And what about *Homo sapiens* ?
*"Dans un oeuf, y'a du blanc et du jaune. Eh bien quand on mélange, il n'y a que du jaune"*. Coluche, Les Vacances (1979)

In the paragraph discussing mammals, I purposefully avoided the difficult subject of the situation of the human race. As we will see in the following paragraphs, there are many aspects whereby what we know of past and current structures of human populations, as well as human instincts appears to fit the model, if only too well for comfort. Indeed, the subjects of our mutation loads and of our species preservation give rise to such grave questions, especially with the spectres of eugenics and Nazism still looming in our not so distant past, that I felt it was best to discuss the data and the situation of *Homo Sapiens* separately.

Today, the human population comprises well over 6 billion people, and this number is predicted to reach 9 billion in about forty years, despite serious uncertainties about the capacity of our planet to sustainably feed that many people. Although it is universally admitted that we all belong to the same species, humans are split into many ethnic groups and races. If one looks at the situation in places where those groups come into close contact with one another, such as in big cities, one does, however, witness a very significant level of intra-racial preferential pairing. Furthermore, offspring of interracial couples, whilst benefitting from high physical fitness, often suffer from reduced social fitness because they find themselves struggling to integrate into either of the groups that their parents came from. In this sense, to highlight once again the difficulty of defining species, if one adopted the same criteria as are often applied to animal or plant species in the wild, one could conclude that speciation has already started occurring in humans. In support of this rather provocative stance, the most distinctive phenotypes to distinguish between ethnies are the colours of skin, hair and eyes, which have progressively gone from dark in our African ancestors to the very pale skin, blond hair and blue eyes seen in Northern European populations. And it is completely fitting with the model that secondary mechanisms of isolation such as xenophobia or racism, should often be asymmetrical, and strongest on the side that expresses the recessive traits. For example, we know that all the "arian" traits that were the basis of the selection criteria of the Nazi doctrine do not actually correspond to real improvements by a gain of a new function, but all correspond to mutations causing losses of function in various pigment genes, which are all either recessive, or co-recessive. Another even darker aspect of the human practices matches certain points discussed in section IV: War is very similar to a sympatric struggle, i.e. a conflict for the occupation of the niche between separate populations. In times of conflict, sexual violence and systematic rape have been used for centuries as a weapon of war (see http://www.unicef.org/sowc96pk/sexviol.htm ). Indeed, in the context of a sympatric struggle, the practice of systematic rape is a very effective strategy to neutralise the reproductive force of the opposing population, and imposes a burden that can last for many years by producing children that are often rejected by both camps. Thankfully, since 1998, the United Nations as decided to consider this abominable practice as genocide, and as a crime against humanity.

The recent discoveries of a few percents of Neanderthal sequences in the genome of Eurasian populations and not in those of sub-Saharan African descent [129] , as well as that of Denisovans specifically in present day Melanesians [130] are also in perfect agreement with this type of scenario: the ancestral population of Homo sapiens, having formed in Africa, came in prolonged contact with Neanderthal or Denisovan populations when it started colonising more northern latitudes, and the two ancestral occupants were most probably out-competed for territory occupation. Under such conditions, it would not be surprising if Haldane's rule applied between Homo sapiens and the older populations, with interspecies mating resulting in hybrid progeny comprised of sterile males and fertile females, for whom further mating with Homo sapiens males would be the most effective way to produce offspring. Over successive generations, genomic DNA from those females would thus have entered the gene pool of the Homo sapiens population during its out of Africa colonising migration, which could actually have proven to be a very effective strategy to acquire sets of genes that were better adapted to the colder and greener territories being colonised, and which the Neanderthal and Denisovan populations had inhabited, and thus adapted to for hundreds of thousands of years.

To date, despite this hybridisation with Neanderthal, and despite the fractionation of Homo sapiens into separate races for tens of centuries, Homo sapiens is still clearly a single species because no population has been described that would be less fertile with another, or that would differ in its overall genetic constitution, for example a fixated chromosomal rearrangement.

Regarding the occurrence of inbreeding in humans, there has been a considerable evolution over the past few decades. For many centuries, the structures of human populations were probably quite similar to those seen in great apes today, being split into groups of a few dozens, with some individuals, most often females, passing from one group to another. Over the centuries, the advent of civilisation resulted in the progressive increase in the size of those groups, driven by a whole range of reasons, among which the most significant were probably i) the conflicts with adjacent groups (with the smaller groups being eliminated) ii) the advent of agriculture, which imposed sedentarity and allowed the sustenance of denser populations iii) the specialisation of individuals into classes of farmers, craftsmen, soldiers, carers … resulting in an increase in the groups' critical mass, i.e. the number of people necessary for having sufficient numbers of the various kinds in each group.

Until the middle ages, the size of most human communities remained small, and average inbreeding coefficients in human populations must thus have been quite significant. In this regard, the recent sequencing of the genome from the hair of a 4000 year old Eskimo gave results consistent with an inbreeding coefficient of 0,06 [131], equivalent to that of the offspring of parents with 0.12 of consanguinity corresponding to the degree shared by first cousins.

Later on, recognising the existence of infectious microbes, leading to the concept of hygiene, did considerably favour the increase in size of cities by decreasing the incidence of epidemics (when visiting the tower of London a few years ago, I learnt from the guide that, if London was the largest city in the world for many years, it was thought to be related to the English's love of tea. Indeed, boiling the water greatly reduced the spreading of water-born pathogens such a typhus, dysentery or cholera). The concept of aseptia also greatly reduced the numbers of deaths during childbirth. Later on, progress in medicine such as vaccination, antibiotics, surgery would increase the survival of individuals, resulting in further swelling of the populations and of the sizes of towns and cities.

Today, nearly 50% of the world population lives in major town and cities [132]. For western populations, one can thus consider that the situation has become progressively panmictic in just a few generations, as testified by the study of regions of extended homozygosity in samples of the North American population, which found that average inbreeding coefficients were above 1% in people born in 1900, but nearing zero in those born around 2000 [133]. These changes in population structures are widely perceived as beneficial because they should result in reduced incidence in the occurrence of rare genetic diseases due to recessive mutations [19, 132]. But, as has been discussed at length in this essay, this could be to be a very short sighted perspective because it equates to, as Muller once put it, "eating all of our cake today" by allowing the recessive mutation load to increase progressively to higher levels, until the rate of elimination by genetic defects once again balances the rate of their accumulation [28].

Incest, the union of individuals sharing half of their genomes (or at least 0.25 in a direct line) is avoided and condemned as taboo in virtually all societies, and this situation probably evolved to counter our natural instincts attracting us to our closest kin, of which the famous Oedipus complex is probably the most striking example. Historically, consanguineous unions have been particularly prominent in rural populations as well as in the upper classes (for example among Egyptians pharaohs or European royalty and aristocracy), whilst stern avoidance of consanguinity is mostly a trait typical of more urban middle classes. Today, the attitudes of various societies and cultures towards consanguinity diverge greatly. Indeed, although first cousin marriage are widely perceived as undesirable in the most developed nations, and are even illegal in 31 of the 50 states of the USA, as well as in China, this is not the case in many other parts of the world such as the middle east or Asia, where weddings between uncles and nieces (degree of consanguinity F=0.25, resulting in offspring with an inbreeding coefficient I=0.125) or between first cousins ( F=0.125, I=0.06) are common, and even sometimes actively encouraged [18, 19]. Today, despite the phenomenal increase of the proportion of the human population living in urban environments, more than 10% of the world's unions are still consanguineous, and this is sometimes transiently reinforced in the communities of recent urban immigrants [18, 19].

Although consanguineous marriages do result in a detectable cost in the fitness and viability of the offspring, this is balanced by various factors such as more stable marriages, better relationships between the members of the extended family, a stronger sense of community, enhanced female autonomy, and importantly, the economic benefits of keeping the family land and belongings together [18]. Fifteen years ago, Bittles and Neel used a meta-analysis of 23 different studies to compare the fate of the offspring from unions between first cousins with those from non-consanguineous parents, and estimated that first cousin marriages resulted approximately in an additional 4% of the offspring dying in the interval between 6 months gestation and ten years of age [16][134]. More recently, a study based on the complete birth records of the Icelandic population over the past 200 years not only confirmed that an evolution towards less consanguinity was also taking place in Iceland, but also showed that couples that were consanguineous at the level of third or fourth cousins produced more grandchildren than those that were either more or less related [20]. The couples that were more closely related had produced at least as many children, but a higher proportion of those had died earlier and/or never reproduced, most probably as a consequence of deleterious recessive mutations.

Today, the situation of human populations is clearly not in a state of equilibrium, but in the process of evolving rapidly. On the one hand, the populations of well developed countries combine low fertility rates with panmictic reproductive strategies that will result in significant increases in mutation loads for the future generations, as well as promoting more and more selfish behaviours. On the other hand, the world's most prolific populations are also the poorest (fertility rates are highest in Latin America, Sub-Saharan Africa and the Middle East), and in many of those, the common occurrence of consanguineous unions should maintain the mutation load to low levels, but this will presumably promote further separation between the various populations of the world in the long run. Indeed, although consanguinity rarely results in very high degrees of inbreeding in humans, I contend that it is only a question of time before a significant chromosomal rearrangement finds itself associated to an advantageous recessive mutation. If such a mutation were to become fixed in an certain portion of the population, which would then have reduced fertility with the rest of the population, the questions of one or more species within the human race would become very real, and lead to extremely serious ethical concerns.

Although circumstances such as wars, water rises due to global warming and food shortages due to overpopulation represent much more pressing threats today than those based on genetic events, the same may not be true for the evolution of the balance between various populations over the next coming decades. Given the differences in fertility rates between the wealthy and poor populations, even if the progress of molecular biology will probably be able to help control genetic loads by pre-natal screening, one does not need to be called Thomas Malthus to see that the situation does indeed look poised for a progressive replacement of the populations descended from those living today in more developed countries by those coming from less developed countries. This may be even amplified further by the well know fact that, when the standards of living first increase in poor populations, this causes the fertility rates first to increase even more for one or two generations, before decreasing dramatically.

The challenge for future generations will be to find a model of society which, at the same time would provide sufficient levels of quality of life to all human beings to curb their fertility, so that economies can be built on sustainable resources, and also promote the right balance between inbreeding and outbreeding: i) enough consanguinity to maintain mutation loads in check, and to nurture the perpetration of traditions and cultures, as well as cooperative behaviours ii) enough outbreeding to allow the shuffling of races, ideas and cultures. Indeed, for ideas and for genes alike, exchanging and mixing is the most effective way to promote the new encounters, the new combinations that result in truly significant innovations and progress, i.e. true evolution. It is, at the same time, also the best way to prevent the phenomenon of speciation. For, even if it results in the awesome natural diversity that surrounds us today, the truth of the matter is that the phenomenon of speciation first and foremost is a downward step since it corresponds to a loss in opportunities for exchange of genetics materials between organisms, which is a direct consequence of the fact that, most of the time, it is initiated by the loss of a pre-existing function rather than by the gain of a new one.

---

[16] Based on the figure of an increase of 4% in the incidence of deaths between 6 months of gestation and 10 years of age in the offspring of first cousins, Beetles & Neel concluded that the average mutation load must be 0.7 lethal equivalent per gamete, and hence 1.4 per zygote. Considering that a large proportion of recessive mutations would probably provoke undetected early abortions, we can presume that the average total load in recessive deleterious mutations was at least twice that figure, and quite possibly somewhere between 5 and 10.

**Table 2:**
**Many of the documented examples of speciation in natural species fit the proposed model.**
(Please refer to text in section V for relevant bibliographic references)

| Species | Nature of the phenotype associated to speciation | Population structure and mutation load |
|---|---|---|
| **Fish** | | |
| Salmonidae | | Highly philopatric Studies on MHC give conflicting results suggesting optimal outbreeding model |
| Cichlids | Bright colours typical of species are recessive (disappear in hybrids) | Close preference for kin, with no detectable inbreeding depression |
| Sticklebacks | EDA mutation (armour plate loss) is completely recessive Pitx1 mutation (loss of pelvic structures) is recessive | Studies on MHC support optimal outbreeding model |
| Panmictic species (cod, macquerel, tuna …) | | Susceptible to large and unpredictable fluctuations in numbers |
| **Birds** | | Migrating birds are highly philopatric |
| Quail | | Preferential mating among cousins (led to Bateson's optimal oubreeding) |
| Darwin's finches | | High inbreeding coefficient due to small size of the niche |
| **Mammals** | | Rate of speciation inversely related to the effective size of populations |
| Mice and rats | | Very fragmented populations correlates with capacity to inbreed |
| Pikas | | Optimal outbreeding |
| **Insects** | | |
| Haplodiploids (bees, ants, termites) | | Very low mutations loads correlate with very high species richness, and global ecological success |
| Drosophila | Mating preferences are recessive (disappear in F1) | Assortative mating, and chromosomal rearrangements are more prominent between populations that are in close contact in the wild. H. Carson highlighted the correlation of speciation with small populations based mostly on data from drosophila. |
| Apple maggot fly | Fruit preference is recessive (disappears in F1) | |
| Heliconius mimetic butterflies | Sexual preference of the males is asymmetric, and linked to the recessive yellow colour | |
| **Plants** | | Selfing plants undergo more speciation, but the species go extinct more quickly |
| Monkey flowers | The red derived phenotype is recessive to the pink ancestral one | |

**Concluding remarks:**

   The ideas developed in this essay are mostly based on rather basic, not to say simplistic, concepts. One of the reasons that kept me from writing up those ideas for several years was the reasoning that, if this model was even partially correct, then one of the many geneticists that have pondered about speciation for the past 150 years should have developed similar ideas before me. And although it took me a long time to identify many of the previously published works most relevant to the ideas presented here, those have turned out to be in line with the models et theories developed by people like Wright, Carson, Shields and Bateson. To date, however, the ideas put forward by these various people have received remarkably little attention from the scientists trying to understand the mechanisms of speciation. I contend that the main reason for this is that speciation has been considered as a phenomenon that should be explained by population genetics. And apart from the fact that recessive phenotypes are much more difficult to integrate into models, the most important factor is probably that most of the grounds for population genetics were laid during the first half of the twentieth century, initially by Fisher, Haldane, and Wright himself, with highly mathematical papers, and later by others such as Dobzhansky, Mayr and Muller, to reach the global concept of what is known as "The modern synthesis". All this groundwork by very intelligent and gifted people took place before the structure of DNA, the genetic code, the digital nature of genetic information and the structure of genes were discovered, which all happened after WW2. But because they did not have access to this molecular knowledge, pre-war geneticists, and Muller in particular [28], considered all mutations as essentially dominant in their calculations. We now know that dominant mutations are usually due to a gain of function, recessive ones to a loss of function, co-dominant ones to a change of function and co-recessive ones to the effect of gene dosage (see table 1). And we now also know that most recessive mutations are indeed truly recessive : having just one functional copy of a particular gene is usually sufficient, and heterozygotes with one wild type and one mutated copy of a particular gene have absolutely no detectable phenotype, and a perfect capacity to reproduce (contrarily to the pre-war assumption of an average effect of 2-5% effect on fitness [28, 135]). As long as one does not recognise that many mutations are truly recessive, one simply cannot venture towards the idea that deleterious ones are driving the absolute requirement for inbreeding, and advantageous ones the initial steps of speciation.

   Another factor that could have contributed to certain geneticists not following the paths I followed in these pages may have been related to the darkness, the political 'incorrectness' of the conclusions that these paths lead to regarding our future, and particularly that of our westernised populations. As Winston Churchill said: *Once in a while you will stumble upon the truth but most of us manage to pick ourselves up and hurry along as if nothing had happened*.

   But, as a geneticist, one should not lose sight of the fact that nature, and particularly the process of natural selection, is not politically correct. Indeed, when one thinks of the survival of the fittest, one often fails to consider the darker side of natural selection and that the counter-balance of the "survival of the fittest" is the "death (or disappearance) of the less fit". As considered at length by both Darwin and Wallace in their respective works, most reproducing organisms in natural populations produce many more than two offspring, and of those, most will not go on to breed and their genes will hence disappear forever. I thus contend that the concept of "mildly deleterious mutations" derived from a very anthropocentric perspective of well fed, wealthy, healthy and secure people. For the vast majority of living organisms, including most human beings on this planet today, there is no such thing as mild natural selection. Under natural conditions, the struggle for existence, as outlined by Darwin himself, is a very tough one in natural populations where only one in ten, hundred or even thousand of conceived zygotes will become a mature organism that goes on to produce offspring. A very recent paper looking at wild population of field crickets reported the very unexpected observation that only one in ten of sexually active adults actually yielded offspring the following year, and this was true for both males and females [136]. If it were not the case, we would not see so much variation, so many new characters being selected for in the first place, and selected against later on, and consequently so many species around us.

   **Acknowledgements:**

   I am employed by INSERM, and I work at the IPBS, which depends both from CNRS and the University of Toulouse. My very special thanks and thoughts go to the late Leigh van Valen, who passed away in October 2010, and had provided me with many bibliographic references and ideas at the early stages of this work, and had also read one of the earliest versions of this essay and provided many helpful suggestions for its improvement. I am also particularly grateful to all those who accepted to read earlier versions of this manuscript for their comments and suggestions that helped me to improve it: Fernando Roch, Patrick Bateson, Chris Grobbelaar, Tanguy Chouard, Pascal Gagneux, Philippe Druet, Gilbert Fournié, Sharat Chandra, William Shields, Justin Leiber and the three referees Pierre Pontarotti, Patrick Nosil, and Eugene Koonin (the first two having also read earlier versions). I am also extremely grateful to the people who run the Google Scholar and JSTOR online services. Without those, I could never have found and accessed the information and bibliography that allowed me to develop and mature the ideas exposed in this essay. I am also grateful to Michael Lynch, Claudia Ricci, Erica Rosenblum and Alan Bittles who sent me copies of their work and replied patiently to my often blunt enquiries.

## Addendum one : Bdelloid Rotifers: A scandal about a ratchet, or a ratchet about a presumed scandal ?

Bdelloid Rotifers, which have been dubbed an "evolutionary scandal" by John Maynard Smith, are the only known example of multi-cellular organisms for which there is absolutely no doubt that they reproduce strictly asexually. They are minuscule females ( < 1mm), who can lay several dozens of parthenogenetic eggs in the course of their 40-day adult lifetime. Bdelloids are found in freshwater and the geological record tells us that they have been around for at least 35 million year. In this sense, they are clearly among the most long lived "taxonomic species" in existence (I specify taxonomic here since the biological species concept only applies to sexual organisms), and they can be found all around the world, testifying of the success of their reproductive strategy. The downside of this is, however, that they probably have extremely limited capacities for evolution, since they have apparently not yielded any more elaborate descendants over that very long period. It therefore seems fair to say that they may well be stuck in an evolutionary dead end, from which more elaborate life forms are extremely unlikely to arise. Outside of their asexual lifestyle, Bdelloids have three very special peculiarities which, I contend, are related to the need to cleanse their diploid genome from recessive mutations:

- They are only found in wet or moist habitats that are prone to successive rounds of desiccation and rehydration. This correlates with what is called anhydrobiosis, i.e. the capacity to survive complete dehydration at any stage of their life cycle.

- They are the most radiation resistant organisms known to date [137], due to an amazing capacity to repair damages to their genomic DNA, which can be explained by the fact that during the dehydration which is part of their natural life cycle, DNA will sustain multiple damages and strand breakage.

- Individuals kept under continuous state of hydration will quite rapidly show reduced fitness compared to individuals undergoing regular cycles of dehydration, which maintain the level of fitness seen in the seeding stock [138]. A very recent study suggests that the main reason for this reduced fitness is presumably due to pathogens such as parasitic yeasts, which the bdelloids are not armed to eliminate. During the desiccation cycles, however, bdelloids will scatter randomly to other locations, where the pathogens will not have followed them, and will then be able to resume their life cycle without the pathogens, at least for a while [139].

My interpretation of the observation that bdelloids are primarily founds in environments that are prone to regular desiccation rather than in permanently hydrated surroundings is that the desiccation cycles could act in place of sexual reproduction to fight off the accumulation of deleterious mutations. Indeed, after DNA has been extensively chopped up by desiccation, DNA repair will involve chromosomal pairing and gene conversion will presumably cause significant homogenisation of the DNA sequences. In addition, I envisage that Bdelloids may have very good DNA repair, but rather low faithfulness in DNA replication. Indeed, this later trait may be required to maintain some level of adaptability in those asexual organisms. The coupled processes of relatively unfaithful DNA replication, together with homogenisation triggered as a result of reiterated DNA damage occurring during desiccation, may thus be replacing sex as a mean to keep some level of adaptability in Bdelloids, whilst fighting off infectious pathogens and Muller's ratchet at the same time. This is in fact exactly equivalent to the lottery that is played by sexual reproduction with a degree of inbreeding, by keeping only those individuals that have at least one good copy of each gene, and eliminating the unlucky ones that get two copies of a bad one. In the short term, this will result in reduced numbers of individuals recovering from desiccation. But given the bdelloid's individual prolificacy, repopulating their environment after a cycle of desiccation does presumably not represent a major challenge, whilst keeping their genome functional must be one ! This is why I suggest that, in Bdelloids, the desiccation cycles would thus act in place of sexual reproduction to fight off the accumulation of deleterious recessive mutations.

And other classes of animals that undergo anhydrobiosis, such as the tardigrades or the darwinulids, may also be taking advantage of desiccation for regular shearing of their DNA to ensure homogenisation of their diploid genomes. Radiations induce damages to DNA that are very similar to those caused by desiccation. In several places on our planet, the use of nuclear power by humans has caused and still causes the natural environment to be exposed to very high levels of radiation. As a rather wild prediction, I would not be surprised if certain asexual forms of life turned out to be able to adapt to those environments, using radiations instead of the cycles of desiccation used by the bdelloids, both to provoke intermittent haploidisation of some of their genome, and to destroy any infectious pathogens.

**Addendum 2 : Three particular examples of the occurrence of haploidy in eukaryotes**:
*How does it feel…. to be on your own ? Bob Dylan*

**Haploid stages**: Within the frame of the biological species concept, the phenomenon of speciation is only relevant to the organisms that can reproduce sexually, i.e. that can go through meiosis. Through the process of meiosis, a diploid cell will become haploid by eliminating half of its chromosomes, and later fuse with another haploid cell to restore a state of diploidy. A critical step in the process leading to meiosis is the pairing of chromosomes, during which many events of recombination occur such as crossing-overs and gene conversion, which ultimately contribute to homogenisation of sequences, and can influence the rate of occurrence of mutations via processes such as biased gene conversion [140]. Depending on the organisms, the haploid state can last for more or less time, and even implicate stages of haploid cell division. Certain classes of organisms, such as yeasts, fungi, algi, many plants and social insects, systematically pass via haploid stages during their life cycles. All these species hence go through the most thorough screen possible for eliminating recessive deleterious mutations, and would not need inbreeding to fight Muller's ratchet. Social insects ( ants, bees, wasps and termites) are known as haplodiploids because the males are haploid, whilst the females are diploid. A proposed explanation for the fact that this strategy has promoted their social behaviour is that, in species where the queen mates with only one male, such as honey bees, the female workers are more related to the offspring of their mother (75%) than to any offspring they would produce themselves if they were to mate ( 50%). Hence, at every generation, half of a social insect's genome goes through a haploid stage that must give rise to a fully fit and sexually active male. The fact that haplodiploid insects do not require inbreeding for the maintenance of their genome is supported by the fact that they actually have a safeguard against inbreeding: according to the complementary allele model, the sex-determining locus of social insects must be heterozygous for the generation of a female [141]. In haploid males, the locus is necessarily hemizygous. If inbreeding takes place, i.e. if the allele of the sex-determining locus carried by the male matches that of one of the two carried by the queen, half of the eggs will be homozygous at the sex-determining locus, and this will give rise to males, but they will be infertile because their offspring would be triploid.

**Sex chromosomes**: Although not all animal species where males and females can be found have sexual chromosomes, this is by far the most common situation. In such species, including ours, the genomes of males and females differ in the chromosomal composition, with either the males being heterogametic (XY, as in mammals, or flies), or the females (ZW, as in certain insects, fish, reptiles and birds). The platypus, the only known egg-laying mammal, carries as many as 10 sex chomosomes (5X and 5Y), which share features with both the mammals and the bird sex chromosomes [142, 143]. Yet another possibility of sex chromosome arrangements is for the males to carry just one copy of the sex chomosome (XO, in certain insects like grasshoppers and roaches), or the females (ZO, in some butterflies), whilst the rest of their genomes is diploid. For all those species, the sex chromosomes they contain will be in a haploid state either all the time (Y and W chromosomes ), or in half of the individuals (X in males and Z in females). For the genes carried by these chromosomes, the accumulation of recessive mutations will hence not be a particular problem. The selective pressures that they are submitted to, and the rate at which these genes evolve has, indeed, been found to differ quite significantly from the genes carried by autosomes ( see [144] for recent review), and the recent comparison of the human and chimpanzee Y chromosomes has revealed an unexpectedly high level of divergence between the two, both in sequence and structure [145]. As developed in section II, the haploid character of sexual chromosomes in heterogametic individuals could be a central factor in allowing selective pressures to give rise to hybrid sterility in those heterogametic individuals whilst remaining silent in homogametic ones (Haldane's rule [58]), thereby favouring the inbreeding that will ultimately lead to speciation.

**Endosymbionts**: Apart from a nuclear envelope, another central characteristic feature of most eukaryotes is that they possess mitochondria, the powerhouses of eukaryotic cells, which provide ATP via respiration. Since Lynn Margulis proposed it in the late 60's, it has been globally accepted that a critical step in the genesis of the ancestral aerobic eukaryote was a symbiotic arrangement whereby an aerobic bacteria, probably related to the rickettsia, was engulfed by the anaerobic ancestor of eukaryotes, which probably helped it to cope with the levels of oxygen which started rising 2.5 billion years ago due to the appearance of photosynthetic cyanobacteria on our planet. This engulfed aerobic bacterium was the ancestor of the mitochondria found in the cytoplasms of virtually all eukaryotes today. On at least three separate occasions, photosynthetic eukaryotes would later arise by the engulfment of cyanobacteria by early eukaryotes, giving rise respectively to the green, red and the glaucophytes'chloroplasts. One remarkable aspect regarding all these endosymbiotic organelles is that they have been maintained as separate entities for billions of years in the cytoplasm of their hosts, where they still replicate by fission, similarly to their bacterial ancestors. And during all that time, although some of their genes have 'migrated' to their hosts' genomes, all those endosymbionts have maintained their own self-replicating circular genomes. Yet, in most metazoan species, the endosymbiotic organelles are inherited from only one parent [146]. The fact that, in yeasts, sexual reproduction results in bi-parental transmission of mitochondria argues in favour of the view that, when sex evolved in the ancestral diploid eukaryote, both parents probably contributed to the offspring's initial stocks of mitochondria. And at first glance, this may seem like a very suitable solution, since having two populations of mitochondria would effectively be equivalent to being diploid, and should hence favour adaptive evolution by promoting the occurrence of new gene combinations. The fact that bi-parental inheritance of mitochondria has almost universally evolved into uni-parental modes (mostly from the mother, but sometimes also from the father) does, however, suggest that bi-parental inheritance of mitochondria must have had more disadvantages than advantages. The first obvious problem would be that it would inevitably lead to Darwinian competition between the two stocks of bacteria, and that the host could end up paying the price of this intestinal wrestling [147]. The second problem is the one related to the subject being discussed here, i.e. the maintenance of the integrity of diploid genomes. Although the main role of mitochondria is respiration, they are also endowed with many other functions such as regulation of cell potential, calcium signalling, apoptosis, and various metabolic pathways. If the stocks of mitochondria were systematically inherited from both parents, they would effectively behave as diploids, and recessive mutations in the genomes of some of them could be tolerated because they would be complemented by the function of the others. But this could not be fixed by recombination between the genomes of the mitochondria because they do not perform sexual reproduction, and hence recombine only rarely. And during mitosis of eukaryotic cells, mitochondria are passed onto daughter cells following simple passive distribution. Over several divisions, many cells will hence end up with only one type of mitochondria. This would not necessarily be very serious for a mono-cellular organism because those unlucky cells inheriting just mutated mitochondria would simply die out and make more room for the others. In certain plants, chloroplasts can be inherited from both parents. In such plants, it is possible to isolate variegated varieties, due to the fact that one of the parents carries mutant chloroplasts that can no longer make chlorophyll. The variegations correspond to areas of the plants that have, randomly, lost the chloroplasts that could make chlorophyll. Such plants are, however, not found in natural environments. For animal mitochondria, it is rather easy to picture how the inheritance of a diploid pool of mitochondria could rapidly become a significant problem rather than an advantage because, for the harmonious development of multi-cellular organisms, if they had inherited a mixed pool of mutated and un-mutated mitochondria, they would end up loosing a significant portion of their cells in certain organs where the mutated mitochondria would have randomly taken over. The final picture that delineates itself from this type of reasoning underlines the close relationship that ties sex and the need to cleanse obligatory diploid genomes off the recessive mutations that they tend to accumulate silently.

**Addendum 3 : The social lifestyle of a lowly amoeba.**
Dictyostelium discoideum (Dd) is an amoeba, which is found in the soil of forests, where it feeds on bacteria. On rare occasions, when Dds of different mating types find themselves growing side by side in conditions of darkness and moderate abundance of nutrients, they undergo sexual reproduction, which involves the formation of a macrocyst [148]. Most of the time, however, Dd amoebas multiply asexually, by mitosis. When food becomes scarce, these unicellular eukaryotes that were until then growing completely independently from one another will gather to form a microscopic slug that can then migrate towards the surface, and form a minuscule plant-like structure, with a stalk and a spore-containing head. Of the 100.000 cells that gathered at the start, around 60 to 70 % will end up as spores, with an increased chance of reaching more suitable environments. But this will be at the cost of 30 to 40 % of the initial stock having sacrificed their chances of survival to differentiate in stalk cells, or other cells types. In the lab, one can see that slugs will form by incorporating amoebae that are not necessarily related to one another, and at first glance, this would seem particularly prone to promote the evolution of selfish behaviour, whereby some individuals would avoid ever becoming stalk cells [149]. This can actually be found under experimental conditions, where the amoebae are grown in bulk, but this is not what is seen in the wild : Dictyostelium amoebae that are found in forest soils are usually all prone to forming well proportioned fruiting bodies, with the optimal proportion of cells sacrificing themselves towards the doomed stalk lineage. I contend that, if Dictyostelids have been able to evolve this social lifestyle, it is because of their capacity to sporulate and disseminate, and hence for single individuals ( or at most a handful of amoebae originating from the same fruiting body) to colonise new isolated niches. The resulting populations must thus be comprised of groups of individuals that are highly related to one another, or even very often clonal. Under such conditions, selfish mutants will be doomed because, when they find themselves on their own, their incapacity to form stalk cells will condemn the fate of their offspring to staying in the same spot. This type of selection can thus be assimilated to group selection, whereby it is not the immediate advantage of an individual withing a group that matters, but the capacity of a group of related individuals to adopt a strategy that will favour the survival of some descendants.

**Referees comments:**

**Eugene V. Koonin:**

This is a very lengthy, very interesting, very provocative essay written with inimitable flare. I believe the main motivation and probably the principal idea of this treatise comes here (quoting from the abstract):"...if so much speciation occurs, it must result from a process of natural selection, whereby it is advantageous for individuals to reproduce preferentially within a group and reduce their breeding with the rest of the population." I plainly refuse to see why wide spread of speciation (in organisms with obligate sexual reproduction, this is the scope of the discussion) implies its adaptive character. Both allopatric and sympatric speciation do not appear to be incompatible with a neutral scenario. This is obvious for the allopatric case but is fully reasonable for the sympatric case as well, e.g., via chromosomal rearrangements caused by spread of mobile elements and other factors that may have nothing to do with selective advantage and adaptation. More generally, I believe that the construction of any evolutionary scenario should start with a neutral null hypothesis. Only when and if the neutral hypothesis is clearly falsified, should one start developing explanations rooted in selection [1-4]. Otherwise, any evolutionary scenario smacks of an adaptationist 'just so story'[1]. This is not meant in a pejorative sense, indeed, this is how evolutionary biology operated for more than a century after Darwin but I believe that in 2011 we are beyond that stage. Again: the task of a work with the claims made here should be not to show that inbreeding 'might' be beneficial and hence speciation 'might' be adaptive (the essay develops perhaps a credible case for that) but rather to show that these processes cannot occur (are highly unlikely) under the neutral model (this is not even attempted in the paper).

The above is by no means intended to deter readers from carefully going through the entire article (its length notwithstanding): it is excellent, thought-provoking reading. Only, I do not accept the conclusions but the work is labelled an essay, so its main value is perhaps not in the conclusions but in stimulating thinking on major and well-explored but still thorny problems such as speciation. In this, the paper truly succeeds.

1. Gould SJ, Lewontin RC: The spandrels of San Marco and the Panglossian paradigm: a critique of the adaptationist programme. Proc R Soc Lond B Biol Sci 1979, 205(1161):581-598.

2. Koonin EV: A non-adaptationist perspective on evolution of genomic complexity or the continued dethroning of man. Cell Cycle 2004, 3(3):280-285.

3. Lynch M: The frailty of adaptive hypotheses for the origins of organismal complexity. Proc Natl Acad Sci U S A 2007, 104 Suppl 1:8597-8604.

4. Koonin EV: The Logic of Chance: The Nature and Origin of Biological Evolution Upper Saddle River, NJ: FT press; 2011.

*Response: I am not only very grateful to Eugene Koonin for his positive appreciation of my work, but even more so for putting his finger with such accuracy on a point of dissention between my views and those of many evolutionary biologist of today. I had indeed not identified this question of 'null hypothesis' previously, and I have therefore not treated it in my essay. Inasmuch as I completely agree with the argument that any sound scientific approach should first aim to disprove the null hypothesis, when it comes to life and evolution, I beg to differ with the view that the null hypothesis rests with things happening simply by chance. Whilst I do completely agree with the views expressed by Gould, Lewontin, Lynch and Koonin in the literature cited above on the ridicule of trying to explain EVERYTHING by means of direct selective advantages, I contend that one should remain careful not to throw the baby out with the bath water. Indeed, as far as I am concerned, what Darwin (and Wallace) established 150 years ago still remains true today: short of a divine intervention, the only way to explain the occurrence of life, and its progressive gain in complexity over time, is via the process of evolution driven by natural selection, and even if this makes me a retrograde conservative in the eyes of some, I stand firmly by my views that the most likely explanation of any evolutionary process (and hence the null hypothesis) lies with the process of natural selection, be it direct or indirect (I would have a lot more to say about the difference between direct vs indirect selection, but this does not seem to be the appropriate place). All in all, I am not saying that neutral mutations cannot reach fixation through non-adaptive processes such as genetic drift and bystander selection and/or cannot lead to the fixation of things like genomic complexity or even certain anatomical features. But when something is found to occur over and over and over again, such as the appearance of reproductive barriers, then I contend that the most likely explanation lies with the possibility of direct natural selection.*

*Incidentally, I have tried to dampen the strength of that particular sentence in the abstract, which now reads:" if so much speciation occurs, the most likely explanation is that there must be conditions where reproductive barriers can be directly selected for."*

**Patrick Nosil** (nominated by Dr Jerzy Jurka):

The manuscript by Joly proposes that the formation of new species occurs by small inbreeding groups budding off from ancestral groups. This process is driven by several advantages of inbreeding, including its ability to purge recessive mutations. This model of speciation is contrasted with more traditional models where new species form by diverging from one another through selection or drift. A number of empirical observations are put forth in support of the inbreeding model. I commented on an earlier version of this article, and many of my smaller suggestions have been incorporated into the submitted article. I thus here focus on a few larger issues, which if considered, would lead to a more balanced (although perhaps not as pointed) article.

1) Disadvantages of inbreeding favoring outbreeding and thus increasing gene flow

The author makes some convincing arguments for some advantages of inbreeding. The author also discusses how inbreeding may often not be as disadvantageous (deleterious) as generally put forth. Although the issues discussed appear not incorrect, it would be fairer to at least provide some discussion of what happens during speciation when indeed inbreeding is deleterious (i.e., reduces fitness). If this is not done, then weaknesses of the proposed model are ignored, leading to a somewhat one-sided treatment of the overall model. For example, when inbreeding does result in reduced fitness, selection could actually favor individuals who outbreed, resulting in an increase in interbreeding (i.e., gene flow) between different populations or species. This inbreeding avoidance mechanism could constrain the divergence of populations, and thus speciation.

Inbreeding avoidance thus might sometimes counteract processes driving divergence and thus stabilize intermediate points in the speciation process. However, this inbreeding mechanism does not apply to allopatric taxa that have no opportunity to outcross, and might not apply during the initial stages of speciation where gene flow is still high and inbreeding depression is not occurring. Thus inbreeding avoidance might affect population divergence, but is not likely to be a universally applicable stabilizing mechanism that always keeps populations at intermediate points in the speciation process.

*Response: Although I am not sure to have followed the above arguments completely, I believe that we are in fact in complete agreement. Whilst I admittedly have gone to great lengths to list and demonstrate the potential advantages of inbreeding, I have also repeatedly tried (and possibly not succeeded ?) to underline its disadvantages, for example in section I,6. If I have not spent so much time in presenting the negative aspects of inbreeding, it is mostly because this has been done so many times before, by so many others, and also because they are rather*

*straightforward to present: once you have said that inbreeding leads to inbreeding depression, and reduces population diversity, I find that there is little else to be said about the disadvantages of inbreeding.*
*As far as inbreeding avoidance is concerned, my point of view is indeed that, although it may contribute to preventing speciation, it is rather the mutation load itself that acts as the main safeguard against rampant speciation. In line with this, in the introductory paragraph of section I, I argue that it is more often incest avoidance than inbreeding avoidance which is being witnessed in natural populations.*

2) Traditional models of speciation

Some of the treatment of the more 'traditional' models of speciation are somewhat inaccurate or slightly misleading (I am not proposing this was done on purpose, but the writing could be modified slightly). This is especially true in the Abstract, which should be modified to clear up a few things. First and foremost, traditional models of speciation did not all see speciation as a passive process, and certainly they did not all propose speciation occurred as a byproduct of random genetic drift. For example, reinforcement speciation is driven by selection against unfit hybrids, which drives the evolution of mating discrimination between populations or species. Even in models of speciation where selection does not favor the evolution of reproductive isolation per se, selection can still play a role. For example, during 'ecological speciation' divergent selection results in divergent adaptation between populations, and these adaptive changes between populations also result in speciation because they happen to generate reproductive isolation. The manuscript, and especially the abstract, should be modified to indicate that previous models of speciation indeed often involved selection, albeit invoked a very different mechanism from the one proposed by Joly.

*Response: I acknowledge that I may have, in many places, overstated the importance of the proposed model beyond what can be proven today. This is because I am personally convinced that most events of speciation do occur via this process, but I have to agree that I am still very far from having proven it. I am therefore very grateful to Patrick Nosil for his help in identifying the places, both in an earlier version of this essay and in this latter one, where changes were advisable or even necessary. I have now modified the abstract, and several other passages in the manuscript to try to present a more tempered view of the possible broad relevance of the model of "speciation by budding". On the subject of reinforcement speciation and unfit hybrids, however, I would like to underline that this picture fits with the model proposed since, if hybrids are unfit, then this suggests that at least one, and possibly several traits of the parents are recessive since they are not maintained in the F1 offspring.*

3) Empirical evidence

The evidence put forth in support of the model strengthen the manuscript, but might be overstated at times. For example, in the abstract it is claimed that 'Most documented cases of speciation in natural populations appear to fit the model proposed....'. Is 'most' really the case, or would 'many' or 'some' be fairer. Without a more formal and quantitative test of the model, it is likely premature to conclude whether most cases of speciation support it.

*Response: Although I personally believe the model proposed here can explain most events of speciation, I fully agree that, at this stage, I have not demonstrated it. I have thus edited the manuscript in several places to replace "most "by" many".*

4) Predictions

In general, more explicit predictions could be put forth that would allow researchers to distinguish the proposed inbreeding model from previously proposed models.

*Response: I have tried to make such predictions in at least two places along the manuscript:*
- *In section IV, which deals with genomic diversity, I predicted that recessive mutations having driven the speciation process should be at the centre of regions of very limited diversity, with the slope of decrease of diversity being even steeper for co-recessive mutations. This prediction should actually become easily testable in the near future with the much expected results of the ongoing 1000 genome project*
- *At the start of section V, I offered the following two predictions i) When speciation occurs, i.e. when a new group gets reproductively isolated from an ancestral population, this should very often be due to one or more recessive advantageous mutation(s) occurring in the new group. ii) Genera that undergo a lot of speciation should be those that carry the lowest mutation loads, correlating to lifestyles favouring inbreeding such as frequent self fertilisation, or very fragmented populations. (and also see footnote 14 on page 28 about the fact that many of these predictions were actually supported by many papers that were already published, but which I discovered only after I had described the model)*

**Pierre Pontarotti:**

This article/review/Hypothesis represents a great amount of work. I am really impressed by the new insights brought by Etienne Joly on a theme that has been extensively debated: Speciation.
I think that the author should go ahead develop his ideas and go for a book.

*Response: Although books have historically been a crucial vector for the dissemination of scientific data and ideas, I personally find that printed books, and copyright restrictions, have now become a major hindrance of scientific progress. With electronic publishing, it is in the interest of all scientists ( both the authors and the readers) that scientific information should be made available for free to anyone who cares to access it. For example, Eugene Koonin's very recent book (see above citation) can be found on Google books, but I only managed to access half the pages. For this reason, I am afraid that I am sternly against the idea of publishing this manuscript in the form of a book, and this is also one of the main reasons why I have chosen Biology Direct, because it is Open Access. As far as I am concerned, the only justification for a printed book nowadays would be if it was aimed at the general public. And I sadly recognise that this is clearly not the case of this lengthy and complicated manuscript which I do not think would make appropriate reading for the layperson on a train journey and even less on the beach.*

Comments: However, I have several criticisms that could be considered to improve the manuscript:
I ) On the form, the paper is especially rich and very difficult to read at the moment, I would organize it a little bit differently:
A) Sympatric versus allopatric speciation
The author starts his paper by stating that most scientists believe that most events of speciation occur via processes of separations and divergences. I do not agree with this statement, please include the classical article : On the origin of species by sympatric speciation by Dieckmann and Doebeli (Nature 1999)
B) Mechanisms leading to sympatric speciation ( note that this chapter has been well developed)

*Response: Although I wish it were not so true, I am afraid that this statement is indeed the reflection of the situation today. As a proof, we can simply turn once more to "Speciation", our preferred reference textbook (C&O , 2004), where the following sentence can be found on page 84 : " While most evolutionists still accept allopatric speciation as the most common mode, others claim that sympatric speciation may be nearly as frequent, or, in some groups, even more frequent". And further along "One can argue that allopatric speciation should be considered the "default" mode of speciation because it is supported by substantial evidence …" .*

*Regarding the suggestion to cite the paper by Dieckmann and Doebeli, the reason I have not included it is because it is a very theoretical paper which I have found to be completely beyond my grasp, and, as explained in the foreword, I make a rule of not citing papers I have not read, or not managed to understand. In the introduction, however, there are already three references to the ongoing up-rise of alternate views advocating the potential importance of sympatric speciation (Via, 2001, Nosil et al. 2005, Mallet et al. 2009), and the paper by Dieckmann and Doebeli is duly referenced in two of those more recent publications. Although the suggestion to rearrange the layout of this manuscript may possibly contribute to making it easier to read, I am afraid that, after more than two years of work on this manuscript, such a major overhaul is well beyond my available capacity today.*

II ) Etienne Joly gave at the end part of his article, many examples arguing for his hypothesis.

The main idea of the author is the Saeptation hypothesis : "initial mutation must occur at some stage which will eventually result in promoting the interbreeding between individuals carrying that mutation rather than with the rest of the population."

When I focus on one of the first example given by Etienne Joly, the Loss of EDA in stickleback, I cannot conclude if this loss has promoted interbreeding between individuals carrying this mutation. Indeed, the inbreeding in that case could be explained by bottleneck and the fixation of the EDA minus genotype by neutral or positive selection. It has been shown that this event (loss of EDA) occurred on a convergent manner, but that still could be explained as bottleneck events.

In this case, the polymorphism decrease link to a speciation events is not due to inbreeding but to a bottleneck events Most of the examples developed here can be challenged this way.

*Response: What is most remarkable in the cases of parallel speciation witnessed in sticklebacks is the existence of reproductive barriers between benthic and limnetic populations within each of several separate lakes, i.e. between groups of fish that independently derived from separate ancestral stocks. Regarding the 'chicken and egg' question, i.e. who came first, it is obviously always the mutation which has to come before saeptation, and in this case it can be affirmed since the very same EDA mutation is found throughout the world, and thus clearly pre-existed the isolation of those fish in their separate lakes. In this sense it is thus not a case of convergent evolution, whilst the loss of a promoter element in the pitx1 gene is (see text).*

*If those events of speciation were just a consequence of tight bottlenecks, some recessive alleles may end up being fixed, but then one would expect that all individuals would carry that one allele, not just the benthic populations. The same reasoning can also apply at the genomic level for the pitx1 gene : if it were just a consequence of bottlenecks, then the genomic diversity should be diminished for the whole genome in isolated populations, but what is found is that dramatically reduced diversity is only found at the level of that particular gene.*

*As is have argued in sections III and IV, bottlenecks such as in the case seen for island (or lake) colonisation will contribute very significantly to promoting speciation because the reduction in the effective population size will result in a decrease in mutation loads, hence lifting the safeguard against further events of speciation. The counter example of this is the case of domestic breeds, which are maintained through repeated bottlenecks, but because they are never subjected to the pressure of their ancestral stocks, they do not develop reproductive barriers against them.*

III ) Concerning the Chapter VI : And What about Homo Sapiens ?

It seems that the author believes in the concept of Race (For example he wrote to date despite this hybridization with Neanderthal and despite the fractionation of race for tens of century )

I recommend the lecture of the following paper: Implications of biogeography of human populations for "Race "and medecine by Tishkoff and Kidd (Nature genetics 2004: 36: S21-27), I do not think that the concept of race can be used for the human species.

*Response: I have read this paper with interest, but I do not think that the authors really question the existence of human races (for example, their figure 4 actually shows a very well supported tree for 37 different such races, based on 80 independent loci). What this paper concludes is that there is too much variation within human genomes, and too much gene flow in modern populations, to make the simplistic classification of people into races useful for biomedicine, for example for the identification of differential risk to disease or pharmaco-sensitivity. This does not, however, rule out the existence of races in human populations.*

**Competing interests:** The author declares that he has no competing interests.

**Bibliographic References:**

1.	van Dyke DL, Weiss L, Roberson JR, Babu VR: **The frequency and mutation rate of balanced autosomal rearrangements in man estimated from prenatal genetic studies for advanced maternal age.** *Am J Hum Genet* 1983, **35:**301-308.
2.	Shields WM: *Philopatry, inbreeding, and the evolution of sex.* State University of New York Press; 1982.
3.	Wright S: **Breeding Structure of Populations in Relation to Speciation.** *The American Naturalist* 1940, **74:**232-248.
4.	Wright S: **The genetical structure of populations.** *Annals of Human Genetics* 1949, **15:**323-354.
5.	Carson HL: **Genetic conditions which promote or retard the formation of species.** *Cold Spring Harb Symp Quant Biol* 1959, **24:**87-105.
6.	Mallet J, Meyer A, Nosil P, Feder JL: **Space, sympatry and speciation.** *J Evol Biol* 2009, **22:**2332-2341.
7.	Nosil P, Vines TH, Funk DJ: **Perspective: Reproductive isolation caused by natural selection against immigrants from divergent habitats.** *Evolution* 2005, **59:**705-719.
8.	Via S: **Sympatric speciation in animals: the ugly duckling grows up.** *Trends Ecol Evol* 2001, **16:**381-390.
9.	Dobzhansky T: **Speciation as a Stage in Evolutionary Divergence.** *The American Naturalist* 1940, **74:**312-321.
10.	Rice WR, Hostert EE: **Laboratory Experiments on Speciation: What Have We Learned in 40 Years?** *Evolution* 1993, **47:**1637-1653.
11.	Fisher RA: **Average excess and average effect of a gene substitution.** *Annals of Eugenics* 1941, **11:**53-63.
12.	Wright S: *Evolution and the Genetics of Populations, Variability Within and Among Natural Populations* Chicago: Univ. of Chicago Press; 1978.
13.	Bateson P: **Sexual imprinting and optimal outbreeding.** *Nature* 1978, **273:**659-660.
14.	Bateson P: **Preferences for cousins in Japanese quail.** *Nature* 1982, **295:**236-237.
15.	Bateson P: **Optimal Outbreeding.** In *Mate Choice*. Edited by Bateson P. Cambridge: Cambridge University Press; 1983: 257-277
16.	Price MV, Waser NM: **Pollen dispersal and optimal outcrossing in Delphinium nelsoni.** *Nature* 1979, **277:**294-297.
17.	Grobbelaar CS: **Crisis inbreeding: rapid evolution in large, ecologically intact populations ?** *Evolutionary Theory* 1989, **8:**365-395.
18.	Bittles AH: **A community genetics perspective on consanguineous marriage.** *Community Genet* 2008, **11:**324-330.
19.	Bittles AH, Black ML: **Evolution in health and medicine Sackler colloquium: Consanguinity, human evolution, and complex diseases.** *Proc Natl Acad Sci U S A* 2010, **107 Suppl 1:**1779-1786.
20.	Helgason A, Palsson S, Gudbjartsson DF, Kristjansson T, Stefansson K: **An association between the kinship and fertility of human couples.** *Science* 2008, **319:**813-816.
21.	Thornhill NW: *The Natural history of inbreeding and outbreeding : theoretical and empirical perspectives*. Chicago: University of Chicago Press; 1993.
22.	Kokko H, Ots I: **When not to avoid inbreeding.** *Evolution* 2006, **60:**467-475.
23.	Fisher RA: *The genetical theory of natural selection*. Oxford: Clarendon Press; 1930.
24.	Nachman MW, Crowell SL: **Estimate of the mutation rate per nucleotide in humans.** *Genetics* 2000, **156:**297-304.
25.	Gould SJ: *Hen's Teeth and Horse's Toes*. New York: W. W. Norton & Company; 1984.
26.	Stephens JC, Reich DE, Goldstein DB, Shin HD, Smith MW, Carrington M, Winkler C, Huttley GA, Allikmets R, Schriml L, et al: **Dating the origin of the CCR5-Delta32 AIDS-resistance allele by the coalescence of haplotypes.** *Am J Hum Genet* 1998, **62:**1507-1515.
27.	Muller HJ: **The relation of recombination to mutational advance.** *Mutation Research/Fundamental and Molecular Mechanisms of Mutagenesis* 1964, **1:**2-9.
28.	Muller HJ: **Our load of mutations.** *Am J Hum Genet* 1950, **2:**111-176.
29.	Kacser H, Burns JA: **The molecular basis of dominance.** *Genetics* 1981, **97:**639-666.
30.	Wright S: **Physiological and Evolutionary Theories of Dominance.** *The American Naturalist* 1934, **68:**24-53.
31.	Swindell WR, Bouzat JL: **Ancestral inbreeding reduces the magnitude of inbreeding depression in Drosophila melanogaster.** *Evolution* 2006, **60:**762-767.
32.	Lynch M: **The origins of eukaryotic gene structure.** *Mol Biol Evol* 2006, **23:**450-468.
33.	Lynch M, Burger R, Butcher D, Gabriel W: **The mutational meltdown in asexual populations.** *J Hered* 1993, **84:**339-344.
34.	Roach JC, Glusman G, Smit AF, Huff CD, Hubley R, Shannon PT, Rowen L, Pant KP, Goodman N, Bamshad M, et al: **Analysis of genetic inheritance in a family quartet by whole-genome sequencing.** *Science* 2010, **328:**636-639.
35.	van Valen L: **A new evolutionary law.** *Evol Theory* 1973, **1:**1-30.
36.	Eldredge N, Gould SJ: **Punctuated equilibria: an alternative to phyletic gradualism**
.	In *Model in Paleobiology*. Edited by Schopg TJM. San Francisco: Freeman, Cooper and Co.; 1972: 82-115
37.	Gould SJ, Eldredge N: **Punctuated Equilibria: The Tempo and Mode of Evolution Reconsidered.** *Paleobiology* 1977, **3:**115-151.
38.	Maynard Smith J: *The Evolution of Sex*. Cambridge: Cambridge Univ. Press; 1978.
39.	Dawkins R: *The selfish gene*. Oxford, UK: Oxford University Press; 1976.

40. Hamilton WD: **The genetical evolution of social behaviour. II.** *J Theor Biol* 1964, **7:**17-52.
41. Hamilton WD: **The genetical evolution of social behaviour. I.** *J Theor Biol* 1964, **7:**1-16.
42. Wilson DS, Wilson EO: **Rethinking the theoretical foundation of sociobiology.** *Q Rev Biol* 2007, **82:**327-348.
43. Queller DC, Strassmann JE: **Beyond society: the evolution of organismality.** *Philos Trans R Soc Lond B Biol Sci* 2009, **364:**3143-3155.
44. Nei M, Maruyama T, Wu CI: **Models of evolution of reproductive isolation.** *Genetics* 1983, **103:**557-579.
45. Lawlor DA, Ward FE, Ennis PD, Jackson AP, Parham P: **HLA-A and B polymorphisms predate the divergence of humans and chimpanzees.** *Nature* 1988, **335:**268-271.
46. Figueroa F, Gunther E, Klein J: **MHC polymorphism pre-dating speciation.** *Nature* 1988, **335:**265-267.
47. Pujol B, Zhou SR, Sanchez Vilas J, Pannell JR: **Reduced inbreeding depression after species range expansion.** *Proc Natl Acad Sci U S A* 2009, **106:**15379-15383.
48. Orr HA: **The population genetics of speciation: the evolution of hybrid incompatibilities.** *Genetics* 1995, **139:**1805-1813.
49. Moyle LC, Nakazato T: **Hybrid incompatibility "snowballs" between Solanum species.** *Science* 2010, **329:**1521-1523.
50. Matute DR, Butler IA, Turissini DA, Coyne JA: **A test of the snowball theory for the rate of evolution of hybrid incompatibilities.** *Science* 2010, **329:**1518-1521.
51. Rieseberg LH: **Chromosomal rearrangements and speciation.** *Trends Ecol Evol* 2001, **16:**351-358.
52. Joly AL, Le Rolle AL, Gonzalez AL, Mehling WJ, Stevens WJ, Coadwell WJ, Hunig JC, Howard JC, Butcher GW: **Co-evolution of rat TAP transporters and MHC class I RT1-A molecules.** *Curr Biol* 1998, **8:**169-172.
53. Jacobs PA: **Mutation rates of structural chromosome rearrangements in man.** *Am J Hum Genet* 1981, **33:**44-54.
54. Oliver-Bonet M, Navarro J, Carrera M, Egozcue J, Benet J: **Aneuploid and unbalanced sperm in two translocation carriers: evaluation of the genetic risk.** *Mol Hum Reprod* 2002, **8:**958-963.
55. Noor MA, Grams KL, Bertucci LA, Reiland J: **Chromosomal inversions and the reproductive isolation of species.** *Proc Natl Acad Sci U S A* 2001, **98:**12084-12088.
56. Nosil P, Crespi BJ, Sandoval CP: **Reproductive isolation driven by the combined effects of ecological adaptation and reinforcement.** *Proc Biol Sci* 2003, **270:**1911-1918.
57. Turelli M, Orr HA: **The Dominance Theory of Haldane's Rule.** *Genetics* 1995, **140:**389-402.
58. Haldane JBS: **Sex-ratio and unisexual sterility in hybrid animals.** *Journal of Genetics* 1922, **12:**101-109.
59. Thorpe RS, Surget-Groba Y, Johansson H: **Genetic tests for ecological and allopatric speciation in anoles on an island archipelago.** *PLoS Genet* 2010, **6:**e1000929.
60. Fitzpatrick MJ: **Pleiotropy and the genomic location of sexually selected genes.** *Am Nat* 2004, **163:**800-808.
61. Price TD, Bouvier MM: **The evolution of F1 postzygotic incompatibilities in birds.** *Evolution* 2002, **56:**2083-2089.
62. Templeton AR: **The theory of speciation via the founder principle.** *Genetics* 1980, **94:**1011-1038.
63. Irwin DE, Irwin JH, Price TD: **Ring species as bridges between microevolution and speciation.** *Genetica* 2001, **112-113:**223-243.
64. Facon B, Hufbauer RA, Tayeh A, Loiseau A, Lombaert E, Vitalis R, Guillemaud T, Lundgren JG, Estoup A: **Inbreeding depression is purged in the invasive insect Harmonia axyridis.** *Curr Biol* 2011, **21:**424-427.
65. Ghaderi D, Springer SA, Ma F, Cohen M, Secrest P, Taylor RE, Varki A, Gagneux P: **Sexual selection by female immunity against paternal antigens can fix loss of function alleles.** *Proc Natl Acad Sci U S A* 2011, **108:**17743-17748.
66. McLean CY, Reno PL, Pollen AA, Bassan AI, Capellini TD, Guenther C, Indjeian VB, Lim X, Menke DB, Schaar BT, et al: **Human-specific loss of regulatory DNA and the evolution of human-specific traits.** *Nature* 2011, **471:**216-219.
67. Nosil P, Funk DJ, Ortiz-Barrientos D: **Divergent selection and heterogeneous genomic divergence.** *Mol Ecol* 2009, **18:**375-402.
68. The_1000_Genomes_Project_Consortium: **A map of human genome variation from population-scale sequencing.** *Nature* 2010, **467:**1061-1073.
69. Navarro A, Barton NH: **Chromosomal speciation and molecular divergence--accelerated evolution in rearranged chromosomes.** *Science* 2003, **300:**321-324.
70. May RM: **How many species inhabit the earth ?** *Scientific American* 1992, **267:**42-48.
71. Makarieva AM, Gorshkov VG: **On the dependence of speciation rates on species abundance and characteristic population size.** *J Biosci* 2004, **29:**119-128.
72. Raup DM: **A Kill Curve For Phanerozoic Marine Species.** *Paleobiology* 1991, **17:**37-48.
73. van Valen L: **Group Selection, Sex, and Fossils.** *Evolution* 1975, **29:**87-94.
74. van Valen L, Sloan RE: **The Extinction of the Multituberculates.** *Systematic Zoology* 1966, **15:**261-278.
75. Jansen VA, van Baalen M: **Altruism through beard chromodynamics.** *Nature* 2006, **440:**663-666.
76. Dettman JR, Anderson JB, Kohn LM: **Divergent adaptation promotes reproductive isolation among experimental populations of the filamentous fungus Neurospora.** *BMC Evol Biol* 2008, **8:**35.
77. Gatenby RA, Frieden BR: **The role of non-genomic information in maintaining thermodynamic stability in living systems.** *Math Biosci Eng* 2005, **2:**43-51.

78.     Venditti C, Meade A, Pagel M: **Phylogenies reveal new interpretation of speciation and the Red Queen.** *Nature* 2010, **463:**349-352.

79.     Silvertown J, Servaes C, Biss P, Macleod D: **Reinforcement of reproductive isolation between adjacent populations in the Park Grass Experiment.** *Heredity* 2005, **95:**198-205.

80.     Barluenga M, Stolting KN, Salzburger W, Muschick M, Meyer A: **Sympatric speciation in Nicaraguan crater lake cichlid fish.** *Nature* 2006, **439:**719-723.

81.     Seehausen O, Alphen JJMv, Witte F: **Cichlid Fish Diversity Threatened by Eutrophication That Curbs Sexual Selection.** *Science* 1997, **277:**1808-1811.

82.     Roberts RB, Ser JR, Kocher TD: **Sexual conflict resolved by invasion of a novel sex determiner in Lake Malawi cichlid fishes.** *Science* 2009, **326:**998-1001.

83.     Thunken T, Bakker TC, Baldauf SA, Kullmann H: **Active inbreeding in a cichlid fish and its adaptive significance.** *Curr Biol* 2007, **17:**225-229.

84.     Landry C, Garant D, Duchesne P, Bernatchez L: **'Good genes as heterozygosity': the major histocompatibility complex and mate choice in Atlantic salmon (Salmo salar).** *Proc Biol Sci* 2001, **268:**1279-1285.

85.     Yeates SE, Einum S, Fleming IA, Megens H-J, Stet RJM, Hindar K, Holt WV, Van Look KJW, Gage MJG: **Atlantic salmon eggs favour sperm in competition that have similar major histocompatibility alleles.** *Proceedings Biological sciences / The Royal Society* 2009, **276:**559-566.

86.     Olsen KH, Grahn M, Lohm J, Langefors A: **MHC and kin discrimination in juvenile Arctic charr, Salvelinus alpinus (L.).** *Anim Behav* 1998, **56:**319-327.

87.     Kelley J, Walter L, Trowsdale J: **Comparative genomics of major histocompatibility complexes.** *Immunogenetics* 2005, **56:**683-695.

88.     Reusch TB, Haberli MA, Aeschlimann PB, Milinski M: **Female sticklebacks count alleles in a strategy of sexual selection explaining MHC polymorphism.** *Nature* 2001, **414:**300-302.

89.     Milinski M, Griffiths S, Wegner KM, Reusch TB, Haas-Assenbaum A, Boehm T: **Mate choice decisions of stickleback females predictably modified by MHC peptide ligands.** *Proc Natl Acad Sci U S A* 2005, **102:**4414-4418.

90.     Frommen JG, Bakker TCM: **Inbreeding avoidance through non-random mating in sticklebacks.** *Biology Letters* 2006, **2:**232-235.

91.     Boughman JW, Rundle HD, Schluter D: **Parallel Evolution of Sexual Isolation in Sticklebacks.** *Evolution* 2005, **59:**361-373.

92.     Rundle HD, Nagel L, Wenrick Boughman J, Schluter D: **Natural selection and parallel speciation in sympatric sticklebacks.** *Science* 2000, **287:**306-308.

93.     Colosimo PF, Hosemann KE, Balabhadra S, Villarreal G, Jr., Dickson M, Grimwood J, Schmutz J, Myers RM, Schluter D, Kingsley DM: **Widespread parallel evolution in sticklebacks by repeated fixation of Ectodysplasin alleles.** *Science* 2005, **307:**1928-1933.

94.     Chan YF, Marks ME, Jones FC, Villarreal G, Jr., Shapiro MD, Brady SD, Southwick AM, Absher DM, Grimwood J, Schmutz J, et al: **Adaptive evolution of pelvic reduction in sticklebacks by recurrent deletion of a Pitx1 enhancer.** *Science* 2010, **327:**302-305.

95.     Abzhanov A, Protas M, Grant BR, Grant PR, Tabin CJ: **Bmp4 and morphological variation of beaks in Darwin's finches.** *Science* 2004, **305:**1462-1465.

96.     Abzhanov A, Kuo WP, Hartmann C, Grant BR, Grant PR, Tabin CJ: **The calmodulin pathway and evolution of elongated beak morphology in Darwin's finches.** *Nature* 2006, **442:**563-567.

97.     Grant BR, Grant PR: **Fission and fusion of Darwin's finches populations.** *Philos Trans R Soc Lond B Biol Sci* 2008, **363:**2821-2829.

98.     Sherborne AL, Thom MD, Paterson S, Jury F, Ollier WE, Stockley P, Beynon RJ, Hurst JL: **The genetic basis of inbreeding avoidance in house mice.** *Curr Biol* 2007, **17:**2061-2066.

99.     Cheetham SA, Smith AL, Armstrong SD, Beynon RJ, Hurst JL: **Limited variation in the major urinary proteins of laboratory mice.** *Physiol Behav* 2009, **96:**253-261.

100.    Roberts SA, Simpson DM, Armstrong SD, Davidson AJ, Robertson DH, McLean L, Beynon RJ, Hurst JL: **Darcin: a male pheromone that stimulates female memory and sexual attraction to an individual male's odour.** *BMC Biol* 2010, **8:**75.

101.    Karn RC, Laukaitis CM: **The Mechanism of Expansion and the Volatility it created in Three Pheromone Gene Clusters in the Mouse (Mus musculus) Genome.** *Genome Biol Evol* 2009, **2009:**494-503.

102.    Logan DW, Marton TF, Stowers L: **Species specificity in major urinary proteins by parallel evolution.** *PLoS One* 2008, **3:**e3280.

103.    Bush GL, Case SM, Wilson AC, Patton JL: **Rapid speciation and chromosomal evolution in mammals.** *Proc Natl Acad Sci U S A* 1977, **74:**3942-3946.

104.    Kerth G: **Causes and Consequences of Sociality in Bats.** *BioScience* 2008, **58:**737-746.

105.    Capanna E, Castiglia R: **Chromosomes and speciation in Mus musculus domesticus.** *Cytogenet Genome Res* 2004, **105:**375-384.

106.    Hauffe HC, Panithanarak T, Dallas JF, Pialek J, Gunduz I, Searle JB: **The tobacco mouse and its relatives: a "tail" of coat colors, chromosomes, hybridization and speciation.** *Cytogenet Genome Res* 2004, **105:**395-405.

107.    Ciszek D: **New colony formation in the "highly inbred" eusocial naked mole-rat: outbreeding is preferred.** *Behav Ecol* 2000, **11:**1-6.

108.    Peacock MM, Smith AT: **Nonrandom mating in pikas Ochotona princeps: evidence for inbreeding between individuals of intermediate relatedness.** *Mol Ecol* 1997, **6:**801-811.

109.    Bretman A, Newcombe D, Tregenza T: **Promiscuous females avoid inbreeding by controlling sperm storage.** *Mol Ecol* 2009, **18:**3340-3345.

110.    Simmons LW, Beveridge M, Wedell N, Tregenza T: **Postcopulatory inbreeding avoidance by female crickets only revealed by molecular markers.** *Mol Ecol* 2006, **15:**3817-3824.

111.    Dobzhansky T: **How Do the Genetic Loads Affect the Fitness of Their Carriers in Drosophila Populations?** *The American Naturalist* 1964, **98:**151-166.

112.    Brown KM, Burk LM, Henagan LM, Noor MA: **A test of the chromosomal rearrangement model of speciation in Drosophila pseudoobscura.** *Evolution* 2004, **58:**1856-1860.

113.    Linn CE, Jr., Dambroski HR, Feder JL, Berlocher SH, Nojima S, Roelofs WL: **Postzygotic isolating factor in sympatric speciation in Rhagoletis flies: reduced response of hybrids to parental host-fruit odors.** *Proc Natl Acad Sci U S A* 2004, **101:**17753-17758.

114.    White MJD: **Models of Speciation.** *Science* 1968, **159:**1065-1070.

115.    White MJD: **Chain Processes in Chromosomal Speciation.** *Systematic Zoology* 1978, **27:**285-298.

116.    Chamberlain NL, Hill RI, Kapan DD, Gilbert LE, Kronforst MR: **Polymorphic butterfly reveals the missing link in ecological speciation.** *Science* 2009, **326:**847-850.

117.    Rieseberg LH, Willis JH: **Plant speciation.** *Science* 2007, **317:**910-914.

118.    Baker HG: **Self-Compatibility and Establishment After 'Long-Distance' Dispersal.** *Evolution* 1955, **9:**347-349.

119.    Heilbuth JC: **Lower Species Richness in Dioecious Clades.** *The American Naturalist* 2000, **156:**221-241.

120.    Goldberg EE, Kohn JR, Lande R, Robertson KA, Smith SA, Igic B: **Species selection maintains self-incompatibility.** *Science* 2010, **330:**493-495.

121.    Baker HG: **Reproductive methods as factors in speciation in flowering plants.** *Cold Spring Harb Symp Quant Biol* 1959, **24:**177-191.

122.    Foxe JP, Slotte T, Stahl EA, Neuffer B, Hurka H, Wright SI: **Recent speciation associated with the evolution of selfing in Capsella.** *Proc Natl Acad Sci U S A* 2009, **106:**5241-5245.

123.    Bradshaw HD, Schemske DW: **Allele substitution at a flower colour locus produces a pollinator shift in monkeyflowers.** *Nature* 2003, **426:**176-178.

124.    Burger WC: **The Species Concept in Quercus.** *Taxon* 1975, **24:**45-50.

125.    Curtu AL, Gailing O, Finkeldey R: **Evidence for hybridization and introgression within a species-rich oak (Quercus spp.) community.** *BMC Evol Biol* 2007, **7:**218.

126.    Rushton B: **Natural hybridization within the genus Quercus L.** *Ann For Sci* 1993, **50:**73s-90s.

127.    Rosenblum EB, Rompler H, Schoneberg T, Hoekstra HE: **Molecular and functional basis of phenotypic convergence in white lizards at White Sands.** *Proc Natl Acad Sci U S A* 2010, **107:**2113-2117.

128.    Rosenblum EB, Harmon LJ: **"Same same but different": replicated ecological speciation at White Sands.** *Evolution* 2011, **65:**946-960.

129.    Green RE, Krause J, Briggs AW, Maricic T, Stenzel U, Kircher M, Patterson N, Li H, Zhai W, Fritz MH, et al: **A draft sequence of the Neandertal genome.** *Science* 2010, **328:**710-722.

130.    Reich D, Green RE, Kircher M, Krause J, Patterson N, Durand EY, Viola B, Briggs AW, Stenzel U, Johnson PLF, et al: **Genetic history of an archaic hominin group from Denisova Cave in Siberia.** *Nature* 2010, **468:**1053-1060.

131.    Rasmussen M, Li Y, Lindgreen S, Pedersen JS, Albrechtsen A, Moltke I, Metspalu M, Metspalu E, Kivisild T, Gupta R, et al: **Ancient human genome sequence of an extinct Palaeo-Eskimo.** *Nature* 2010, **463:**757-762.

132.    Campbell H, Rudan I, Bittles AH, Wright AF: **Human population structure, genome autozygosity and human health.** *Genome Med* 2009, **1:**91.

133.    Nalls MA, Simon-Sanchez J, Gibbs JR, Paisan-Ruiz C, Bras JT, Tanaka T, Matarin M, Scholz S, Weitz C, Harris TB, et al: **Measures of autozygosity in decline: globalization, urbanization, and its implications for medical genetics.** *PLoS Genet* 2009, **5:**e1000415.

134.    Bittles AH, Neel JV: **The costs of human inbreeding and their implications for variations at the DNA level.** *Nat Genet* 1994, **8:**117-121.

135.    Haldane J: **The cost of natural selection.** *Journal of Genetics* 1957, **55:**511-524.

136.    Rodriguez-Munoz R, Bretman A, Slate J, Walling CA, Tregenza T: **Natural and sexual selection in a wild insect population.** *Science* 2010, **328:**1269-1272.

137.    Gladyshev E, Meselson M: **Extreme resistance of bdelloid rotifers to ionizing radiation.** *Proc Natl Acad Sci U S A* 2008, **105:**5139-5144.

138.    Ricci C, Caprioli M, Fontaneto D: **Stress and fitness in parthenogens: is dormancy a key feature for bdelloid rotifers?** *BMC Evol Biol* 2007, **7 Suppl 2:**S9.

139. Wilson CG, Sherman PW: **Anciently asexual bdelloid rotifers escape lethal fungal parasites by drying up and blowing away.** *Science* 2010, **327:**574-576.
140. Duret L, Galtier N: **Biased gene conversion and the evolution of mammalian genomic landscapes.** *Annu Rev Genomics Hum Genet* 2009, **10:**285-311.
141. van Wilgenburg E, Driessen G, Beukeboom LW: **Single locus complementary sex determination in Hymenoptera: an "unintelligent" design?** *Front Zool* 2006, **3:**1.
142. Grutzner F, Rens W, Tsend-Ayush E, El-Mogharbel N, O'Brien PC, Jones RC, Ferguson-Smith MA, Marshall Graves JA: **In the platypus a meiotic chain of ten sex chromosomes shares genes with the bird Z and mammal X chromosomes.** *Nature* 2004, **432:**913-917.
143. Rens W, Grutzner F, O'Brien P C, Fairclough H, Graves JA, Ferguson-Smith MA: **Resolution and evolution of the duck-billed platypus karyotype with an X1Y1X2Y2X3Y3X4Y4X5Y5 male sex chromosome constitution.** *Proc Natl Acad Sci U S A* 2004, **101:**16257-16261.
144. Qvarnstrom A, Bailey RI: **Speciation through evolution of sex-linked genes.** *Heredity* 2009, **102:**4-15.
145. Hughes JF, Skaletsky H, Pyntikova T, Graves TA, van Daalen SKM, Minx PJ, Fulton RS, McGrath SD, Locke DP, Friedman C, et al: **Chimpanzee and human Y chromosomes are remarkably divergent in structure and gene content.** *Nature* 2010, **463:**536-539.
146. Neiman M, Taylor DR: **The causes of mutation accumulation in mitochondrial genomes.** *Proc Biol Sci* 2009, **276:**1201-1209.
147. Hastings IM: **Population genetic aspects of deleterious cytoplasmic genomes and their effect on the evolution of sexual reproduction.** *Genet Res* 1992, **59:**215-225.
148. Chang MT, Raper KB: **Mating types and macrocyst formation in Dictyostelium.** *J Bacteriol* 1981, **147:**1049-1053.
149. Pal C, Papp B: **Selfish cells threaten multicellular life.** *Trends Ecol Evol* 2000, **15:**351-352.